%% file: main.tex
\documentclass[onecolumn,sort&compress,numbers]{els-mrw} 

\usepackage{amsmath,amssymb,amsfonts,amsthm,makeidx,graphicx}
\usepackage{txfonts}
\usepackage{helvet}
\usepackage{graphicx} 
\usepackage{subcaption}
\usepackage{slashed}
\usepackage{bbold}
\usepackage[hidelinks]{hyperref}

\def\als{\alpha_s}
\def\mus{\mu^2}
\def\ycut{y_{\mathrm{cut}}}
\newcommand{\nn}{\nonumber}

\def\rd{\mathrm{d}}
\def\eps{\varepsilon}
\def\muf{\mu_{\mathrm{F}}}
\def\mur{\mu_{\mathrm{R}}}
\def\flux{\frac{1}{2\hat{s}}}

\newcommand\tr[1]{\mathrm{Tr}\left\{ {#1} \right\}}

\newcommand{\secref}[1]{Sec.\,\,\ref{#1}}  
\newcommand{\figref}[1]{Fig.\,\,\ref{#1}}  
\newcommand{\tabref}[1]{Table\,\,\ref{#1}}  
\newcommand{\equref}[1]{Eq.\,\,(\ref{#1})}  

\begin{document}


\chapter{Perturbative QCD}\label{chap1}

\author[1]{Gudrun Heinrich}%
\author[2]{Anton Olsson}%

\address[1]{\orgname{Karlsruhe Institute of Technology}, \orgdiv{Institute for Theoretical Physics}, \orgaddress{Wolfgang-Gaede-Str.~1, 76131 Karlsruhe, Germany}}

\articletag{Chapter Article tagline: update of previous edition, reprint.}

\maketitle

\begin{abstract}[Abstract]
We give an introduction to perturbative Quantum Chromodynamics, focusing on a pedagogical description of concepts and methods to calculate cross sections measured at high energy colliders. After introducing basic concepts that allow for a perturbative expansion, such as factorisation and asymptotic freedom, we introduce loop integrals and the treatment of ultraviolet and infrared divergences in QCD. The definition of jets and event shape observables is also discussed. Finally, we give a brief overview of the current state of the art.
\end{abstract}

\begin{keywords}
 	Perturbation theory\sep QCD \sep loop integrals\sep infrared divergences \sep higher order corrections
\end{keywords}



\section*{Objectives}
\begin{itemize}
	\item  The QCD Lagrangian is introduced and the factorisation of perturbative and non-perturbative contributions to hadronic cross sections is described.
    \item The perturbative expansion of hard scattering cross sections is introduced.
	\item It is explained how scattering amplitudes and cross sections are constructed from Feynman rules.
	\item The calculation of perturbative corrections is described, with special emphasis on the treatment of infrared divergences in QCD.
    \item Jets and event shapes are introduced.
    \item The current state of the art is briefly reviewed.
\end{itemize}

\section{Introduction}\label{intro}

\input{intro.tex}

\section{The QCD Lagrangian}
\label{sec:QCD}
QCD is a non-Abelian gauge theory described by the Lagrangian \cite{Ellis:1996mzs}
\begin{equation}
\label{eq:L_QCD}
    \mathcal{L}_{\mathrm{QCD}} = -\frac{1}{4} F^a_{\mu\nu}F^{a\,\mu\nu} + \sum_f \bar{q}_f(i\gamma_{\mu} D^{\mu} -m_f)q_f - \frac{1}{2\xi} \partial_{\mu} A^{a\,\mu} \partial^{\nu} A^a_{\nu} + \partial_{\mu} \Bar{c}^a (\delta^{ac}\partial_{\mu} + g_s f^{abc}A^b_{\mu}) \, c^c,
\end{equation}
where the field-strength tensor and covariant derivative are respectively defined by
\begin{equation}
     F^a_{\mu\nu} = \partial_{\mu} A_{\nu}^a - \partial_{\nu} A_{\mu}^a - g_s f^{abc} A^b_{\mu}A^c_{\nu}, \quad D_{\mu} = \partial_{\mu} + ig_sA_{\mu}^a t^a.
\end{equation}
The $t^a$ are generators of $SU(3)$ in the fundamental representation. They are defined by the commutation relation $[t^a,t^b] = i f^{abc} t^c$, where $f^{abc}$ are the totally antisymmetric structure constants.
The first term in $\mathcal{L}_{\mathrm{QCD}}$ describes the pure gluon dynamics. It involves a factor $g_s f^{abc} A^b_{\mu}A^c_{\nu}$ which encodes a characteristic feature of non-Abelian theories, namely the presence of self-interactions among the gauge bosons. The second term is a sum over quark flavours, where $m_f$ is the mass of the quark of flavour $f$. It includes the covariant derivative, which generates the interactions between gluons and quarks through the $ig_sA_{\mu}^a t^a$ term. The symbol $\gamma^{\mu}$ denotes the Dirac matrices which are defined by the anti-commutation relation $\{ \gamma^{\mu}, \gamma^{\nu} \} = 2 g^{\mu\nu}$ (Clifford algebra).

The last two terms in \equref{eq:L_QCD} are related to the treatment of redundant degrees of freedom of the theory, since physical gluons only have two degrees of freedom (the transverse polarisations). The third term is a gauge-fixing term and $\xi$ is a so-called gauge parameter. Its value is arbitrary and must not affect physical predictions. A common choice is the Feynman gauge where $\xi = 1$, since this leads to a simple form of the gluon propagator. The fourth term involves the so-called Faddeev-Popov ghost fields \cite{Faddeev:1967fc}, which is a gauge dependent term that is necessary to cancel unphysical degrees of freedom. Ghost fields are unphysical and only appear as virtual states. They are constructed such that they exactly cancel the unphysical degrees of freedom corresponding to longitudinal and time-like polarisations of gluons. Additionally, the ghost fields make the QCD Lagrangian invariant under a global transformation parametrised by a Grassmann-valued parameter $\theta$, so-called BRST-transformations, 
discovered by Becchi, Rouet, Stora and Tyutin~\cite{Becchi:1974md,Tyutin:1975qk}. The BRST invariance is important for the proofs of unitarity and
renormalisability of QCD.
The ghost fields decouple in axial (physical) gauges, however this leads to a more complicated gluon propagator and therefore increases computational complexity. 

$\mathcal{L}_{\mathrm{QCD}}$ is used to derive the Feynman rules of QCD. They are the building blocks of scattering amplitudes and are used to construct Feynman diagrams, see \secref{sec:perturbation_theory}. The vertex and propagator rules of QCD can be seen in \tabref{tab:feynman_rules1} and \tabref{tab:feynman_rules2} respectively. 
\begin{table}[b]
    \centering
    \begin{tabular}{c|c|c|c}
         \raisebox{-0.45\totalheight}{\includegraphics[width=0.2\textwidth]{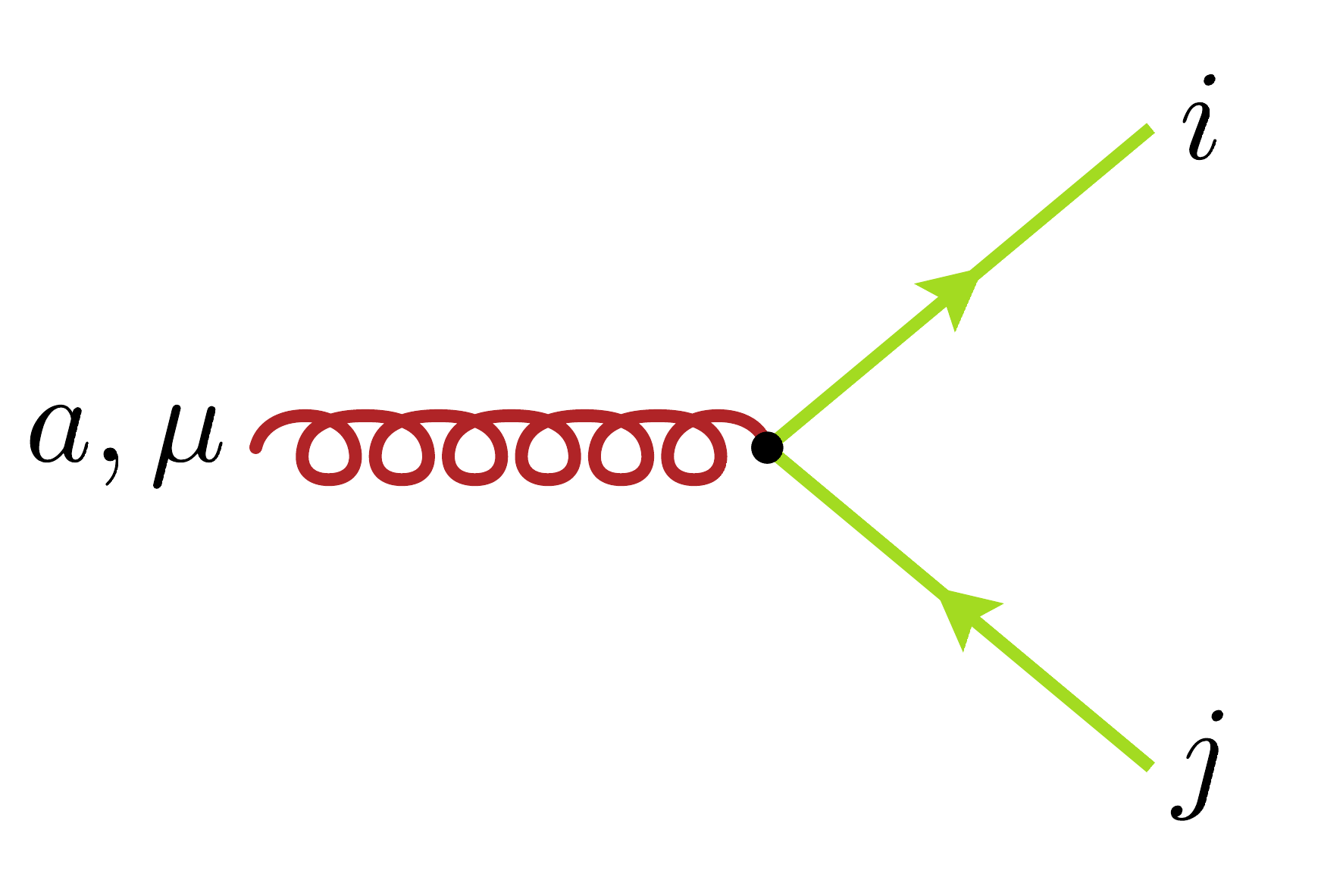}} & \raisebox{-0.45\totalheight}{\includegraphics[width=0.2\textwidth]{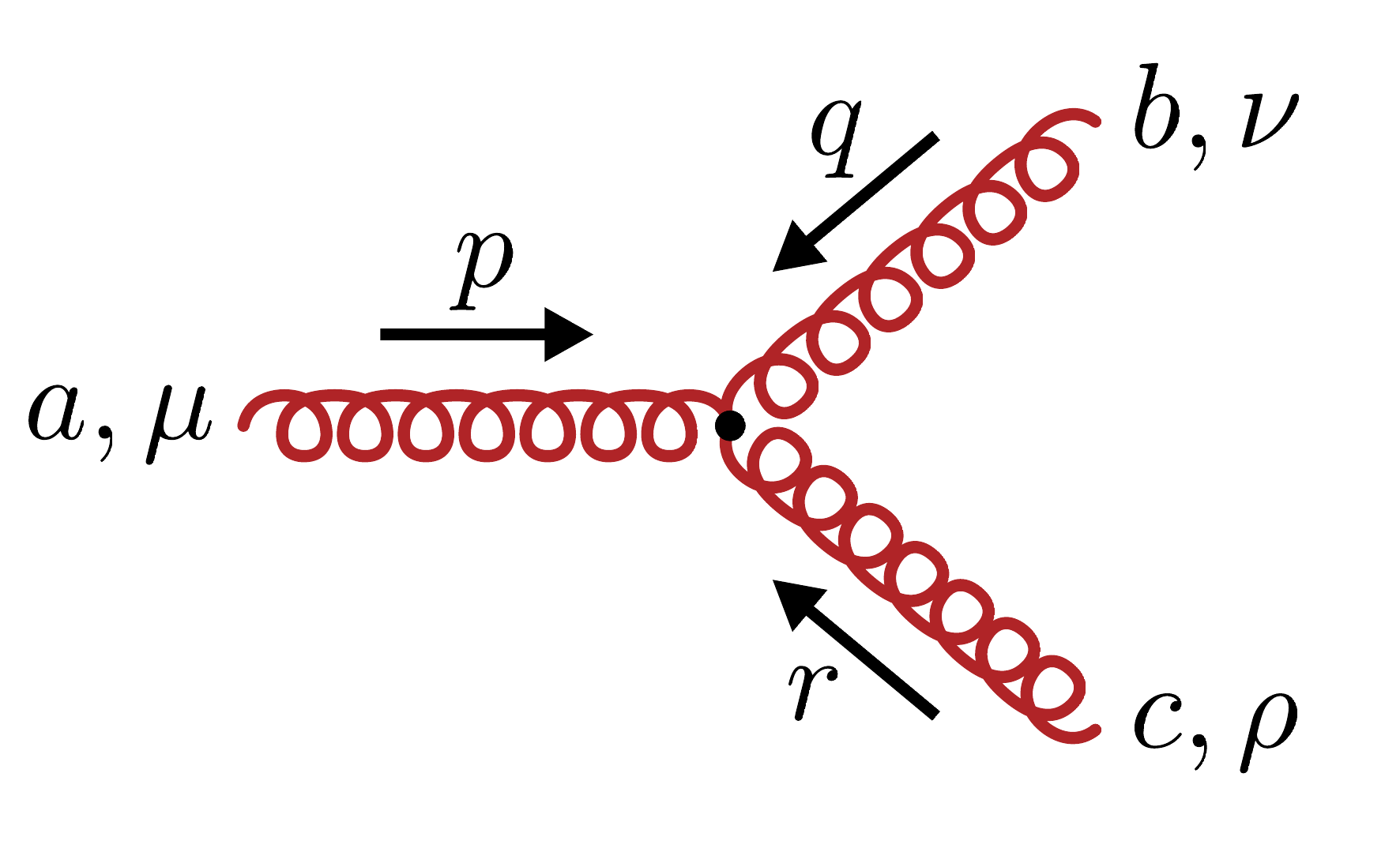}} & \raisebox{-0.45\totalheight}{\includegraphics[width=0.2\textwidth]{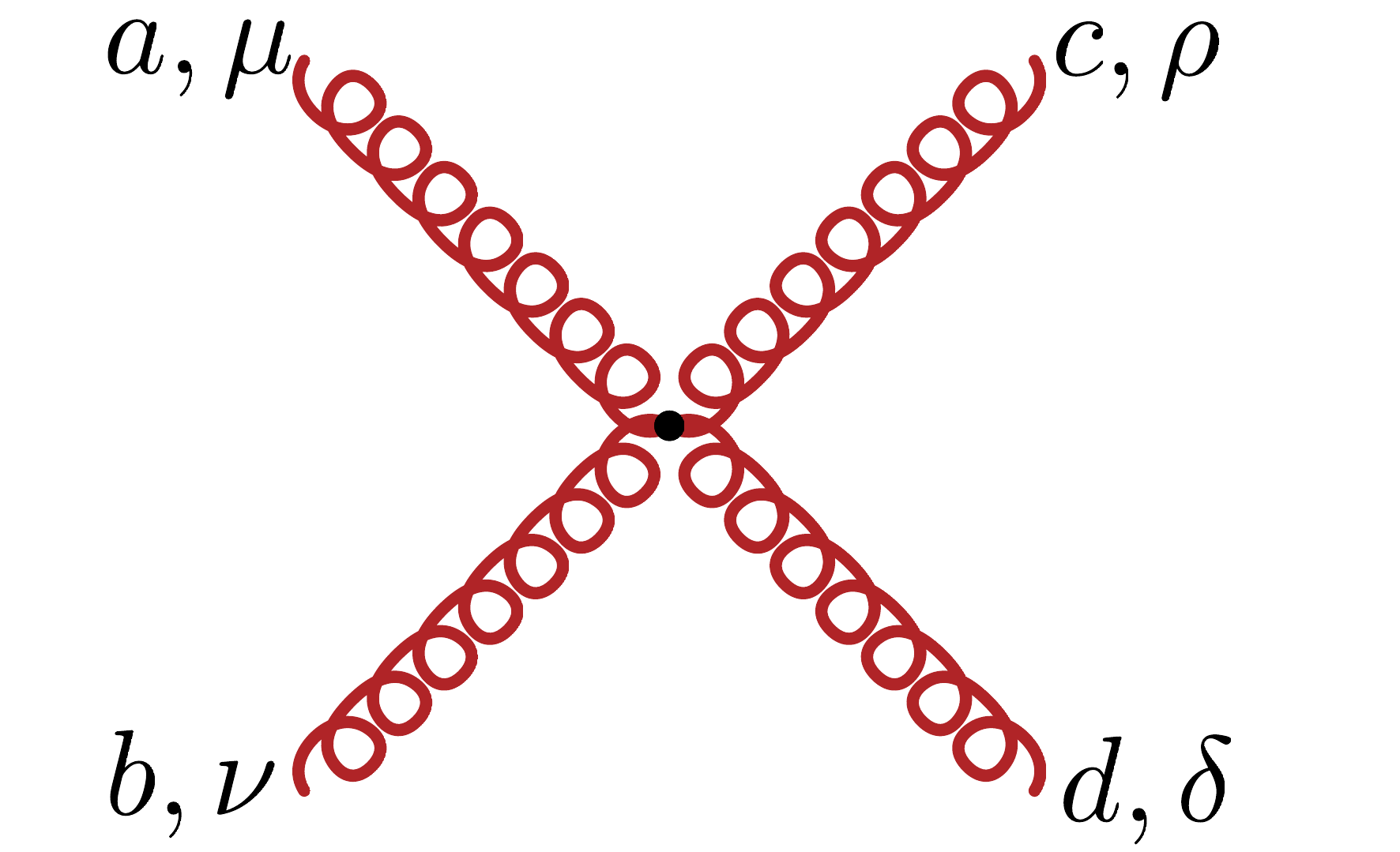}} & \raisebox{-0.45\totalheight}{\includegraphics[width=0.2\textwidth]{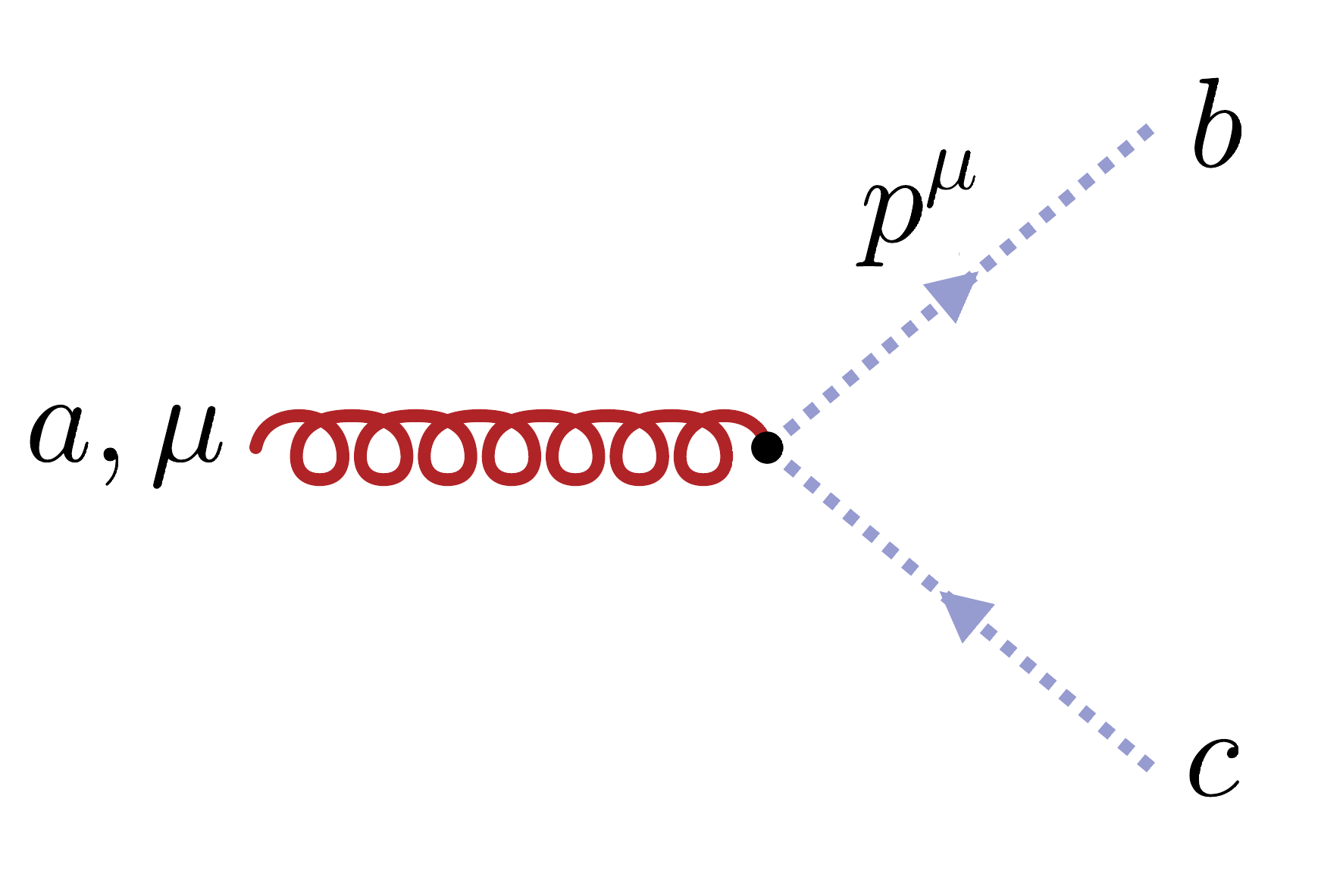}} \\
         \midrule
        $-i g_s \gamma_{\mu} (t^a)_{ij}$ & $-g_s f^{abc} \begin{bmatrix}
             + g^{\mu \nu}(p-q)^{\rho} \\
             + g^{\nu \rho}(q-r)^{\mu} \\ 
             + g^{\rho \mu}(r-p)^{\nu}
         \end{bmatrix}$ & $-ig_s^2\begin{bmatrix}
             +f^{xac}f^{xbd}(g_{\mu\nu} g_{\rho\delta} - g_{\mu\delta} g_{\nu\rho}) \\ +f^{xad}f^{xcb}(g_{\mu\rho} g_{\nu\delta} - g_{\mu\nu} g_{\rho\delta}) \\ +f^{xab}f^{xdc}(g_{\mu\delta} g_{\nu\rho} - g_{\mu\rho} g_{\nu\delta})
         \end{bmatrix}$ & $g_s (f^a)_{bc}p^{\mu}$\\
         \midrule
         quark-gluon & three-gluon & four-gluon & ghost-gluon
    \end{tabular}
    \caption{QCD vertex rules. The red curly lines are gluons, the solid green lines are quarks and the dashed grey lines are ghosts. The $a,b,c$ are colour indices and $i,j$ are spinor indices. The $f^{abc}$ are the totally antisymmetric structure constants of $SU(N)$ and $g^{\mu\nu}$ is the Minkowski metric. The Greek letters ($\mu, \nu, \dots$) are Lorentz indices and the $\gamma^{\mu}$ are Dirac matrices. The convention for the momentum directions is all-incoming. All diagrams in this chapter have been drawn using \textsc{FeynGame} \cite{Harlander:2020cyh,Harlander:2024qbn,Bundgen:2025utt}.} 
    \label{tab:feynman_rules1}
\end{table}

\begin{table}[b]
    \centering
    \begin{tabular}{c|c|c}
         \raisebox{-0.45\totalheight}{\includegraphics[width=0.2\textwidth]{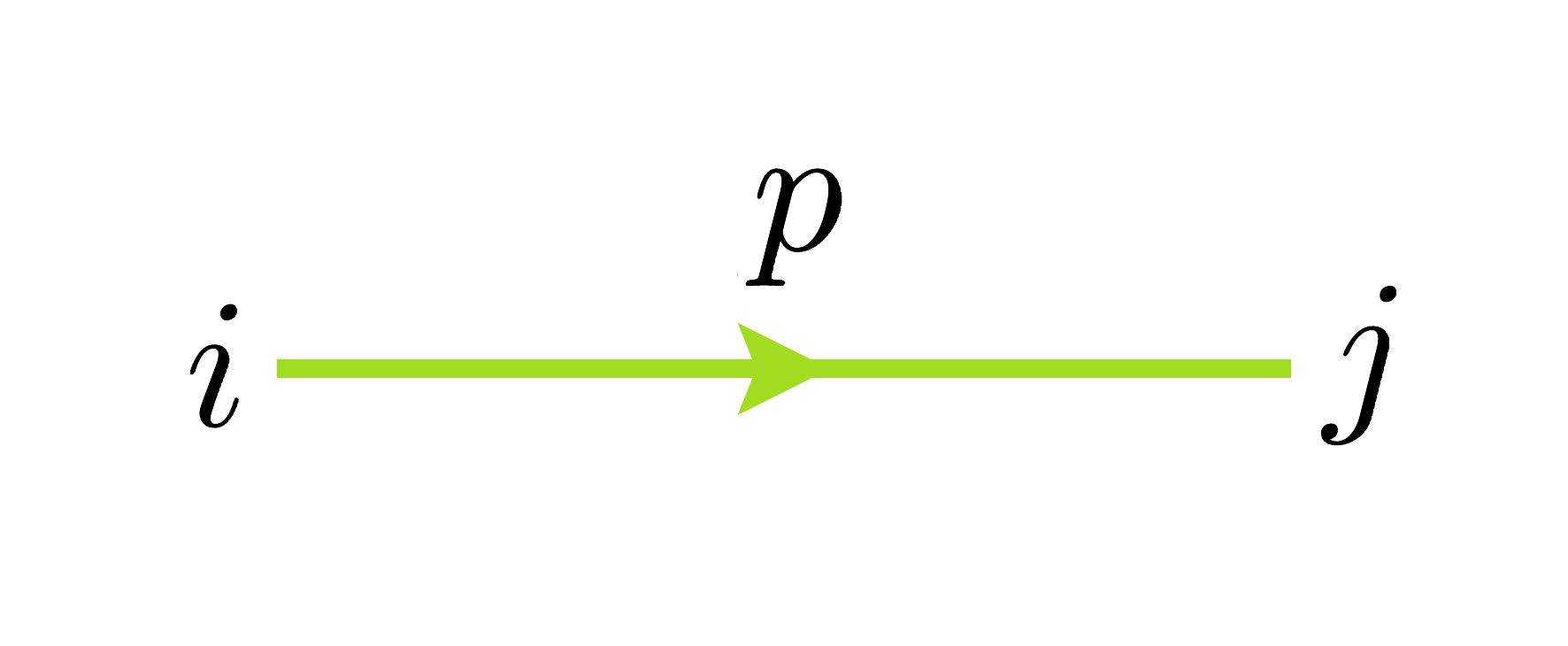}} & \raisebox{-0.45\totalheight}{\includegraphics[width=0.2\textwidth]{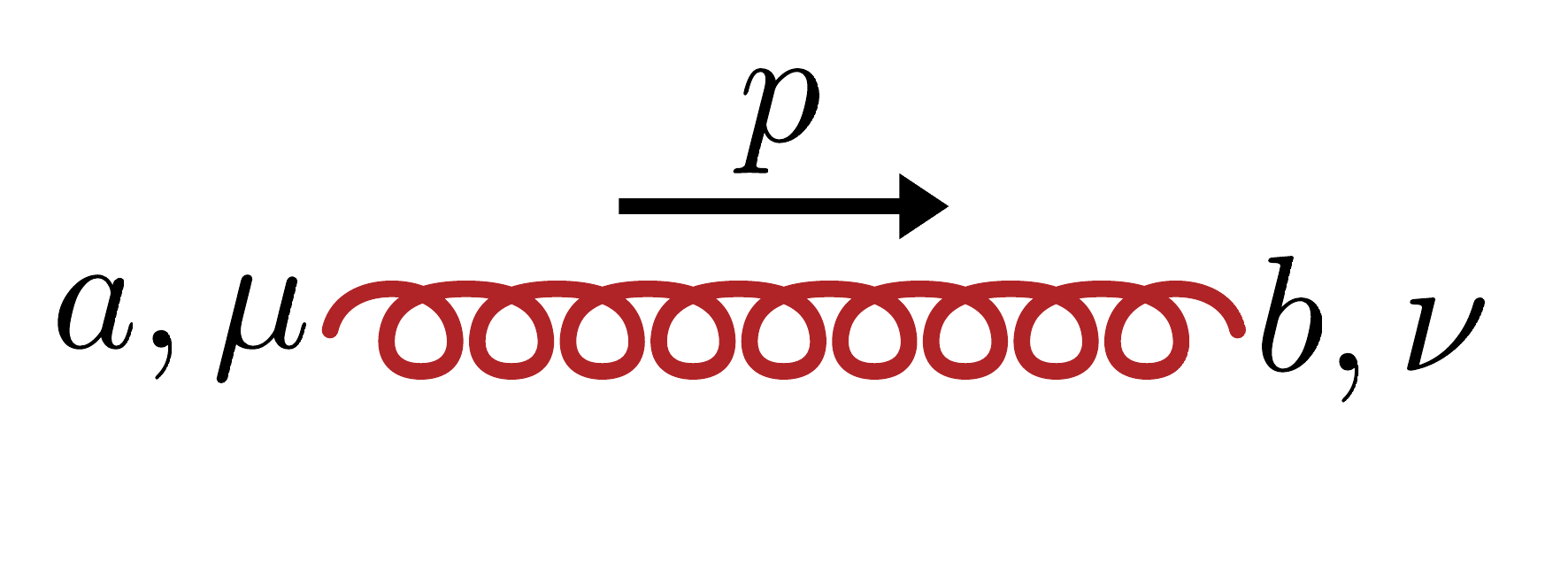}} & \raisebox{-0.45\totalheight}{\includegraphics[width=0.2\textwidth]{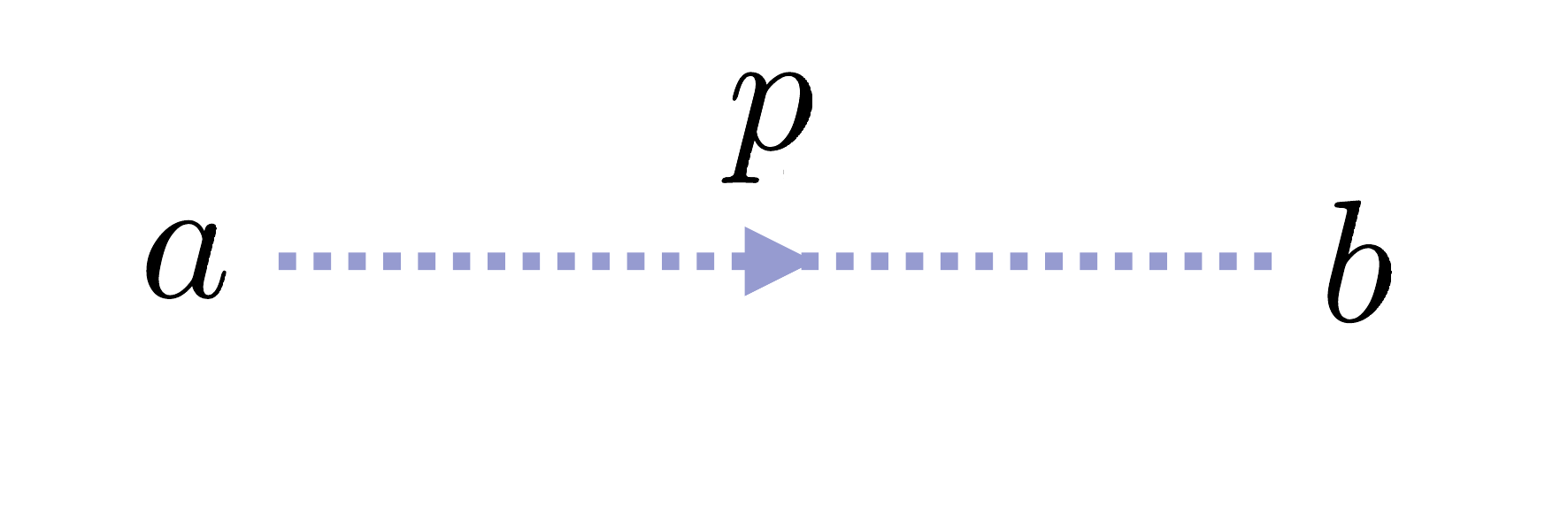}} \\ 
         \midrule
         $\delta^{ij} \frac{i(\slashed{p}+m)}{p^2-m^2+i\delta}$ & $\frac{-i\delta_{ab}}{p^2+i\delta} \left( g^{\mu\nu} - (1-\xi) \frac{p^{\mu}p^{\nu}}{p^2} \right)$ & $\delta^{ab} \frac{i}{p^2 + i\delta}$ \\
         \midrule
        quark & gluon & ghost
    \end{tabular}
    \caption{QCD propagator rules. The gluon propagator is given in covariant gauge and $\xi$ is a gauge parameter. Setting $\xi = 1$, for example, would correspond to the Feynman gauge. The $\delta^{ab}$ is the Kronecker-Delta symbol and the convention for the indices is the same as in \tabref{tab:feynman_rules1}. The $i\delta$ in the denominators is the causal prescription for the Feynman propagators.}
    \label{tab:feynman_rules2}
\end{table}

\section{Perturbation Theory}
\label{sec:perturbation_theory}
A very successful framework to calculate observable quantities from $\mathcal{L}_{\mathrm{QCD}}$ is perturbation theory. It is particularly useful to make predictions for scattering processes measured at high energy collider experiments such as the LHC at CERN. 
Lattice gauge theory is another framework to make predictions based on $\mathcal{L}_{\mathrm{QCD}}$. It does not rely on perturbation theory and therefore is particularly suited to calculate non-perturbative quantities such as hadron masses.

This section reviews the concepts of factorisation, scattering amplitudes and their perturbative expansion in terms of Feynman diagrams.
\subsection{Factorisation}
In hadron-hadron collisions, cross sections $\sigma$ for $2 \to n$ scattering are computed through the factorisation formula 
\begin{equation}
\label{eq:cross_section}
    \sigma = \sum_{a,b} \int_0^1 \rd x_a \, \rd x_b \, f_{a/h_1}(x_a) \, f_{b/h_2}(x_b) \, \rd \hat{\sigma}_{ab \to n}\;+\; {\cal O}\left(\frac{\Lambda_{\rm{QCD}}}{Q}\right)^p,
\end{equation}
where $\hat{\sigma}_{ab \to n}$ is the partonic cross section that describes the interaction between partons $a$ and $b$ taking place at a high energy, also called ``hard scattering".
The  functions $f_{i/h}(x)$ are Parton Distribution Functions (PDFs) that encode the long-range interactions in the hadron. At leading order, $f_{a/h}(x_a)$ describes the probability of finding parton $a$ in hadron $h$ with a longitudinal momentum fraction $x_a$ of the total hadron momentum, where it is assumed that the parton taking part in the hard interaction is collinear to the parent hadron.
The low energy scale of the long-range interactions in the hadron means the PDFs are inherently non-perturbative objects. They can thus not be computed in the framework of perturbation theory and must instead be fitted from experimental data. Factorization holds up to the so-called power corrections of order $\left(\Lambda_{\rm{QCD}}/Q\right)^p$, where the power $p$ is process- and observable-dependent and usually larger than one, $Q$ is a typical energy scale of the scattering process, and $\Lambda_{\rm{QCD}}\approx$ 250\,MeV.
For comprehensive reviews on factorisation in QCD, we refer to Refs.~\cite{Collins:1989gx,Agarwal:2021ais}.
Exceptions are subject of current studies, see e.g. Refs.~\cite{Beneke:1998ui,Makarov:2024ijn,Chen:2024nyc} for power corrections and Refs.~\cite{Dixon:2019lnw,Nabeebaccus:2024mia,Becher:2024kmk,Becher:2024nqc,Duhr:2025cye} about more exclusive final states, Glauber gluons and multiple collinear limits.

Assuming factorisation holds, the PDFs are process independent. They can thus be fitted with data from precisely known processes, that are easy to compute and measure, and then be applied to other processes. The evolution of PDFs between different energy scales can be calculated perturbatively using the Dokshitzer–Gribov–Lipatov–Altarelli–Parisi (DGLAP) equations \cite{Altarelli:1977zs,Dokshitzer:1977sg,Gribov:1972ri}, see \secref{sec:PDFs}.

\subsection{Partonic cross sections and perturbative expansions}
\label{sec:partonic}
A high-energy collision between elementary particles, such as the partons coming out of a hadron, is known as a hard interaction, and is described by a partonic cross section according to
\begin{equation}
\label{eq:partonic_cross_section}
    \hat{\sigma}_{ab \to n} = \flux \int \rd \Phi_n \left|\mathcal{M}_{ab \to n}(p_1,\dots,p_n)\right|^2,
\end{equation}
where $\rd \Phi_n$ is the $n$-particle Lorentz-Invariant-Phase-Space (LIPS) defined as
\begin{equation}
\label{eq:LIPS}
    \rd \Phi_n = (2\pi)^4 \, \delta^4 (q_a+q_b-\sum_i p_i) \prod_{i=1}^n \frac{\rd^4 p_i}{(2\pi)^4} \, 2\pi \, \delta(p_i^2-m_i^2) \, \Theta(p_i^{(0)}).
\end{equation}
The $\delta^4 (q_a+q_b-\sum\nolimits_i p_i)$ imposes momentum conservation between the initial and final states, $\delta(p_i^2-m_i^2)$ is an on-shell condition for the final-state particles and $\Theta(p_i^{(0)})$ ensures that the final-state particles have positive energy. The prefactor in \equref{eq:partonic_cross_section} is known as the flux factor and is related to the centre-of-mass energy $s$ of the underlying hadron collision by $\hat{s} = x_a x_b s$. The expression $\mathcal{M}_{ab \to n}(p_1,\dots,p_n)$ is a central object in perturbative calculations and is known as the Feynman amplitude, sometimes also called {\em  matrix element}. It is the non-trivial part of the $S$-matrix~\cite{ELOP} that describes the transition probability between an initial state $i$ and a final state $f$\footnote{We always use the shorthand $\langle f \, | \mathcal{M} \, | \, i \, \rangle = \mathcal{M}_{i\to f}$ and often also $\mathcal{M}_{i\to f} = \mathcal{M}$ when the context makes it obvious which process we are referring to.}
\begin{equation}
    \langle f \, |S - \mathbb{1} | \, i \, \rangle = i \, (2\pi)^4 \, \delta^4 (q_a+q_b-\sum\nolimits_i p_i) \, \mathcal{M}_{i\to f}(p_1,\dots,p_n).
\end{equation}
It is a complex-valued function and its square can be interpreted as a probability density that, when integrated over a phase-space region, describes the probability of producing the final state $f$ in that region. In QCD the amplitude depends on the strong coupling $g_s$ and we can make a perturbative expansion 
\begin{equation}
\label{eq:perturbative_expansion_amplitude}
    \mathcal{M} = \sum_{k=0}^{\infty} g_s^{2k} \mathcal{M}_k,
\end{equation}
where $\mathcal{M}_0$ may or may not contain QCD couplings already, and the higher order terms are suppressed by increasing powers of the coupling. The first non-zero term in this expansion is referred to as the Leading-Order (LO) amplitude, the second the Next-to-Leading-Order (NLO) contribution to the amplitude, and so on. In terms of Feynman diagrams, the $\mathcal{M}_k$ in \equref{eq:perturbative_expansion_amplitude} can be interpreted as the sum of all diagrams containing $k$ loops (or the radiation of up to $k$ extra particles). We will use Feynman diagrams to study the $\mathcal{M}_0$ contribution to $q\bar{q} \to gg$ in \secref{sec:diagrams}. The expansion of $\mathcal{M}$ suggests that the cross section can also be decomposed order-by-order as
\begin{equation}
\label{eq:perturbative_expansion_cross_section}
    \hat{\sigma} = \hat{\sigma}_{\mathrm{LO}} + \alpha_s \, \hat{\sigma}_{\mathrm{NLO}} + \alpha^2_s \, \hat{\sigma}_{\mathrm{NNLO}} + \dots,
\end{equation}
where $\alpha_s = g^2_s/4\pi$. \secref{sec:IR} describes in detail the amplitude ingredients that must enter the cross section at the different orders. For now we state that the higher-order terms, that can be interpreted as quantum corrections to the Born-level scattering process, increase precision at the cost of being more complex to calculate. In practice, the sum has to be truncated at a finite order. This gives rise to dependence on the unphysical scales $\mur$ and $\muf$ for both the amplitude and the cross section, such that
\begin{align*}
    \mathcal{M}(p_1,\dots,p_n) &\to \mathcal{M}(p_1,\dots,p_n;\muf,\mur), \\
    \hat{\sigma} &\to \hat{\sigma}(\muf,\mur).
\end{align*}
The subscripts refer to factorisation and renormalisation scales, which are both discussed in \secref{sec:higher_orders}. The presence of $\mur$ and $\muf$ implies that there is an uncertainty due to the choice of the unphysical scales on the cross section, associated with the truncation of the perturbative expansion. The more terms that we are able to compute in \equref{eq:perturbative_expansion_cross_section}, the smaller the scale uncertainty becomes. 

\subsection{Tree-level amplitudes}
\label{sec:diagrams}
\begin{figure}[t]
    \centering
    \includegraphics[width=0.3\textwidth]{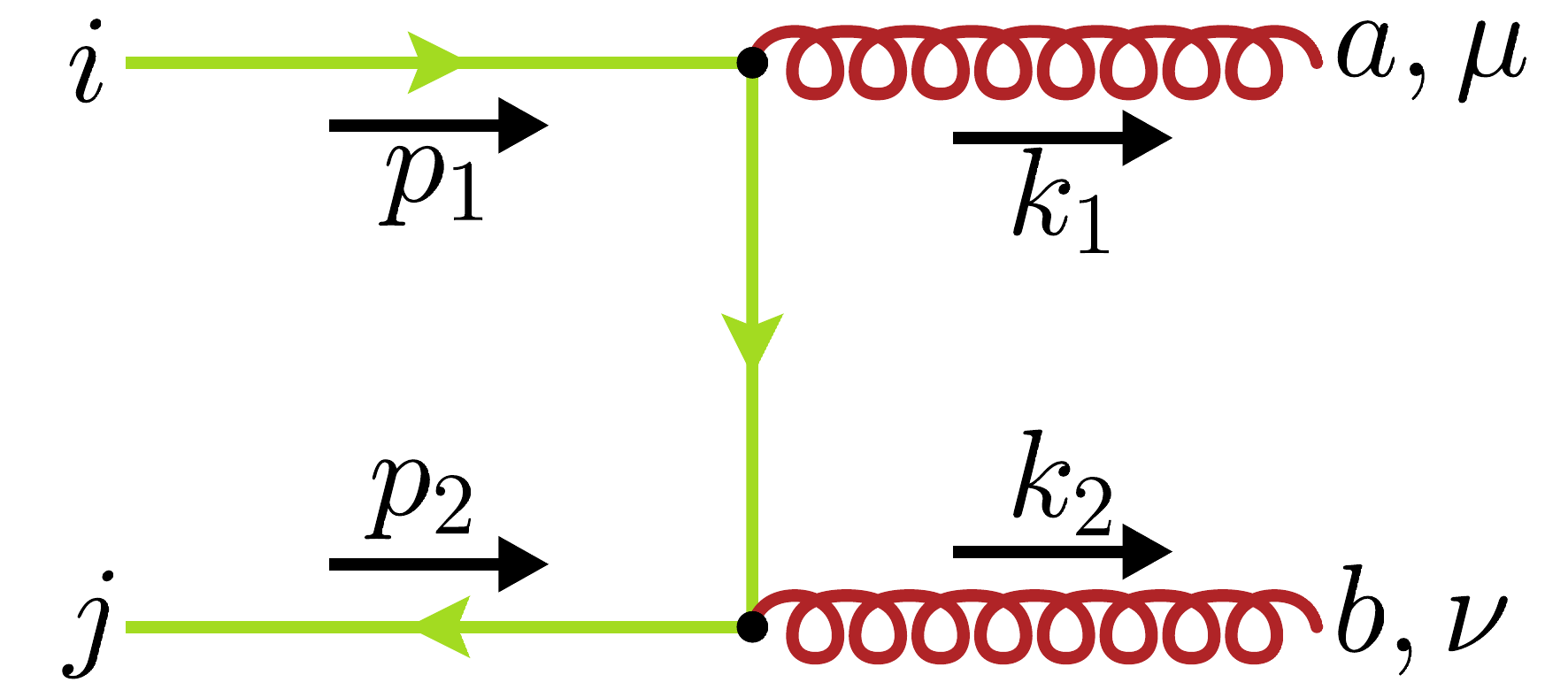}
    \hfill
    \includegraphics[width=0.3\textwidth]{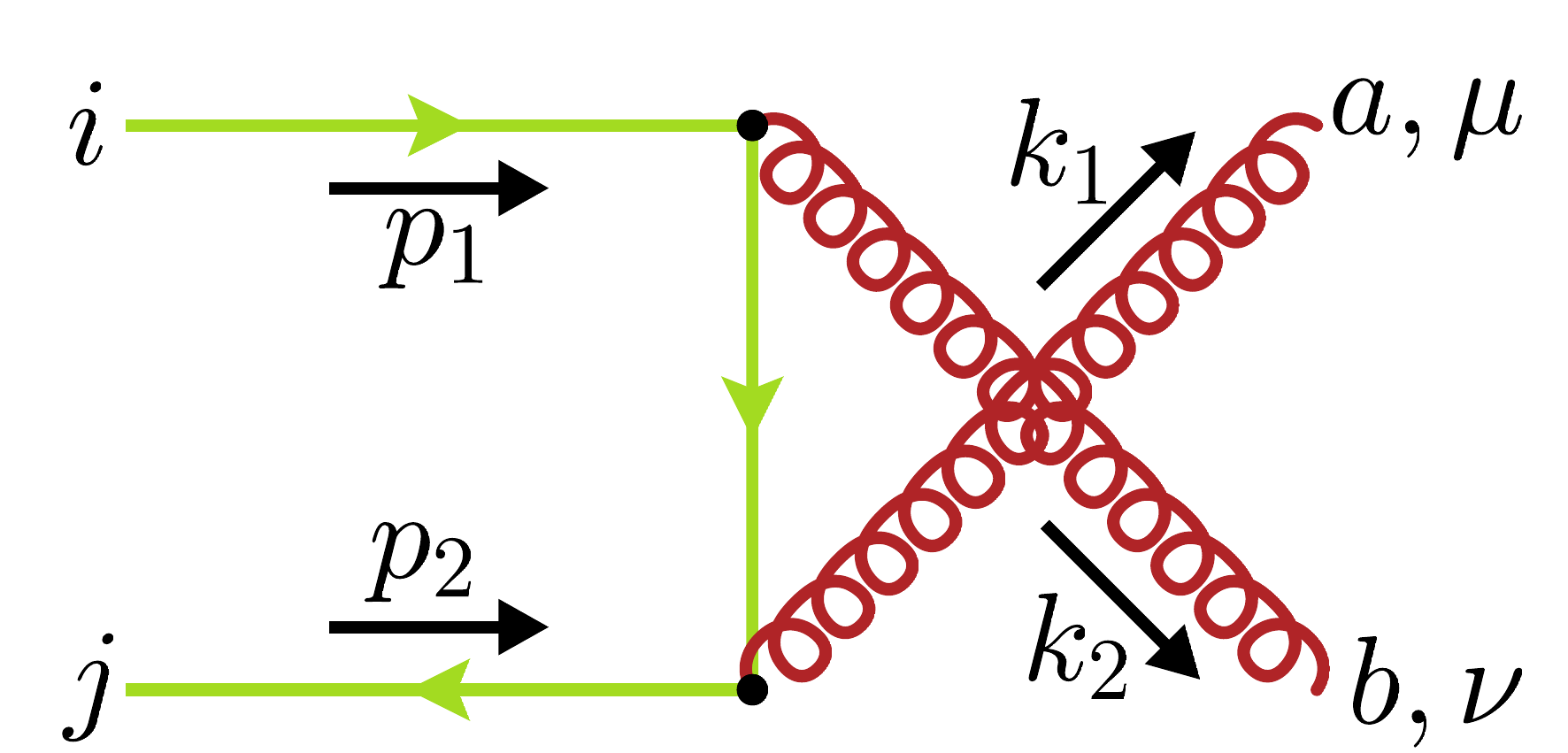}
    \hfill
    \includegraphics[width=0.3\textwidth]{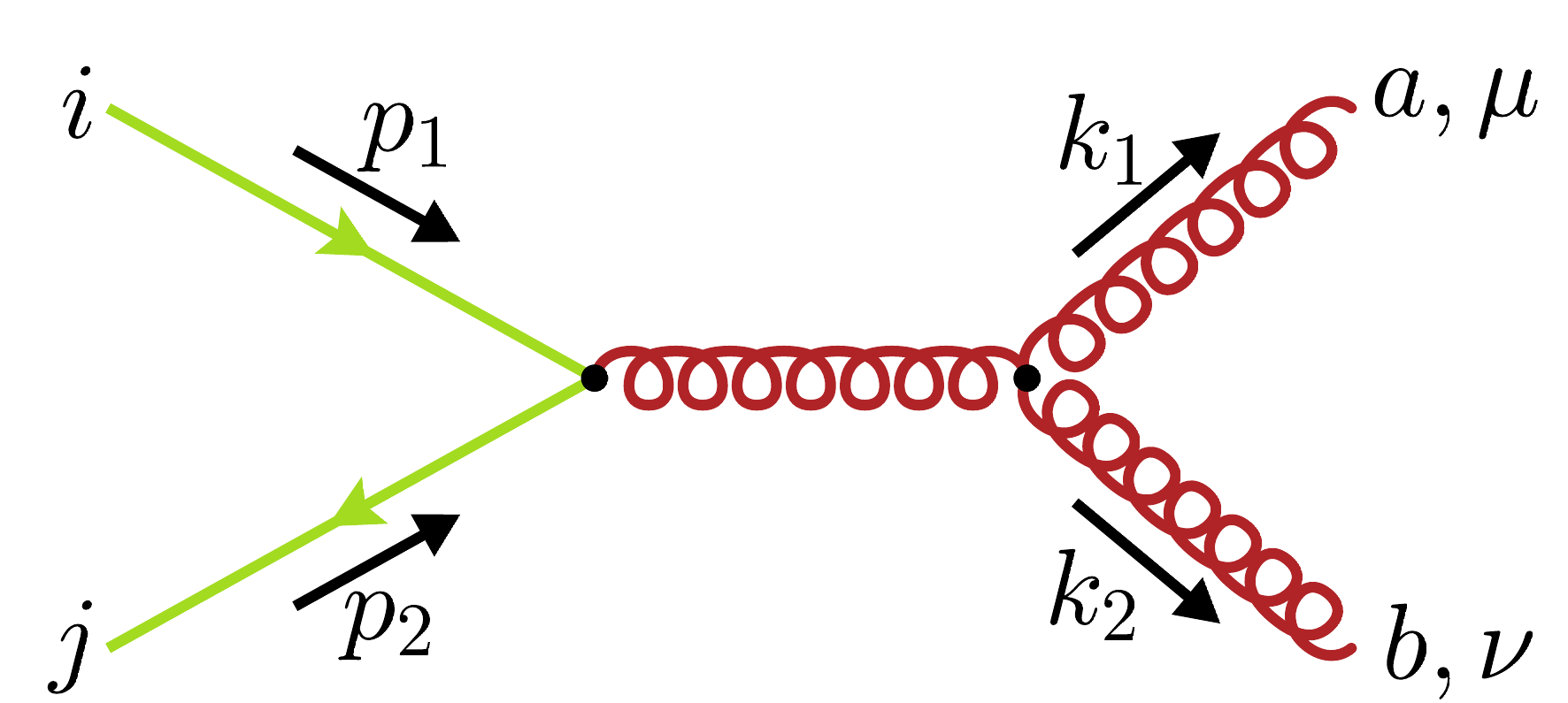}
    \caption{%
        Tree-level diagrams (respectively of $t-$, $u-$ and $s$-channel type) contributing to the LO $q\bar{q} \to gg$ amplitude. The $s$-channel type diagram features a three-gluon vertex and is thus a consequence of the non-Abelian nature of QCD.
    }
    \label{fig:ggqq-LO}
\end{figure}
To understand what the terms in \equref{eq:perturbative_expansion_amplitude} are, we consider their pictorial representation in terms of Feynman diagrams. We use the example of $q\bar{q} \to gg$ at LO to demonstrate how this works. The LO amplitude consists of three \textit{tree-level} diagrams, see \figref{fig:ggqq-LO}. Applying the Feynman rules to each diagram we obtain the following amplitude
\begin{equation}
    i\mathcal{M}_{q\bar{q} \to gg} = -ig_s^2 \epsilon_{1,\lambda_1}^{\mu}(k_1) \epsilon_{2,\lambda_2}^{\nu}(k_2) M_{\mu\nu}, \quad M_{\mu\nu} = (t^at^b)_{ij} M_{\mu\nu}^{(t)} + (t^bt^a)_{ij} M_{\mu\nu}^{(u)} + i\,f^{abc} t^c M_{\mu\nu}^{(s)},
\end{equation}
where
\begin{gather}
    M_{\mu\nu}^{(t)} = \bar{\nu}^{s_2}(p_2) \gamma_{\nu} \frac{\slashed{p}_1 - \slashed{k}_1}{(p_1 - k_1)^2} \gamma_{\mu} u^{s_1}(p_1), \quad \quad M_{\mu\nu}^{(u)} = \bar{\nu}^{s_2}(p_2) \gamma_{\mu} \frac{\slashed{p}_1 - \slashed{k}_2}{(p_1 - k_2)^2} \gamma_{\nu} u^{s_1}(p_1), \\[4pt] M_{\mu\nu}^{(s)} = \bar{\nu}^{s_2}(p_2) \left[g^{\mu\nu}(\slashed{k}_2 - \slashed{k}_1) - \gamma^{\nu}(k_1 + 2k_2)^{\mu} + \gamma^{\mu}(2k_1 + k_2)^{\nu}\right] u^{s_1}(p_1),
\end{gather}
where $s_1$ and $s_2$ are the spin labels of the quark and antiquark respectively and spinor indices have been left implicit. We can make an analogy with an Abelian theory, such as QED, where the $s$-channel diagram is absent due to the lack of gauge-boson self interactions. Using $[t^a,t^b] = i f^{abc} t^c$ the amplitude can be written as 
\begin{equation}
\label{eq:qed_like}
    M_{\mu\nu} = (t^at^b)_{ij} \left[M_{\mu\nu}^{(t)} + M_{\mu\nu}^{(u)}\right] + i f^{abc} t^c \left[M_{\mu\nu}^{(s)} - M_{\mu\nu}^{(u)}\right].
\end{equation}
In the case of an Abelian theory (considering e.g. $e^+e^-\to \gamma\gamma$), the amplitude would be just the first bracket (without the colour factors) in \equref{eq:qed_like}.\\
In \equref{eq:partonic_cross_section} we need the squared amplitude $|\mathcal{M}|^2 = \mathcal{M} \mathcal{M}^{\dagger}$. If the experiment does not measure polarisation, we have to sum over all possible polarisations $\lambda$ in the final state. Since colour cannot be measured we have to sum over all colours in the final state as well. Moreover, if the incoming beams are not polarised, we must average over the spins in the initial state. For this example, the physically useful quantity to calculate is therefore the spin- and colour-averaged, polarisation- and colour-summed squared amplitude
\begin{equation}
    |\mathcal{M}|^2 \to \overline{\sum} |\mathcal{M}|^2 = \prod_{\mathrm{initial}} \frac{1}{N_{\mathrm{spins}} N_{\mathrm{cols}}}\sum_{\mathrm{cols}} \sum_{\substack{s_1,s_2,\\{\lambda_1,\lambda_2}}} |\mathcal{M}|^2.
\end{equation}
In the case of $q\bar{q} \to gg$ we have $N_{\mathrm{spins}} = 2$, $N_{\mathrm{cols}} = 3$ in the initial state and $N_{\mathrm{pols}} = 2$, $N_{\mathrm{cols}} = 8$ in the final state. The spin sums follow from the completeness relations. In the present example they take the form
\begin{equation}
    \sum_{s_1} u^{s_1}(p_1) \bar{u}^{s_1}(p_1) = \slashed{p}_1 + m, \quad \quad \sum_{s_2} v^{s_2}(p_2) \bar{v}^{s_2}(p_2) = \slashed{p}_2 - m,
\end{equation}
where $m$ is the mass of the incoming (anti)quark, which is usually taken to be $0$.
The polarisation sum for a gluon with momentum $k^{\mu}$ is 
\begin{equation}
\label{eq:pol_sum}
     \sum_{\lambda} \epsilon^{\mu}_{\lambda}(k) \epsilon^{\star\nu}_{\lambda}(k) = -g^{\mu\nu} + \frac{k^{\mu}n^{\nu} + k^{\nu}n^{\mu}}{k\cdot n}\;, \quad n^2 = 0, 
\end{equation}
where $n^{\mu}$ is a light-like vector, dual to $k^{\mu}$, $k \cdot n \neq 0$. 


\section{Higher order corrections}
\label{sec:higher_orders}

\subsection{Loop corrections and UV divergences}
\label{sec:UV}

\begin{figure}[t]
  \centering
\includegraphics[width=5cm]{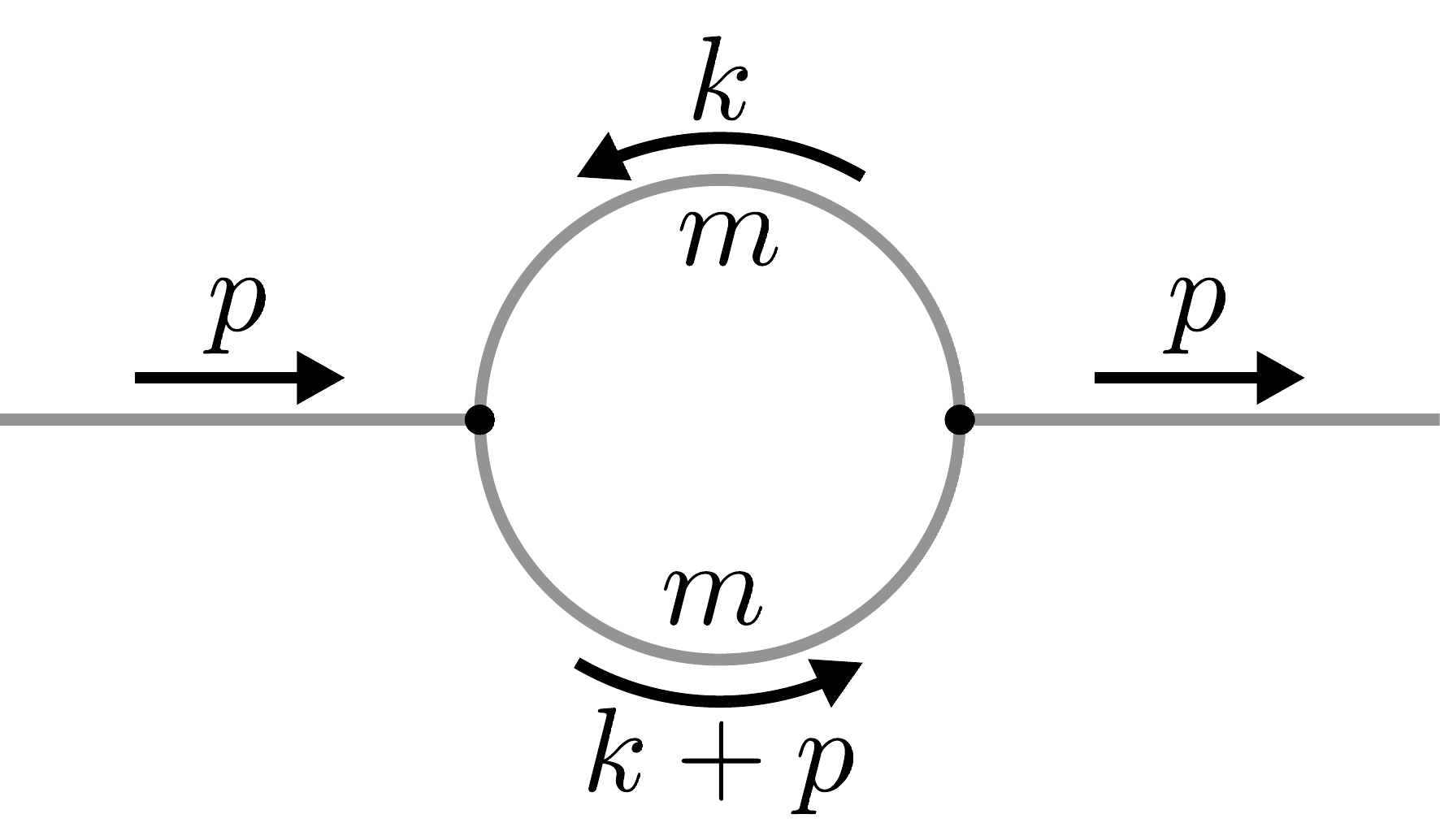}
 \caption{One-loop two-point function (``bubble'').}
  \label{fig:bubble1L}
\end{figure}

As an example of how ultraviolet (UV) divergences arise, we consider the integral $I_2$ shown diagrammatically in Fig.~\ref{fig:bubble1L}, also called {\em one-loop 2-point function} because the diagram has two external legs. The expression for this loop integral naively would be
\begin{equation}
I_2=\int_{-\infty}^{\infty}\frac{d^4 k}{(2\pi)^4}\,\frac{1}{[k^2-m^2+i\delta][(k+p)^2-m^2+i\delta]}\;.
\end{equation}
If we are only interested in the behaviour of the integral for $|k|\to \infty$ we can neglect the masses, 
transform to polar coordinates and obtain
\begin{equation}
I_2\sim \int\rd\Omega_3\int_0^\infty \rd |k| \frac{|k|^3}{|k|^4}\;.
\label{eq:UVdiv}
\end{equation}
This integral is clearly not well-defined. If we introduce an upper cutoff $\Lambda$ 
(and a lower limit $|k|_{\rm{min}}$ because we neglected the masses and $p^2$) it is regulated: 
\begin{equation}
I_2\sim \int_{|k|_{\rm{min}}}^\Lambda \rd |k| \frac{1}{|k|}\sim \log\Lambda\;.
\end{equation}
The integral has a logarithmic UV divergence.
The problem with the regulator $\Lambda$ is that it is neither Lorentz invariant nor gauge invariant.
A regularisation method which preserves the symmetries is {\em dimensional regularisation}.

\subsubsection{Dimensional regularisation}
\label{sec:dim-reg}
In modern precision computations, the standard regularisation procedure is \textit{dimensional regularisation} \cite{tHooft:1972tcz,Bollini:1972ui}. The main reason for it being so prominent is that calculations in this framework have turned out to be the simplest \cite{Weinzierl:2022eaz}. In particular, dimensional regularisation resolves divergences originating from both the ultraviolet and infrared regimes. 

The mechanism of dimensional regularisation is to shift the number of space-time dimensions to $d = 4 - 2 \epsilon$. Usually, $\epsilon$ is assumed to be real, but for purposes of analytic continuation in $d$, $\epsilon$ can also be complex. The behaviour of UV divergences is better if $\epsilon > 0$ while for IR divergences it is better to have $\epsilon < 0$. One can immediately see from \equref{eq:UVdiv} that lowering the space-time dimension would decrease the power of the loop momentum in the numerator and therefore improve the convergence for 
$|k|\to \infty$.

In practice, the renormalisation constants are computed first with the assumption that $\epsilon > 0$. After cancelling all UV divergences, the rest of the computation can be performed with the assumption that $\epsilon < 0$. Loop integrals in $d$ dimensions are well defined and divergences manifest as poles in $\epsilon$. The original theory is restored upon taking $\epsilon \to 0$ after the singularities have been subtracted through renormalisation or cancelled with other parts of the calculation. 
We do not discuss renormalisation in more detail here, since this subject is treated in the chapter on renormalisation in this volume.

On a technical level, most objects and operations behave similarly when extended to $d$ dimensions. The action integral is $d$-dimensional
\begin{equation}
\label{eq:action_d}
    S = \int \rd^{d-1} x \, \rd t \, \mathcal{L},
\end{equation}
which necessitates $\left[ \mathcal{L} \right] = d$ to preserve that  $\left[ S \right] = 0$. It is conventional to make parameter redefinitions such as
\begin{equation}
    g_s \to \mur^{\frac{4-d}{2}} g_s,
\end{equation}
to prevent the couplings from acquiring a non-integer dimensionality. Each loop thus receives a prefactor $\mur^{4-d}$, and the integration over loop momenta is in $d$ dimensions, i.e. the integration measure is $\int\frac{\rd^d k}{(2\pi)^d}$ for each loop. How to perform such an integration is described in more detail in \secref{sec:loops}.

A common source of confusion with dimensional regularisation is the difference between regularisation schemes and $\gamma^5$ schemes. We give a brief overview of both topics and point to Refs.~\cite{Catani:1996pk,Gnendiger:2017pys,Belusca-Maito:2023wah,Chen:2023lus,OlgosoRuiz:2024dzq,Chen:2024hlv,vonManteuffel:2025swv} for a more in-depth discussion.

In all variants of dimensional regularisation the loop momenta must be continued into $d\neq4$ to ensure that the loop integrals are well defined. There is however freedom in the treatment of other Lorentz objects, such as $\gamma$-matrices and vector fields (gluons in QCD). This corresponds to different \textit{regularisation schemes}, or \textit{variants}. We will consider four variants, usually grouped into two classes. It is helpful to introduce three vector spaces; the strictly $4$-dimensional space (4S), the quasi-$d$-dimensional space (QDS) and the quasi-$4$-dimensional space (QD$_s$S). The latter two are formally infinite-dimensional vector spaces, with certain $d$-dimensional and $4$-dimensional properties, respectively~\cite{Signer:2008va}. 
What matters mainly is the following relation between the vector spaces
\begin{equation}
    \mathrm{4S} \subset \mathrm{QDS} \subset \mathrm{QD_sS}.
\end{equation}
Additionally, in the language of Refs.~\cite{Gnendiger:2017pys,Belusca-Maito:2023wah} we differentiate between \textit{singular} gluons, that appear either in divergent loops or as external propagators in phase-space regions that lead to infrared singularities, and \textit{regular} gluons that live strictly outside singular phase-space regions.

Now we can define the first class of variants, comprised of conventional dimensional regularisation (CDR) and the 't Hooft-Veltman scheme (HV). In CDR all Lorentz objects are treated in QDS, including the regular gluons. In HV on the other hand, regular gluons are treated in 4S. An important point is that in both CDR and HV all Lorentz objects appearing in the Feynman rules are treated in $d$ dimensions. We thus require a $d$-dimensional interpretation of the Dirac algebra, for example. 

The second class of variants consists of dimensional reduction (DRED) and the four-dimensional helicity scheme (FDH). In this case, the Lorentz objects in the Feynman rules are strictly four-dimensional (except those that appear with a loop momentum). The difference between DRED and FDH is analogous to that between CDR and HV. In DRED both singular and regular gluons are treated in QD$_s$S, while in FDH regular gluons are allowed to live in 4S. The difference between CDR and HV, and the difference between DRED and FDH, only starts at $\mathcal{O}(\epsilon)$. This means that in pure one-loop calculations, CDR and HV, as well as DRED and FDH, are equivalent. 

A common way of distinguishing the four regularisation schemes is by defining the metric tensors associated to each vector space. We use $\bar{g}^{\mu\nu}$ for 4S, $g^{\mu\nu}$ for QDS and $\Tilde{g}^{\mu\nu}$ for QD$_s$S\footnote{Caution: this notation varies wildly between different references.}. The dimensionalities are then given by \cite{Belusca-Maito:2023wah}
\begin{equation}
    \bar{g}^{\mu\nu} \bar{g}_{\mu\nu} = 4, \quad \quad g^{\mu\nu} g_{\mu\nu} = d, \quad \quad \Tilde{g}^{\mu\nu} \Tilde{g}_{\mu\nu} = d_s,
\end{equation}
where $d_s$ is the dimensionality of QD$_s$S.
The relations between the vector spaces imply the following projections \cite{Signer:2008va}
\begin{equation}
    \Tilde{g}^{\mu\nu} g_{\nu}^{\text{~}\text{~}\rho} = g^{\mu\rho}, \quad \quad \Tilde{g}^{\mu\nu} \bar{g}_{\nu}^{\text{~}\text{~} \rho} = \bar{g}^{\mu\rho}, \quad \quad g^{\mu\nu} \bar{g}_{\nu}^{\text{~}\text{~}\rho} = \bar{g}^{\mu\rho}.
\end{equation}
The projections encode what happens when Lorentz objects with indices of different dimensionality interact with each other. This is particularly relevant when working in HV (FDH) where the four-dimensional treatment of regular vector fields generate $\bar{g}^{\mu\nu}$ that may interact with $g^{\mu\nu}$ ($\Tilde{g}^{\mu\nu}$) coming from the Dirac algebra in the loop. 
With these definitions we compactly encode the differences in the treatment of gluons between regularisation schemes in \tabref{tab:reg_schemes}.

\begin{table}[t]
    \centering
    \begin{tabular}{c|cccc}
      & CDR & HV & DRED & FDH \\ 
      \hline 
      Singular gluons & $g^{\mu\nu}$ & $g^{\mu\nu}$ & $\Tilde{g}^{\mu\nu}$ & $\Tilde{g}^{\mu\nu}$ \\
      Regular gluons & $g^{\mu\nu}$ & $\bar{g}^{\mu\nu}$ & $\Tilde{g}^{\mu\nu}$ & $\bar{g}^{\mu\nu}$
    \end{tabular}
    \caption{Treatment of gluons, i.e. definition of the metric tensor in the propagators and polarisation sums, in four different variants of dimensional regularisation.}
    \label{tab:reg_schemes}
\end{table}

Now we move on from regularisation variants and consider instead $\gamma^5$ schemes. The treatment of $\gamma^5$ in dimensional regularisation is a well-known problem, related to the extension of the Dirac algebra to $d$ dimensions. The basic interpretation is a set of $d$ four-dimensional matrices, $\gamma^0,\gamma^1,\dots,\gamma^{d-1}$ that satisfy the anti-commutation relation
\begin{equation}
    \{ \gamma^{\mu}, \gamma^{\nu} \} = 2 g^{\mu\nu}.
\end{equation}
The problem is that the four-dimensional definition $\gamma^5~=~i \gamma^0 \gamma^1 \gamma^2 \gamma^3$, is in $d=4-2\epsilon$ not compatible with preserving cyclicity of traces while also satisfying~\cite{Jegerlehner:2000dz}
\begin{equation}
\label{eq:g5_properties}
    \{ \gamma^{\mu}, \gamma^{5} \} = 0 \quad \mathrm{and} \quad \tr{\gamma_{\mu} \gamma_{\nu} \gamma_{\rho} \gamma_{\delta} \gamma_{5}} = 4i\epsilon_{\mu\nu\rho\delta}.
\end{equation}
There are thus various \emph{$\gamma^5$ schemes} that correspond to extensions where subsets of the above three properties are fulfilled. The most standard one is the Breitenlohner-Maison-’t Hooft-Veltman (BMHV) scheme, which gives up the anti-commutation property of \equref{eq:g5_properties} and defines $\gamma^5$ as in four dimensions. It is the most well-defined and mathematically consistent scheme in the sense that it is compatible with unitarity and causality of the theory \cite{Belusca-Maito:2023wah}. In this case we have
\begin{equation}
\label{eq:BMHV_commutation}
    \{ \hat{\gamma}^{\mu}, \gamma^{5} \} = 2 \hat{\gamma}^{\mu} \gamma^5 \quad \mathrm{and} \quad \{ \bar{\gamma}^{\mu}, \gamma^{5} \} = 0,
\end{equation}
where the Dirac matrices have been split up into a strictly $4$-dimensional part $\bar{\gamma}_{\mu}$ and a $(d-4)$-dimensional part $\hat{\gamma}_{\mu}$, such that $\gamma_{\mu} = \bar{\gamma}_{\mu} + \hat{\gamma}_{\mu}$. The first relation implies $[\hat{\gamma}^{\mu}, \gamma^5]=0$. 
Other options include the Larin scheme \cite{Larin:1993tq} and the Kreimer scheme \cite{Korner:1991sx}. In the former $\gamma^5~=~\frac{i}{4!} \epsilon_{\mu\nu\rho\delta}\gamma^{\mu} \gamma^{\nu} \gamma^{\rho} \gamma^{\delta}$ but the anti-commutation property is dropped. In the latter we do have $\{ \gamma^{\mu}, \gamma^{5} \} = 0$, but the cyclicity of traces involving an odd number of $\gamma^5$ matrices is lost. For more technical details on $\gamma^5$-schemes, we refer to the reviews in Refs.~\cite{Belusca-Maito:2023wah,Gnendiger:2019vnp,Gnendiger:2017pys}.

\subsubsection{The running coupling}
\label{sec:running}
For the perturbative expansion in \equref{eq:perturbative_expansion_amplitude} to be well defined and converge quickly, a necessary\footnote{But not sufficient. Additionally, the higher-order amplitudes $\mathcal{M}_k$ must also be small enough to not spoil convergence. For example, if each power in $\als$ is accompanied by large logarithms,  this necessitates all-order resummation. Another consideration is that the number of terms at each order must not grow too fast. In fact, the number of diagrams is known to grow factorially, which means the suppression due to higher powers of $\alpha_s$ is eventually overtaken \cite{Beneke:1998ui}. Fortunately, this does not occur at phenomenologically relevant orders in QCD.} requirement is that the strong coupling $\alpha_s$ must be small enough. The whole machinery with Feynman rules and diagrams is built on the assumption that including at most a few $\mathcal{M}_k$, yields a sufficiently accurate approximation of $\mathcal{M}$ for phenomenological applications. 

\begin{figure}[t]
    \centering
    \includegraphics[width=0.75\textwidth]{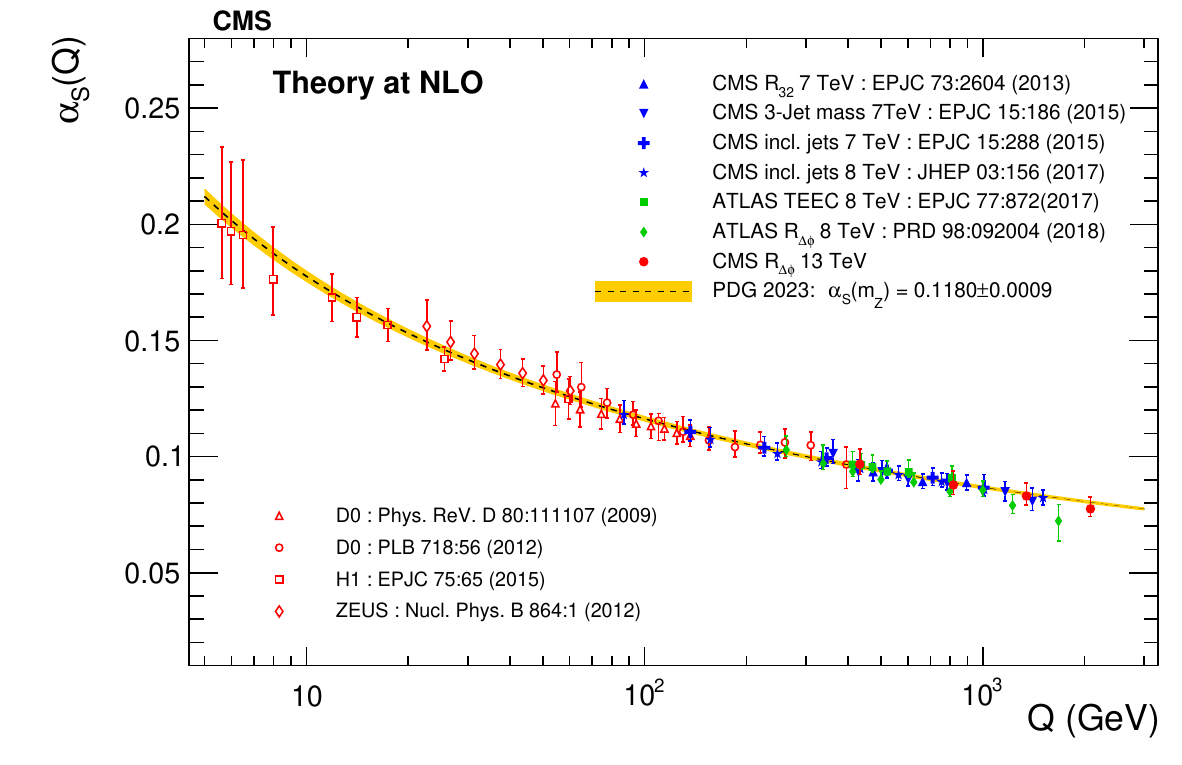}
    \caption{
    Experimental determination from the CMS collaboration of the strong coupling $\alpha_s$ as a function of the scale $Q$. Figure taken from Ref. \protect\cite{CMS:2024hwr}.
    }
    \label{fig:alpha_s}
\end{figure}

It is explained in \secref{sec:UV} how loop corrections produce UV divergences that necessitate a regularisation and renormalisation procedure. The result is that a dependence on an unphysical renormalisation scale is induced to the coupling such that $\alpha_s := \alpha_s(\mur)$. The scale dependence of $\alpha_s(\mur)$ is referred to as the \textit{running} of the coupling~\cite{Gross:1973id,Politzer:1973fx,Deur:2025rjo}.
It is thus implied that the validity and rate of convergence of the perturbative expansion may depend on the energy scale of the interaction. The value of the strong coupling has been measured at different energy scales by various experiments. The most recent results from the CMS collaboration are shown in \figref{fig:alpha_s}. A standard reference point is the value at the $Z$-pole; the current world average value is $\als(m_Z) = (0.1180\pm 0.0009)$~\cite{Huston:2023ofk}.

\begin{figure}
    \centering
    \begin{subfigure}[t]{0.23\textwidth}
        \includegraphics[width=\linewidth]{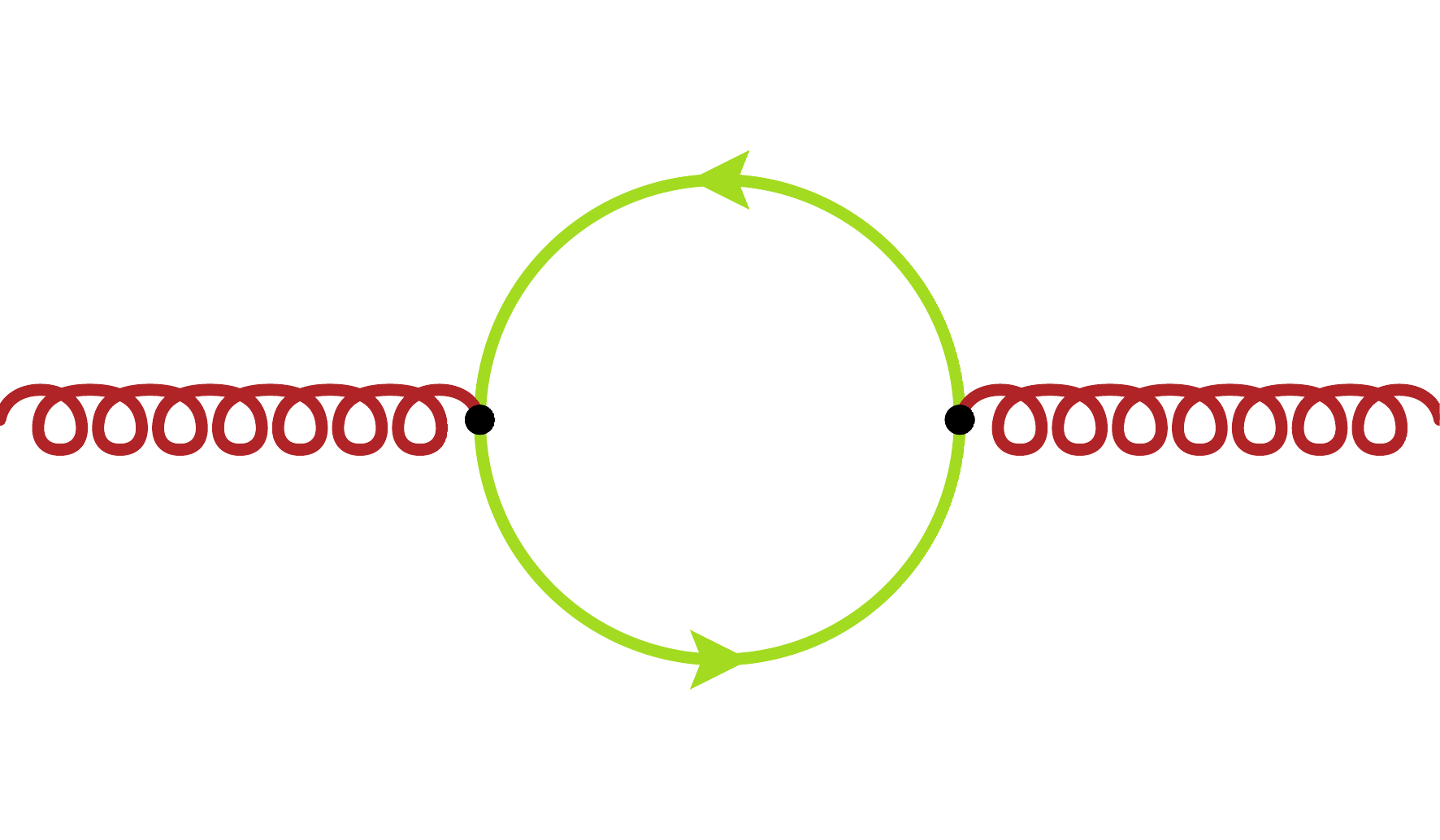}
        \caption{Quark loop}
    \end{subfigure}
    \hfill
    \begin{subfigure}[t]{0.23\textwidth}
        \includegraphics[width=\linewidth]{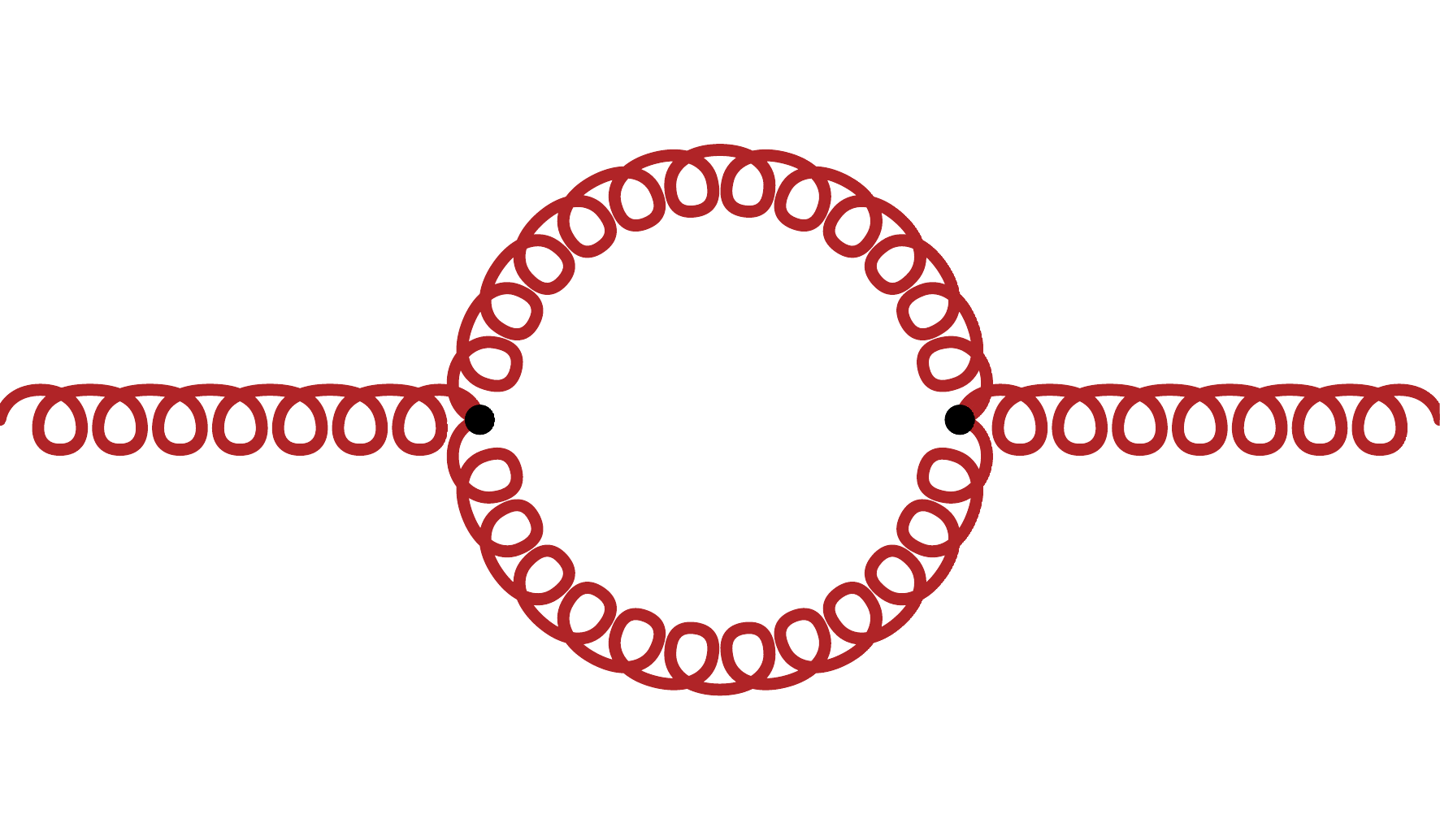}
        \caption{Gluon loop}
    \end{subfigure}
    \hfill
    \begin{subfigure}[t]{0.23\textwidth}
        \includegraphics[width=\linewidth]{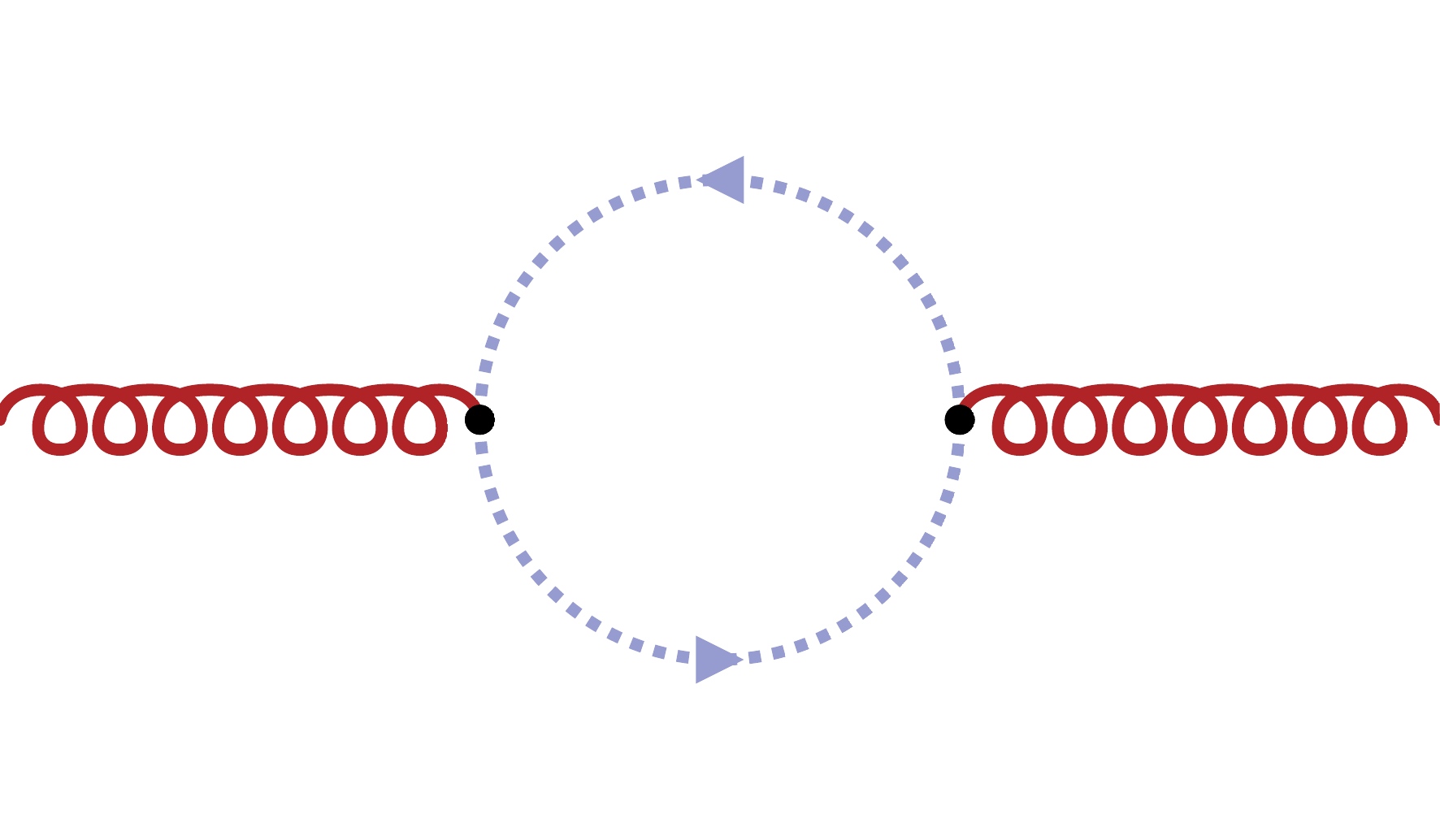}
        \caption{Ghost loop}
    \end{subfigure}
    \hfill
    \raisebox{0.24mm}{
    \begin{subfigure}[t]{0.23\textwidth}
        \includegraphics[width=\linewidth]{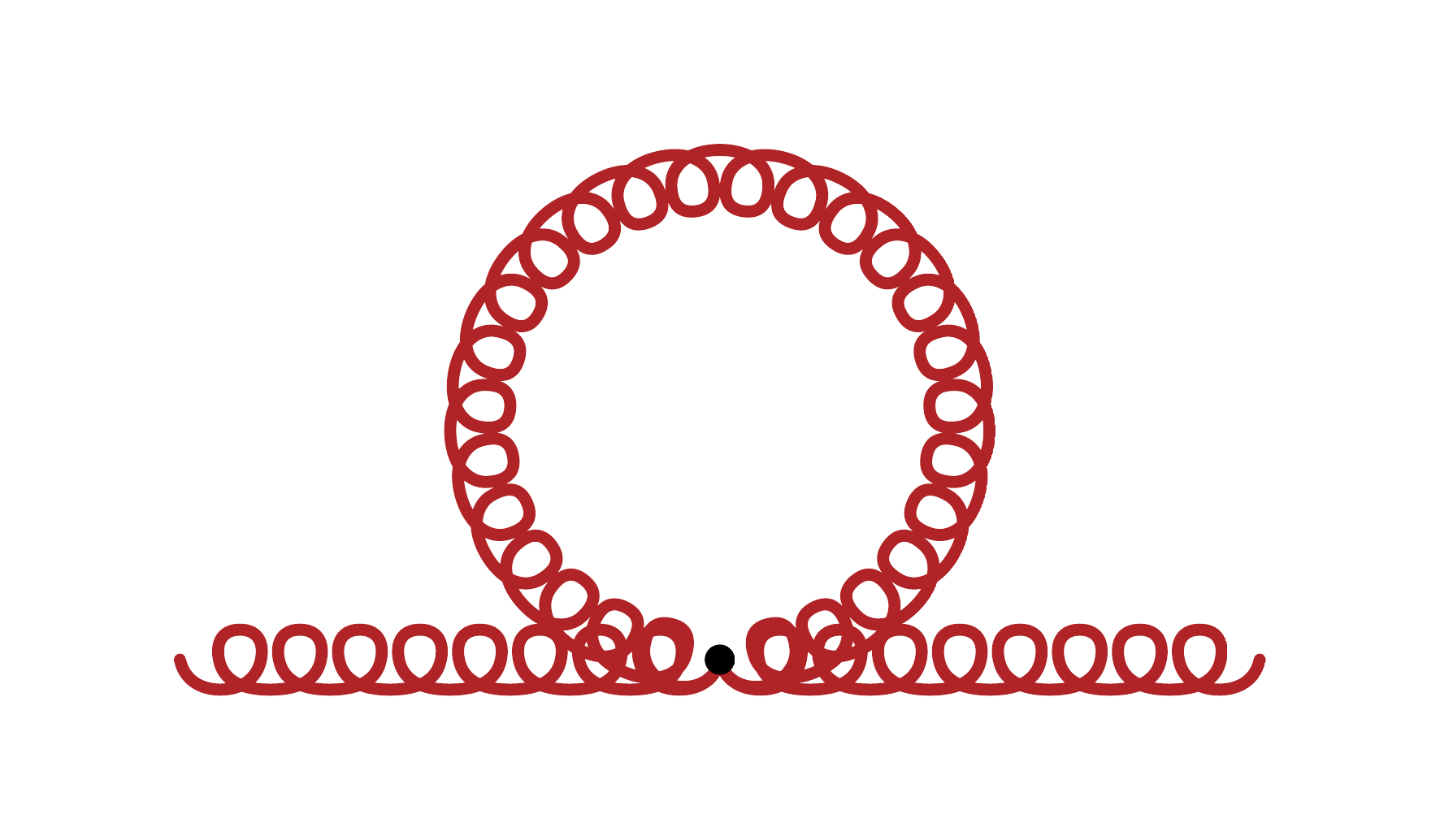}
        \caption{Gluon snail}
    \end{subfigure}}
    \caption{One-loop diagrams contributing to the correction of the gluon propagator and the running of $\alpha_s$. They comprise the first $\beta$-function coefficient, $b_0$. The quark loop contribution is proportional to $n_f$, the number of active flavours. }
  \label{fig:running_1L}
\end{figure}

The running is described by a renormalisation group equation (RGE), which for QCD takes the form \cite{Deur:2025rjo,ParticleDataGroup:2024cfk}
\begin{equation}
\label{eq:RGE}
    \mur^2\frac{\rd \alpha_s}{\rd \mur^2} = \beta(\alpha_s) = -\alpha_s^2\sum_{n=0}^{\infty} \alpha_s^n \, b_n,
\end{equation}
where $\beta(\alpha_s)$ is known as the QCD $\beta$-function. The second equality is a perturbative ansatz for the $\beta$-function and the $b_n$ are the $(n+1)$-loop $\beta$-function coefficients. They have been computed numerically up to five loops in the $\overline{\mathrm{MS}}$ renormalisation scheme \cite{Gross:1973id,Politzer:1973fx,Caswell:1974gg,Jones:1974mm,Egorian:1978zx,Larin:1993tp,vanRitbergen:1997va,Czakon:2004bu,Baikov:2016tgj,Luthe:2016ima,Herzog:2017ohr,Luthe:2017ttg,Chetyrkin:2017bjc}. The two first coefficients, $b_0$ and $b_1$, are renormalisation-scheme independent and have the form \cite{ParticleDataGroup:2024cfk}
\begin{equation}
\label{eq:b0_b1}
    b_0 = \frac{33-2n_f}{12\pi}, \quad b_1 = \frac{153-19n_f}{24\pi^2},
\end{equation}
where $n_f$ is the number of active quark flavours contributing to the running. The diagrams that contribute at one loop, i.e. to $b_0$, are shown in \figref{fig:running_1L}. In \equref{eq:b0_b1} both coefficients are positive for the number of quark flavours observed in nature, which means that the QCD $\beta$-function is negative. This predicts two characteristic properties of QCD, namely that the coupling decreases at higher energies (short distances) and increases at lower energies (long distances). The former is known as asymptotic freedom and the latter predicts the formation of QCD bound states \cite{ParticleDataGroup:2024cfk}. This can be seen explicitly from the solution of the RGE. At leading order it involves only $b_0$ and takes the form \cite{Schwartz:2014sze}
\begin{equation}
\label{eq:RGE_sol}
    \alpha_s(\mu_R) = \frac{2\pi}{b_0} \frac{1}{\ln \frac{\mu_R}{\Lambda_{\mathrm{QCD}}}}.
\end{equation}
This means that the perturbative regime for QCD, where $\alpha_s$ is small enough for the expansion in \equref{eq:perturbative_expansion_amplitude} to converge quickly, is the energy region above some low-energy cutoff, which is usually said to be $\Lambda_{\mathrm{QCD}}$.

\subsubsection{Loop integrals}
\label{sec:loops}

\input{loops.tex}

\subsection{Real radiation and infrared divergences}
\label{sec:IR}

\input{4b_realradiation.tex}

\subsubsection{Parton distribution functions and DGLAP evolution}

\input{PDFs.tex}

\label{sec:PDFs}

\subsection{Jets and event shape observables}

\input{4c_jets.tex}

\subsection{Estimation of theory uncertainties}

\input{4d_scaledep.tex}

\section{Current state of the art}

\input{5_state_of_art.tex}
\label{sec:state_of_art}


\section{Conclusions}
\label{sec:conclusions}

In this chapter, we have introduced basic concepts of perturbative QCD and outlined how calculations of higher perturbative orders are organised and how infrared singularities due to soft or collinear massless particles can be handled.
The depth is kept at a level that may serve beginning graduate students in entering the subject, giving also suggestions for further reading and some insight into the current state of the art with regard to precision calculations in perturbative QCD.

As the LHC experiments are progressing towards the high-luminosity phase, and in view of future colliders that will achieve even higher precision, the calculation of higher-order corrections in QCD will certainly continue to be one of the main pillars of the theoretical particle physics program. On the other hand, it is clear that only a multi-pronged approach can lead to better theory predictions overall: the limitations of the perturbative approach has to be carefully assessed, 
and better control of non-perturbative ingredients (such as PDFs, fragmentation functions, power corrections, effects of multi-parton scattering, hadronisation), of parton shower uncertainties and of parametric uncertainties (couplings, quark masses, etc.) should be part of the precision wishlist.
Furthermore, electroweak corrections will be of paramount importance, in particular at future lepton colliders. This also relates to the question how far analytic approaches can be pushed and whether analytic expressions are needed in case numerical approaches would lead to results of similar accuracy and speed, be it through ``traditional" methods or assisted by deep-learning approaches.
In any case, we should keep in mind that deeper insights into the mathematical structure of scattering amplitudes and radiation patterns are important drivers of conceptual progress, and the latter eventually leads to progress in physical applications.


\begin{ack}[Acknowledgments]%
  This research was supported by the Deutsche Forschungsgemeinschaft (DFG, German Research Foundation) under grant 396021762--TRR 257. 
\end{ack}


\bibliographystyle{Numbered-Style} 
\bibliography{references_pQCD}

\end{document}

%% file: intro.tex
Quantum Chromodynamics (QCD) is the theory of the strong interactions
between quarks, antiquarks and gluons, also called 
{\em partons}, after the  parton model that was introduced by Richard Feynman to describe the internal structure of hadrons (such as protons and neutrons), thus explaining the results of deep-inelastic scattering experiments.
In the 1960s, the parton model was complementary to the quark model developed by Gell-Mann, Zweig and others. Only later it was recognized that partons correspond to quarks and gluons.

The interactions are called  ``strong'' since they are the strongest of
the four known fundamental forces at a length scale a bit larger
than the proton radius.
At a distance of 1fm = $10^{-15}$m, which can be roughly associated with the radius of the proton, its strength is approximately 137 times higher than the electromagnetic force, approximately $10^{6}$ times higher than the weak force, and
about $10^{38}$ times higher than the gravitational force.
However, the strong coupling is not constant, it varies with
energy. The higher the energy at which we probe the interaction (i.e. the smaller the distance between
the partons, the weaker it will be.
%
This phenomenon is called {\em  asymptotic freedom}.
However, at large distances between the  quarks and gluons,
the interaction (i.e. the coupling) becomes very
strong. Therefore, they cannot be observed as isolated
particles. They are {\em confined} in hadrons, which are bound states
of several partons.

Why {\em Chromodynamics}?  In addition to the well-known quantum numbers
like electromagnetic charge, spin or parity, quarks carry an additional
quantum number called {\em colour} (the name was introduced by Murray Gell-Mann, reminiscent of the three primary colours red, green and blue). Bound
states are colour singlets, which means they are colour neutral or ``white''.
Quarks come in six different {\em flavours}, called $u, d, c, s, t, b$ (up, down, charm, strange, top, bottom). 
The top quark is the heaviest elementary particle known so far. A compelling reason why the quark masses of different flavours are so different has not been found yet.

Quarks are fermions, therefore, without the colour quantum number, a bound state consisting
of three quarks of the same type, e.g. three $u$-quarks (called $\Delta^{++}$) would violate the Pauli
exclusion principle if there was no additional quantum number to distinguish them.
 
The emergence of QCD from the quark model~\cite{GellMann:1964nj,Zweig:352337,Fritzsch:1973pi} started
more than 50 years ago, for a review see e.g. Ref.~\cite{Gross:2022hyw}.
QCD as the theory of strong interactions is nowadays well established, and experiments at high energy colliders have delivered an impressive
amount of high quality data in the last decades. This went hand-in-hand with
enormous progress in the calculation of perturbative QCD corrections
to scattering processes. However, there are still many open questions, and keeping up with the increasing experimental accuracy expected at the high-luminosity phase of the Large Hadron Collider (LHC) at CERN and at future colliders that are currently discussed is a challenge for perturbative QCD that will keep boosting the field of precision calculations.

There are various approaches to make theoretical predictions based on QCD. 
They can be put into two broad categories: (i) perturbative QCD (requires small coupling), (ii) non-perturbative QCD (e.g. ``Lattice QCD'').
We will focus on perturbative QCD in this Chapter.

The intention of the following sections is to
provide a pedagogical and concise introduction to the concepts and methods underlying perturbative calculations in QCD, aimed at persons that already have some basic knowledge of quantum field theory.
For further reading on the subjects of Sections \ref{sec:QCD}--\ref{sec:higher_orders}, textbooks such as Refs.~\cite{Ellis:1996mzs,Campbell:2017hsr,Dissertori:2003pj,Collins:2011zzd} can be useful.

In Section~\ref{sec:QCD}, the QCD Lagrangian is introduced. This section is rather short since there is the Chapter ``Introduction to QCD" to cover this topic in more detail. 
Section \ref{sec:perturbation_theory} is dedicated to basic concepts such as factorisation, the perturbative expansion of partonic cross sections and how to construct tree-level amplitudes from Feynman rules.
Section \ref{sec:higher_orders} represents the core of the chapter, discussing higher-order corrections in perturbation theory.  The running coupling is introduced, as well as loop integrals and dimensional regularisation, \secref{sec:IR} explains the treatment of soft and collinear singularities in QCD. At the end of Section~\ref{sec:higher_orders}, more phenomenological subjects are discussed, such as jets and event shapes and the estimation of theoretical uncertainties. Finally, in Section~\ref{sec:state_of_art}, the current state of the art in perturbative QCD is briefly reviewed.

%% file: loops.tex


In this section we describe how to turn integrals over loop momenta into parametric integrals and discuss some properties of loop integrals. For more details we refer to the chapter on Feynman diagrams in this volume, Ref. \cite{Weinzierl:2025blg}, and to Refs.~\cite{Smirnov:2004ym,Weinzierl:2022eaz}.
 
An integral with $L$ loops in $d$ dimensions, with $N$ propagators $P_j$, raised to the power $\nu_j$,  
can be written as
\begin{eqnarray}\label{eq0}
G(\nu_1\ldots\nu_N)  &=&
\int_{-\infty}^{\infty} \prod\limits_{l=1}^{L} \frac{\rd^d k_l}{i\pi^{\frac{d}{2}}}\;
\prod\limits_{j=1}^{N} 
\frac{1}{P_{j}^{\nu_j}(\{k,p\},m_j^2)}\;.
\end{eqnarray}
The propagators $P_{j}(\{k,p\},m_j^2)$ 
depend on the  loop momenta $k_{l}$, the external momenta
$\{p_1,\dots p_E\}$ and the (not necessarily non-zero) masses $m_j$. 
 Here we will restrict ourselves to the case where all propagator powers are positive, $\nu_j>0$. The factor $i\pi^{\frac{d}{2}}$ in the denominator is introduced for convenience, integrating over the loop momenta will cancel it.


To combine products of denominators of the type 
$P_j^{\nu_j}=[q_j^2(\{k,p\})-m_j^2+i\delta]^{\nu_j}$ into one single denominator, we can use 
the identity
\begin{equation}
\frac{1}{P_1^{\nu_1}P_2^{\nu_2}\ldots P_N^{\nu_N}}=
\frac{\Gamma(\sum_{i=1}^N\nu_i)}{\prod_{i=1}^N\Gamma(\nu_i)}
\int_0^\infty \prod_{i=1}^N \rd x_i\,x_i^{\nu_i-1}
\frac{\delta(1-\sum_{j=1}^N x_j)}{[x_1P_1+x_2P_2+\ldots+x_NP_N+i\delta]^{\sum_{i=1}^N\nu_i}}\label{eq:Feynpar}
\end{equation}
The integration parameters $x_i$ are called {\it Feynman parameters}. 
For generic one-loop diagrams we have $\nu_i=1\;\forall i$. The
propagator powers $\nu_i$ are also called {\it indices}.
We introduce the short-hand notation $N_{\nu}=\sum_{i=1}^N\nu_i$.
Using Eq.~(\ref{eq:Feynpar}) for each propagator, irrespective of which loop momenta the propagator involves,
leads to the following form:
\begin{align}
G &=  \frac{\Gamma(N_\nu)}{(i\pi)^{L\frac{d}{2}}}\int \prod\limits_{j=1}^{N}\,dx_j\,x_j^{\nu_j-1}
 \,\, \delta(1-\sum_{i=1}^N x_i)
\, \int_{-\infty}^{\infty} \rd k_1\dots \rd k_L 
\left[ 
       \sum\limits_{j,l=1}^{L} k_j\cdot k_l \, M_{jl} - 
                                        2\sum\limits_{j=1}^{L} k_j\cdot Q_j +J +i\delta\right]^{-N_\nu}\;,
  \label{eq:Feynparmultiloop}\end{align}
where we have used
\begin{align}
  \sum_{i=1}^N x_i P_i=\sum\limits_{j,l=1}^{L} k_j\cdot k_l \, M_{jl} - 
2\sum\limits_{j=1}^{L} k_j\cdot Q_j +J +i\delta\;,\end{align}
and $k_j\cdot k_l$ denotes the scalar product of two $d$-dimensional Lorentz-vectors. The matrix $M$ has the Feynman parameters as entries that multiply the bilinear terms in the loop momenta, $Q$ is an array of dimension $L$, where each entry contains the combination of Feynman parameters and external momenta that multiply the term linear in the corresponding loop momentum $k_j$, and $J$ collects the terms that do not involve loop momenta.

The benefit of this procedure lies in the fact that,
after the shift $k_j=l_j+M_{jl}^{-1}Q_l$ that eliminates the linear term, we arrive at a quadratic form in the loop momenta, and the loop momentum integration in $L\times d$ dimensions can be carried out after using Wick rotation 
and Gaussian integration. This leads to
  \begin{align}
 G&=(-1)^{N_{\nu}}\frac{\Gamma(N_{\nu}-Ld/2)}{\prod_{j=1}^{N}\Gamma(\nu_j)}
\int\limits_{0}^{\infty} \prod\limits_{j=1}^{N}\,\rd x_j\,x_j^{\nu_j-1} 
\delta(1-\sum_{i=1}^N x_i)\,\frac{{\cal U}^{N_{\nu}-(L+1) d/2}}
{{\cal F}^{N_{\nu}-L d/2}}\;,\label{EQ:param_rep}\\
&\nn\\
{\cal U}&=\det(M)\quad , \quad 
{\cal F}=\det(M)\,\left[ \sum_{i,j=1}^L Q_i M_{ij}^{-1}Q_j-J-i\delta \right] \;.\nn
\end{align}
The functions ${\cal U}$ and ${\cal F}$ are also called first and second {\em Symanzik polynomial}, respectively.
A general representation for tensor integrals is straightforward, it can be found e.g. in Ref.~\cite{Heinrich:2008si}.

\vspace*{3mm}

\noindent {\bf Discussion of singularities}

A necessary condition
for the presence of infrared divergences is 
${\cal F} = 0$.
The function ${\cal U}$ cannot lead to infrared divergences of the 
graph, since giving a mass to all external legs would not change ${\cal U}$.
Apart from the fact that the graph may have an overall UV divergence contained in
the overall $\Gamma$-function in Eq.~(\ref{EQ:param_rep}), UV subdivergences
may also be present beyond one loop. A necessary condition for the latter is  
that ${\cal U}$ is vanishing. The exponent of ${\cal U}$ decreases with the number of loops and dimensions and therefore a negative power of  ${\cal U}$ points to a potential  UV divergence.

The function  ${\cal F}$ can  vanish within the integration 
region on a hyper-surface given by  solutions of the Landau
equations~\cite{Landau:1959fi,ELOP,Collins:2020euz}, corresponding for example to physical thresholds or to endpoint singularities. In momentum space, the Landau equations can be formulated as follows. If the $N$ propagators are denoted by $P_i=q_i^2(\{k,p\})-m_i^2+i\delta$ and  $x_i$ are the Feynman parameters associated with propagator $P_i$, they read
\begin{align}
\begin{split}
x_i\,(q_i ^2(\{k,p\})-m_i^2)&=0\quad \forall\, i\in \{1,\ldots,N\}\\
\frac{\partial}{\partial k_l^\mu}\sum_{i\in {\rm loop}\, l}\, x_i\,(q_i ^2(\{k,p\})-m_i^2)&=0\quad \forall\, l \in \{1,\ldots,L\}\;.
\end{split}
\end{align}
The Landau equations are necessary, but in general not sufficient conditions for a singularity to be produced.
The first condition contains endpoint singularities ($x_i=0$) as well
as kinematic singularities, related to a propagator  going on-shell, ($q_i^2=m_i^2$).
In Feynman parameter space the Landau equations translate to 
\begin{align}
&  {\cal F} = 0 \quad \mbox{ and }\quad
\left(\mbox{ either } x_i = 0 \;
\mbox{ or } \frac{\partial}{\partial x_i} {\cal F} = 0\; \right)\,\; \forall i\;\;.
\end{align}
A singularity with all $x_i\not=0$ is called {\em leading Landau singularity}. 
Subleading singularities with $x_i = 0$ for a subset of the parameters $x_i$  correspond to singularities of subgraphs.

\vspace*{3mm}

\noindent {\bf Example for the construction of Symanzik polynomials from  propagators}

\vspace*{3mm}

\begin{figure}[htb]
\centering
\includegraphics[width=0.3\linewidth]{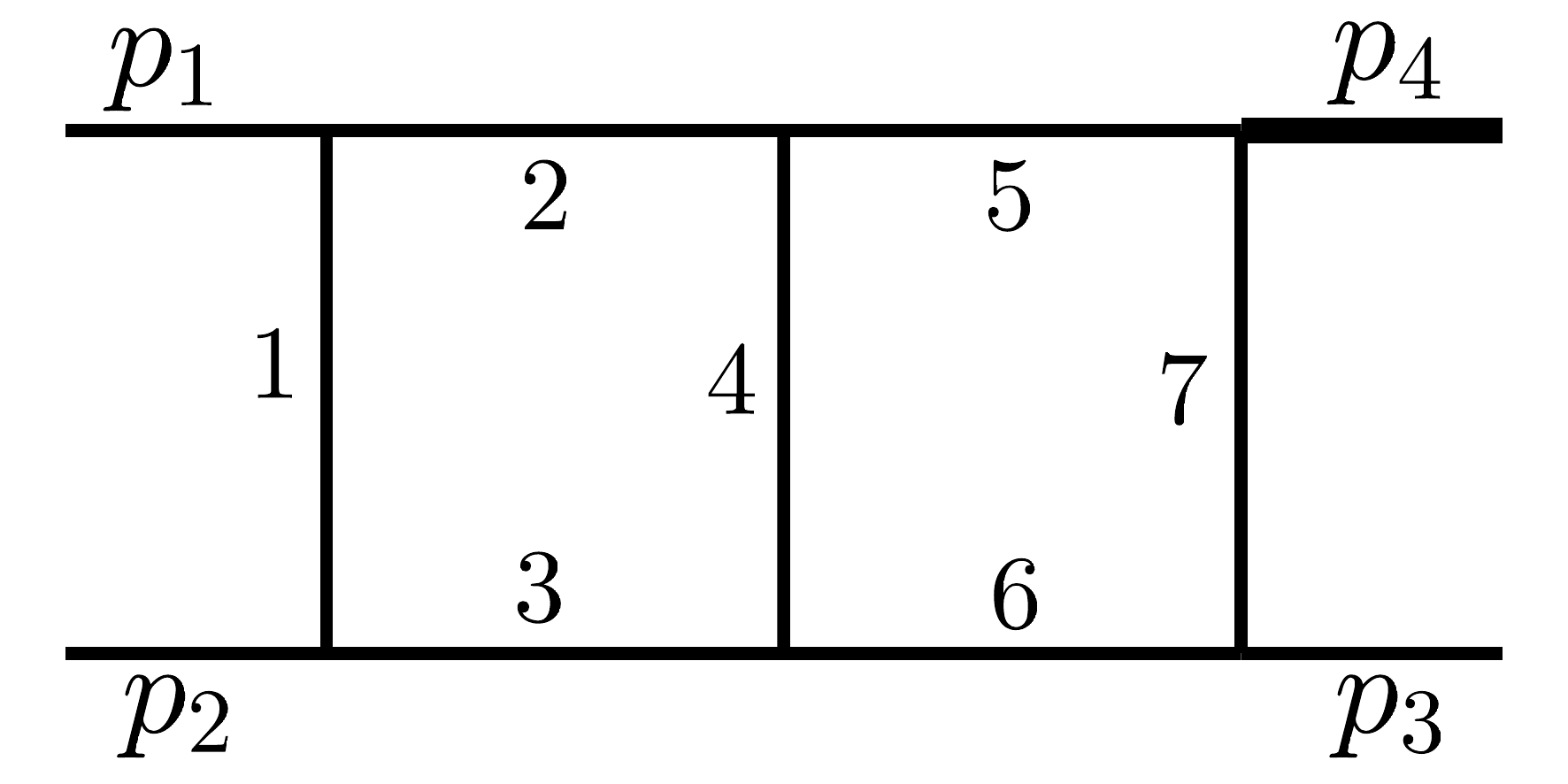}
\caption{Labelling for the planar two-loop box example with $p_4$ off-shell.}
\label{fig:two_loop_box_off_shell}
\end{figure}

\noindent As an example we consider a planar two-loop box integral with $p_1^2=p_2^2=p_3^2=0, \,p_4^2\not=0$, see \figref{fig:two_loop_box_off_shell}.
Using $k_1=k,k_2=l$ and labelling $1/(k^2+i\delta)$ as propagator number one, 
the denominator, after Feynman parametrisation, can be written as
\begin{align}
{\cal D}&=x_1\,k^2+x_2\,(k-p_1)^2+x_3\,(k+p_2)^2+x_4\,
(k-l)^2+x_5\,(l-p_1)^2
+x_6\,(l+p_2)^2+x_7\,(l+p_2+p_3)^2+i\delta\nn\\
&=(\,k,l\,)\left(\begin{array}{ll}
x_{1234} &-x_4\\
-x_4&x_{4567}\end{array}\right)\left(\begin{array}{c}k\\l\end{array}\right)
-2\,\left(Q_1,Q_2\right)\,
\left(\begin{array}{c}k\\l\end{array}\right)\;+\;x_7\,(p_2+p_3)^2+i\delta\nn\\
Q&=(Q_1,Q_2)=\left(x_2p_1-x_3p_2, x_5p_1-x_6p_2-x_7(p_2+p_3)\right)
\;,\nn
\end{align}
where we have used the short notation
$x_{ijk\ldots}=x_i+x_j+x_k+\ldots$.	
We find
\begin{align}
{\cal U} &= x_{123} x_{567} + x_{4} x_{123567} \\
{\cal F} &= (-s_{12})\, (x_2 x_3 x_{4567} + x_5 x_6 x_{1234} 
                         + x_2 x_4 x_6 + x_3 x_4 x_5) 
            +(-s_{23})\, x_1 x_4 x_7  
            + (-p_4^2) \,x_7 ( x_2 x_4 + x_5 x_{1234} ) -i\delta\;.  \nn     
\end{align} 
Another possibility to construct ${\cal F}$ and ${\cal U}$ is from topological rules, this is explained e.g. in Ref.~\cite{Heinrich:2020ybq}.


\subsubsection{Scattering Amplitudes}

The loop integrals, of course, form just one building block of scattering amplitudes. 
 The typical workflow to calculate an amplitude beyond one loop is the following:
\begin{enumerate}[align=left]
\item amplitude generation, for example in terms of Feynman diagrams, 
\item reduction of the occurring integrals to a minimal set, the so-called {\em master integrals},
\item calculation of amplitude as a linear combination of the master integrals.
 \end{enumerate}
For the reduction to master integrals, powerful automated and publicly available programs exist, such as 
{\sc Fire}~\cite{Smirnov:2008iw,Smirnov:2023yhb},
{\sc Reduze}~\cite{Studerus:2009ye,vonManteuffel:2012np}, 
{\sc LiteRed}~\cite{Lee:2012cn,Lee:2013mka},
{\sc Kira}~\cite{Maierhofer:2017gsa, Klappert:2020nbg,Lange:2025fba},
{\sc Blade}~\cite{Guan:2024byi} or {\sc NeatIBP}~\cite{Wu:2023upw}.  The use of finite-field techniques, as implemented in {\sc FireFly}~\cite{Klappert:2019emp,Klappert:2020aqs}, {\sc FiniteFlow}~\cite{Peraro:2019svx} or
{\sc RaTracer}~\cite{Magerya:2022hvj} can be used to speed up the functional reconstruction of the coefficients of the master integrals.

The analytic calculation of multi-loop integrals today is mostly based on differential equations~\cite{Kotikov:1990kg,Remiddi:1997ny,Gehrmann:1999as,Henn:2013pwa} rather than direct
integration in Feynman parameter space. The main idea of the DE method is to take derivatives of a given integral with respect to kinematic invariants and/or masses, which relates them to other integrals of a given family. This leads to a system of differential equations for the master integrals which can be solved given appropriate boundary conditions, see e.g. Refs.~\cite{Argeri:2007up,Henn:2014qga} for a pedagogical introduction.
In the presence of several mass scales, a fully analytic solution of the differential equations is hard to obtain; then the use of generalised series expansions as implemented in
\textsc{DiffExp}~\cite{Hidding:2020ytt}, \textsc{SeaSyde}~\cite{Armadillo:2022ugh} or \textsc{Line}~\cite{Prisco:2020kyb} is very useful.
The method of Auxiliary Mass Flow~\cite{Liu:2017jxz,Liu:2020kpc,Liu:2021wks}, implemented in \textsc{AMFlow}~\cite{Liu:2022chg}, can also be used for high precision numerical evaluations of master integrals after reduction.

Numerical calculations of multi-loop integrals are only meaningful if potential UV and IR singularities are isolated and subtracted beforehand. In Feynman parameter space, this can be achieved for example via sector decomposition~\cite{Hepp:1966eg,Roth:1996pd,Binoth:2000ps}.
Modern tools to perform the numerical integration of multi-loop integrals in Feynman parameter space are e.g. {\sc Fiesta}~\cite{Smirnov:2015mct,Smirnov:2021rhf}, {\sc pySecDec}~\cite{Borowka:2017idc,Borowka:2018goh,Heinrich:2021dbf,Heinrich:2023til} or {\sc FeynTrop}~\cite{Borinsky:2023jdv}.

Scattering amplitudes are at the core of any
perturbative calculation of a physical quantity relevant to particle
interactions in collider experiments.
The calculation of scattering amplitudes beyond the leading order in
perturbation theory has advanced immensely in the last decade,
which led to a deeper mathematical understanding of the structure of both tree- and
loop amplitudes, and opened the door to many important
phenomenological applications. For further reading we refer to Refs.~\cite{Heinrich:2020ybq,Agarwal:2021ais,Huss:2025nlt}, see also \secref{sec:multi-loop}.

%% file: 4b_realradiation.tex
One of the advantages of dimensional regularisation is the fact that it can regulate both, UV and IR divergences. Conceptually, however, the treatment of these two types of divergences is very different. 
While the UV divergences are subtracted through a renormalisation procedure, the IR divergences cancel under certain conditions between real and virtual higher order corrections. Initial-state collinear singularities in hadronic collisions do not cancel, but can be absorbed into the ``bare" parton distribution functions. The latter procedure is very similar to renormalisation.

\subsubsection{The KLN-Theorem}

\begin{figure}
    \centering
    \begin{subfigure}[t]{0.2\textwidth}
        \includegraphics[width=\textwidth]{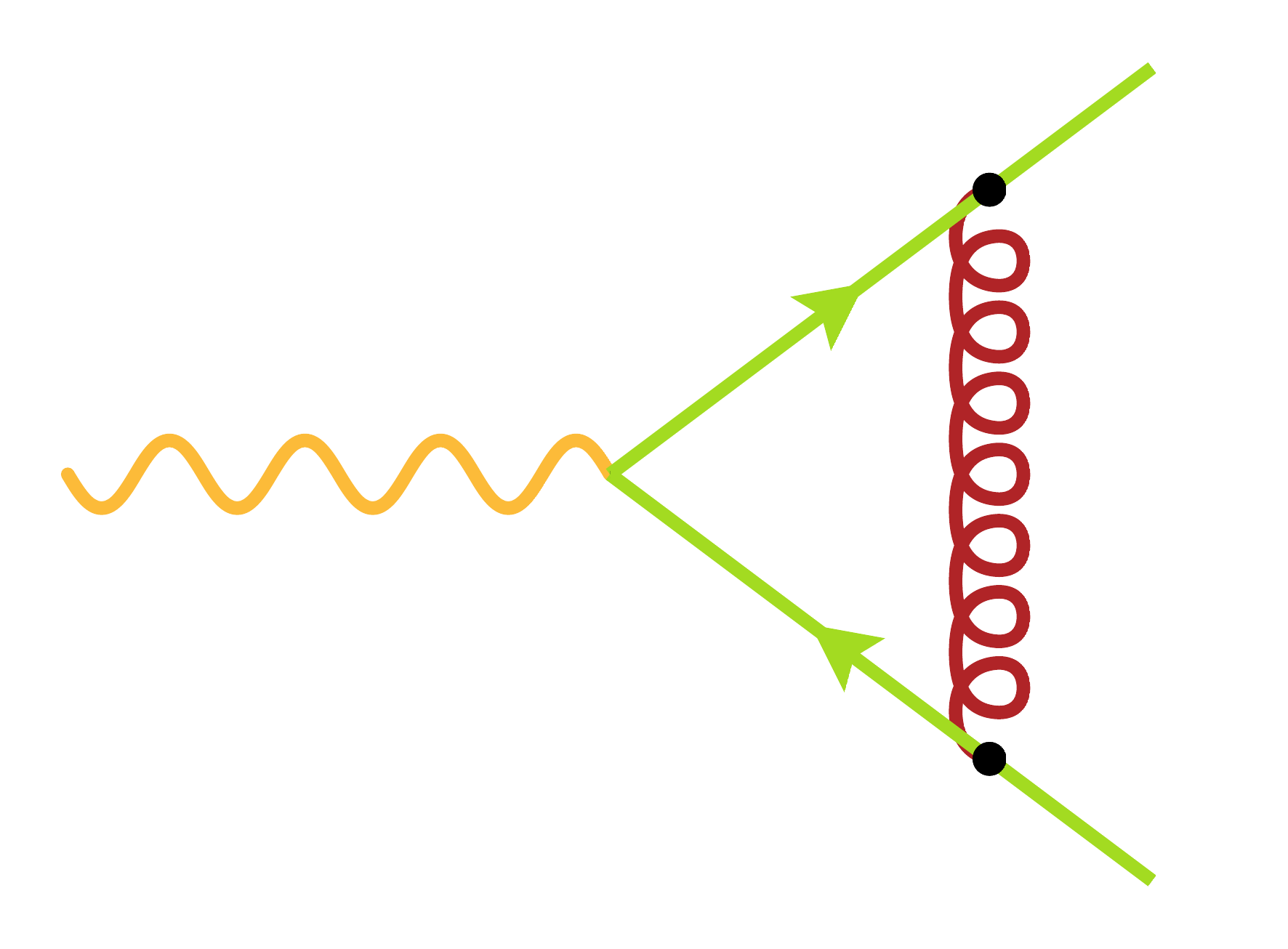}
    \end{subfigure}
    \begin{subfigure}[t]{0.2\textwidth}
        \includegraphics[width=\textwidth]{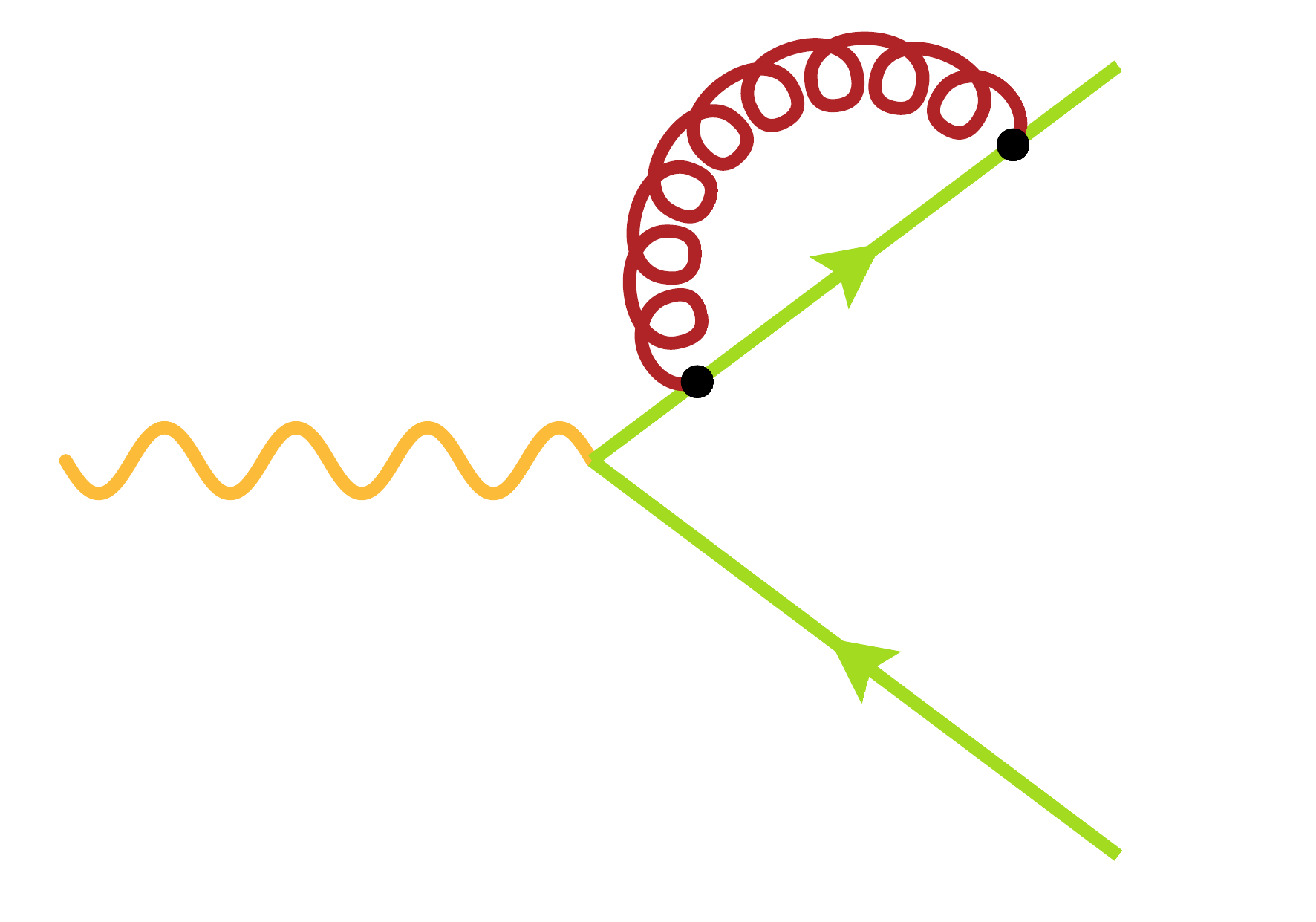}
    \end{subfigure}
    \begin{subfigure}[t]{0.2\textwidth}
        \includegraphics[width=\textwidth]{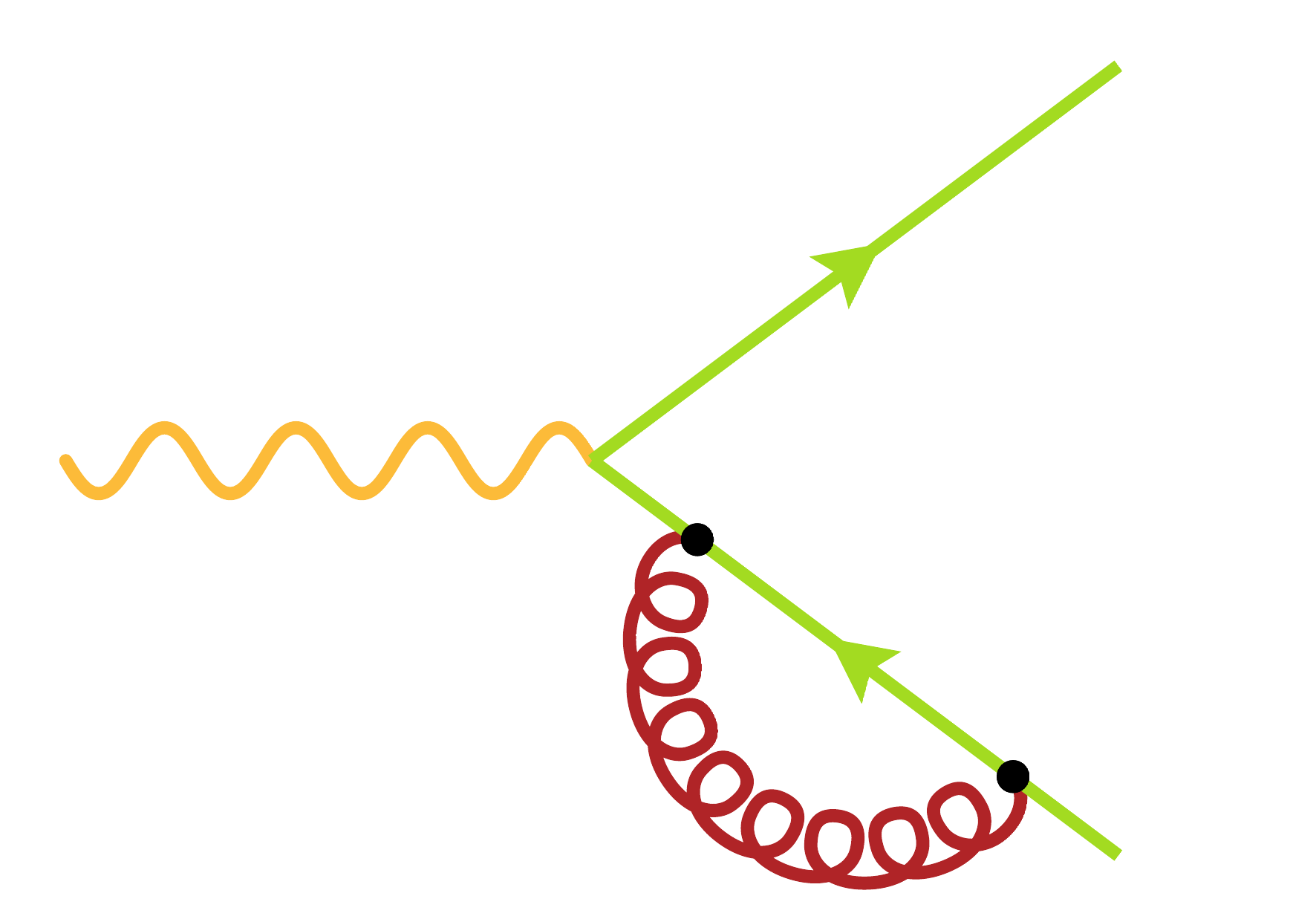}
    \end{subfigure}
    
    \begin{subfigure}[t]{0.2\textwidth}
        \includegraphics[width=\textwidth]{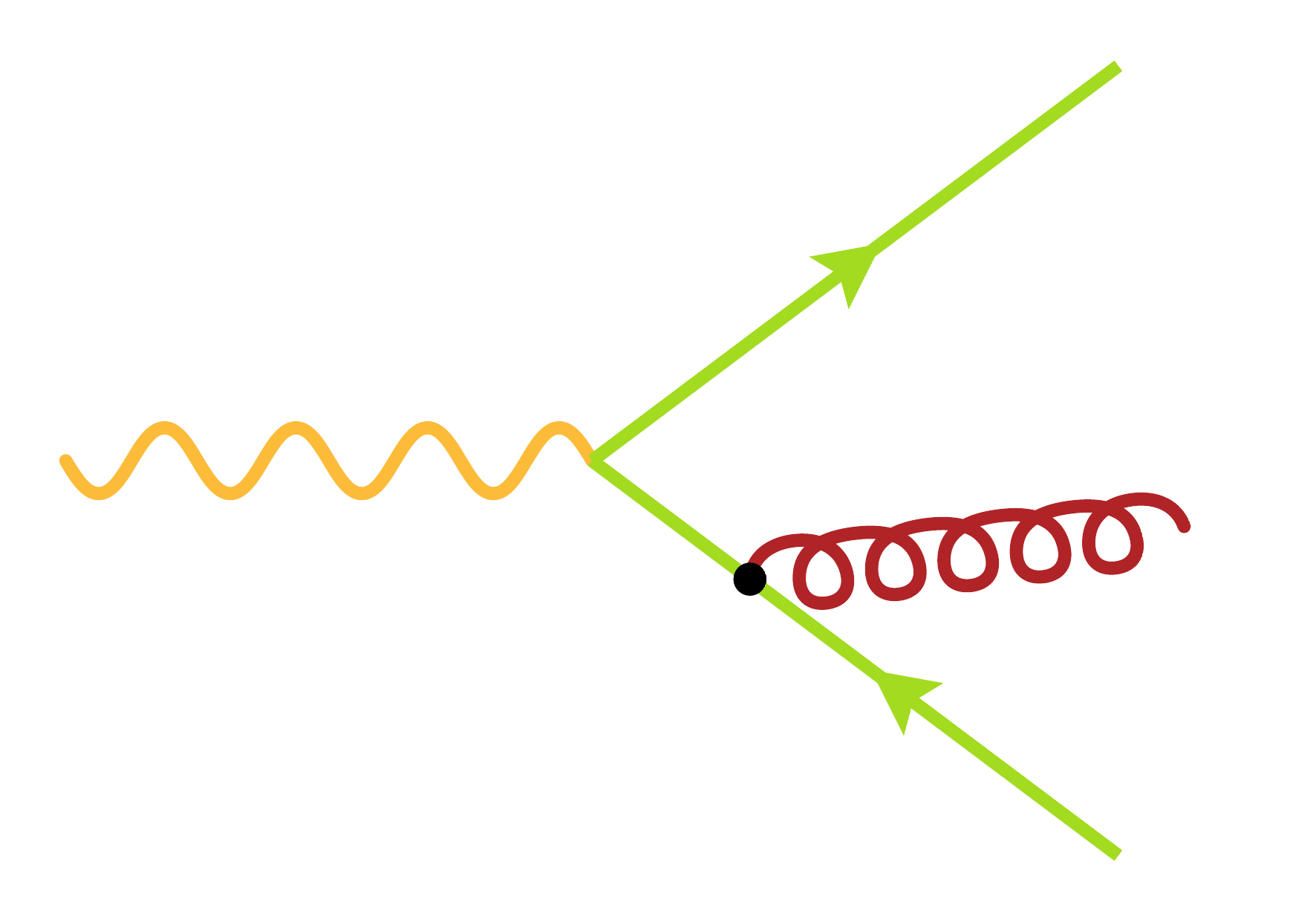}
    \end{subfigure}
    \begin{subfigure}[t]{0.2\textwidth}
        \includegraphics[width=\textwidth]{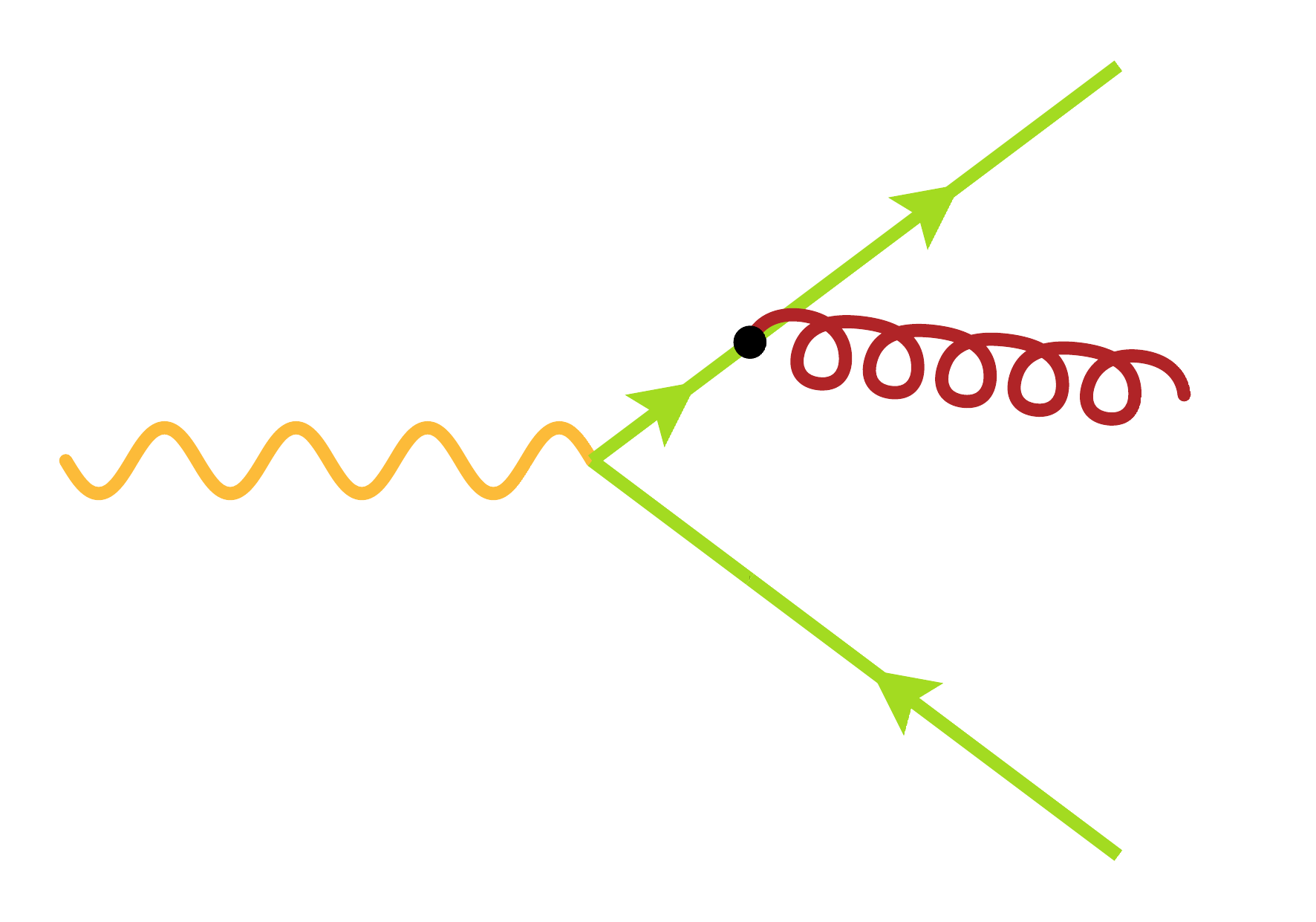}
    \end{subfigure}
    \caption{The virtual (first line) and real (second line) NLO QCD contributions to $\gamma^*\to q\bar{q}$.}
    \label{fig:real_virtual}
\end{figure}

To illustrate the mechanisms of cancellation and subtraction of IR singularities, let us consider as an example 
the ${\cal O}(\als)$ real and virtual contributions to $\gamma^*\to q\bar{q}$, wich can be considered as the hadronic part of $e^+e^-\to q\bar{q}$. The corresponding diagrams are shown in Fig.~\ref{fig:real_virtual}.


If ${\cal M}_0$ is the leading order amplitude  and ${\cal M}_{\rm{virt}}, {\cal M}_{\rm{real}}$ are the
virtual and real NLO amplitudes as shown in Fig.~\ref{fig:real_virtual}, 
the corresponding cross section is given by 
\begin{align}
\sigma^{NLO}=\underbrace{\int \mathrm{d}\Phi_2 \left|{\cal M}_0
\right|^2}_{\sigma^{LO}}+\underbrace{\int_R \mathrm{d}\Phi_3 \left|{\cal
M}_{\rm{real}}\right|^2}_{\sigma_{R}} +\underbrace{\int_V \mathrm{d}\Phi_2 \,2 \mathrm{Re}\left({\cal
M}_{\rm{virt}} {\cal M}_0^*\right)}_{\sigma_{V}} \;.
\label{eq:sigNLO}
\end{align}
The sum of the integrals $\int_R$ and $\int_V$ above is finite.
However, this is not true for the individual contributions. The real
part contains divergences due to soft and collinear radiation of
gluons. While ${\cal M}_{\rm{real}}$ itself is a tree
level amplitude and thus finite, the divergences emerge upon
integration over the 3-particle phase space $\mathrm{d}\Phi_3$.
In contrast, for $\int_V$ the phase
space $\rd \Phi_2$ is the same as for the Born amplitude, but the loop
integrals  in ${\cal M}_{\rm{virt}}$ contain explicit IR
singularities stemming from the integration over the loop momentum, as the latter can also become soft, or collinear to an external momentum.

\noindent Let us anticipate the answer, which we will (partly) calculate later. 
We find: 
\begin{align}
\sigma_R&=\sigma^{\rm{LO}}\tilde{H}(\eps)\,C_F\frac{\als}{2\pi}\left(\frac{2}{\eps^2}+\frac{3}{\eps}+\frac{19}{2} \right)\;,\label{eq:sigmar}\\
\sigma_V&=\sigma^{\rm{LO}}H(\eps)\,C_F\frac{\als}{2\pi}\left(-\frac{2}{\eps^2}-\frac{3}{\eps}-8\right)\;,\nn
\end{align}
where 
$H(\eps)=\left(\frac{4\pi\mu^2}{-Q^2}\right)^\eps\frac{\Gamma(1+\eps)\Gamma^2(1-\eps)}{\Gamma(1-2\eps)}$
and
$\tilde{H}(\eps)=H(\eps)+{\cal O}(\eps^3)$.
The exact $\eps$-dependence of $H(\eps)=1+{\cal O}(\eps)$
is irrelevant after summing up real and virtual contributions,
because the poles in $\eps$ all cancel. \\
This must be the case according to the
{\bf KLN  theorem} (Kinoshita-Lee-Nauenberg)~\cite{Kinoshita:1962ur,Lee:1964is}. 
It says that

\vspace*{1mm}

 {\it IR singularities must cancel when summing the
    transition rate over all degenerate (initial and final) states.}

\vspace*{1mm}

\noindent In our example, we do not have initial-state singularities. 
However, in the final state we can have a massless quark accompanied by
a soft gluon, or a collinear quark-gluon pair.
Such a state cannot be distinguished from just a quark state,
and therefore these two configurations are ``degenerate''. 
Only when summing over all the final-state multiplicities contributing to the cross section at a given
order in $\als$, the divergences cancel.
Initial-state radiation is more difficult, because the initial state is typically fixed by the experiment. In addition, for hadronic collisions, it is impossible to determine all quark and gluon configurations in the proton, as this is a non-perturbative bound state. Therefore, initial-state singularities in hadronic collisions are absorbed in ``bare" parton distribution functions (PDFs) to obtain the PDFs that are determined from data.
PDFs are introduced in a dedicated chapter in this volume, see also Ref.~\cite{Lorce:2025aqp}.

\begin{figure}[htb]
  \centering
  \includegraphics[width=0.7\textwidth]{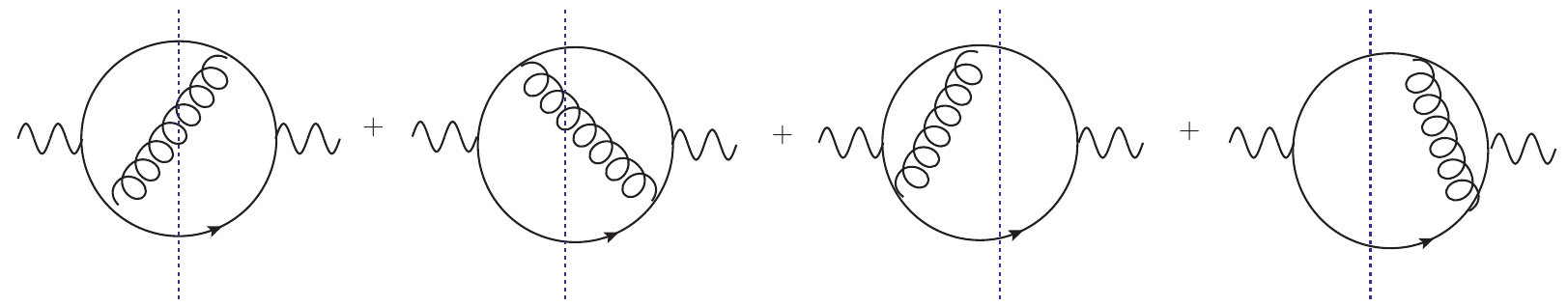}
  \caption{The sum over cuts of the amplitude squared shown above is
  finite according to the KLN theorem.}
  \label{fig:cutkosky}
\end{figure}

Another way of stating the cancellation mechanism of (final state) IR divergences 
is by looking at the squared amplitude at order $\als$ and considering all cuts of $|{\cal M}|^2$,  see
Fig.~\ref{fig:cutkosky}. This notation makes use of the optical theorem~\cite{Schwartz:2014sze}, the cut propagators denote the on-shell final-state particles. Self-energy contributions, which are zero for massless quarks, are not shown.
The KLN theorem states that {\it the sum of all diagrams resulting from cuts that lead to physical
final states is free of IR poles.}

The cancellations between $\sigma_R$ and $\sigma_V$ in
\equref{eq:sigNLO} are non-trivial, because the phase-space
integrals contain a different number of particles in the final state.


\vspace*{3mm}

\noindent {\bf Phase-space integrals in $d$ dimensions}

To see how the cancellation works for {\em inclusive} quantities such as the total cross
section, let us consider the real radiation contribution to $e^+e^-\to
2$\,jets at NLO (corresponding to the second line in Fig.~\ref{fig:real_virtual}) in more detail. For this purpose we need phase-space integrals in $d$ dimensions.

The general formula for a $1\to n$ particle phase space $\mathrm{d} \Phi_n$
with $Q\to p_1\ldots p_n$ is given by
\begin{equation}\label{eq:phin}
\mathrm{d}\Phi_{1\to n} = (2\pi)^{ n - d (n-1)} 
\Big[ \prod\limits_{j=1}^{n} \rd^d p_j \, \delta(p_j^2-m_j^2) \, \Theta(E_j)\Big] \,
\delta\Bigl(Q-\sum\limits_{i=1}^{n} p_i \Bigr)\;.
\end{equation}
In the following we will stick to the massless case $m_j=0$.
We use
\begin{equation}
\rd^d p_j \, \delta(p_j^2) \, \Theta(E_j)=\rd E_j
\, \rd^{d-1}\vec{p}_j \, \delta(E_j^2-\vec{p}_j^2) \, \Theta(E_j)=
\frac{1}{2E_j}\rd^{d-1}\vec{p}_j\Big|_{E_j=|\vec{p}_j|}
\end{equation}
for $j=1,\ldots,n-1$ to arrive at
\begin{align}
\mathrm{d}\Phi_{1\to n} = (2\pi)^{ n - d (n-1)} 2^{1-n}
 \prod\limits_{j=1}^{n-1}\frac{\mathrm{d}^{d-1}\vec{p}_j}{|\vec{p}_j|}\delta\Bigl(p_n^2 \Bigr)\Big|_{p_n=Q-\sum\limits_{i=1}^{n-1} p_i}\;,  
\end{align}
 where we have used the last $\delta$-function in \equref{eq:phin} to eliminate $p_n$.
We further use
\begin{align}
&\frac{\mathrm{d}^{d-1}\vec{p}}{|\vec{p}|}\,f(|\vec{p}|)=\mathrm{d}\Omega_{d-2}\,\mathrm{d} |\vec{p}|\,|\vec{p}|^{d-3}\,f(|\vec{p}|)\;,\\
&\int \mathrm{d}\Omega_{d-2}=\int \mathrm{d}\Omega_{d-3}\int_0^\pi\mathrm{d}\theta
(\sin\theta)^{d-3}=\int_0^\pi\mathrm{d}\theta_{1}
(\sin\theta_{1})^{d-3}\int_0^\pi\mathrm{d}\theta_{2}
(\sin\theta_{2})^{d-4}\ldots \int_0^{2\pi}d\theta\;,\nn\\
&\int\limits_{S_{d-2}} \mathrm{d}\Omega_{d-2} = 
 V(d-1) = \frac{2\,\pi^{\frac{d-1}{2}}}{\Gamma(\frac{d-1}{2})}\;,\nn
\end{align}
to obtain
\begin{align}
\mathrm{d}\Phi_{1\to n} = (2\pi)^{ n - d (n-1)} 2^{1-n}
 \left(\prod\limits_{j=1}^{n-1}\mathrm{d}\Omega_{d-1-j}\mathrm{d} |\vec{p}_j|\,|\vec{p}_j|^{d-3}\right)\,\delta\left(\Bigl(Q-\sum\limits_{i=1}^{n-1} p_i\Bigr)^2\right) \;.
\end{align}

\vspace{3mm}

\noindent {\bf Example $1\to 3$:} \\
For $n=3$ one can choose a coordinate frame such that 
\begin{align}
Q   = (E,\vec 0^{(d-1)});      
p_1 = E_1\, (1,\vec 0^{(d-2)},1);
p_2 = E_2\, (1,\vec 0^{(d-3)},\sin\theta,\cos\theta);
p_3 &= Q-p_2-p_1\;,
\end{align}
leading to
\begin{align}
\mathrm{d}\Phi_{1\to 3} =\frac{1}{4} (2\pi)^{3- 2\,d} \; \mathrm{d} E_1 \mathrm{d} E_2 
\mathrm{d}\theta_1 \left(E_1 E_2\sin\theta\right)^{d-3}
\mathrm{d}\Omega_{d-2} \; \mathrm{d}\Omega_{d-3}\,
\Theta(E_1)\,\Theta(E_2)\,\Theta(E-E_1-E_2)\,\delta(p_3^2)\Big|_{p_3=Q-p_1-p_2}\;.  
\label{eq:phi13}
\end{align}
In the following a parametrisation in terms of the Mandelstam variables
$s_{ij}=2\,p_i\cdot p_j$
will be useful, therefore we make the transformation 
$E_1, E_2, \theta \to s_{12},s_{23},s_{13}$.
To work with dimensionless variables we define $y_1=s_{12}/Q^2$, $y_2=s_{13}/Q^2$, $y_3=s_{23}/Q^2$
which leads to 
\begin{align}
\mathrm{d}\Phi_{1\to 3} &= (2\pi)^{3- 2\,d}\, 2^{-1-d}
(Q^2)^{d-3}\;\mathrm{d}\Omega_{d-2} \; \mathrm{d}\Omega_{d-3}\,\mathrm{d} y_{1}\, \mathrm{d} y_{2}\,
\mathrm{d} y_{3}\,
 \left(y_1\,y_2\,y_3\right)^{d/2-2}\, \Theta(y_1)\, \Theta(y_2)\, \Theta(y_3) \,\delta(1-y_1-y_2-y_3)\;.\nonumber
\end{align}
 Now we are in the position to calculate the full real radiation
 contribution.
The matrix element (for one quark flavour with charge $q_f$) in the
variables defined above, where $p_3$ is the gluon, is given by 
\begin{align}
|{\cal M}|^2_{\rm{real}}=C_Fe^2q_f^2g_s^2\,8\,(1-\eps)\,\left\{\frac{2}{y_2y_3} +\frac{-2+(1-\eps)y_3}{y_2}+\frac{-2+(1-\eps)y_2}{y_3}-2\eps\right\}\;.\label{eq:Mreal}
\end{align}
In our variables, soft singularities mean $p_3\to 0$ and therefore
both $y_2$ and $y_3\to 0$, while $p_3\parallel p_1$ means $y_2\to 0$
and $p_3\parallel p_2$ means $y_3\to 0$. Combined with the factors
$\left(y_2\,y_3\right)^{d/2-2}$ from the phase space it is clear that
the first term in the bracket of \equref{eq:Mreal} will lead to a
$1/\eps^2$ pole, coming from the region in phase space where soft and
collinear limits coincide.
%
The integrals can be expressed in terms of Euler-Beta functions and lead
to the result quoted in \equref{eq:sigmar}.

\subsubsection{Infrared safety}

If we want to calculate a prediction for a certain observable, based on an $n$-particle final state,
we need to multiply the amplitude by a {\it measurement function} $J(p_1 \ldots p_n)$ that specifies the observable.
The measurement function can contain for example a jet definition, or the definition of an event-shape observable,
or it defines observables such as the transverse momentum distribution of a final-state particle.
Schematically, the structure of the NLO cross section
 is the following. In the real radiation part, we have $n+1$ particles in the final state.
Therefore the measurement function in the real radiation part must depend on $n+1$ particles.
Let us consider the case where we have an IR
pole if the variable $x$, describing for example the energy of an extra
gluon with momentum $p_{n+1}$ in the real radiation part, goes to zero.
 If we define
\begin{align}
{\cal B}_n&=\int  \mathrm{d}\Phi_n\left|{\cal M}_0\right|^2 =\int \mathrm{d}\Phi_n B_n\nn\\
{\cal V}_n&=\int \mathrm{d}\Phi_n\,2 \mathrm{Re}\left({\cal M}_{\rm{virt}} {\cal M}_0^*\right)=\int \mathrm{d}\Phi_n\,\left(\frac{V_n}{\eps}+V_{\rm{fin}}\right)\nn\\
{\cal R}_n&=\int\mathrm{d}\Phi_{n+1} \left|{\cal
M}_{\rm{real}}\right|^2=\int \mathrm{d}\Phi_n\int_0^1 \mathrm{d} x\,\left(x^{-1-\eps}\,R_n(x)+R_{\rm{fin}}\right)
\end{align}
and a measurement function $J(p_1 \ldots p_n, p_{n+1})$ we have 
\begin{align}
\sigma^{NLO}&=\int \mathrm{d}\Phi_n\left\{\left(B_n+\frac{V_n}{\eps}+V_{\rm{fin}}\right)\,J(p_1 \ldots p_n, 0) + \int_0^1 \mathrm{d} x\,\left(x^{-1-\eps}\,R_n(x)+R_{\rm{fin}}\right)\,J(p_1 \ldots  p_{n+1})\right\}\;.
\end{align}
In the inclusive case (calculation of the total cross section) we have $J\equiv 1$. The integration over $x$
leads to the explicit $1/\eps$ poles which must cancel with the virtual
part:
\begin{align}
\int_0^1 \mathrm{d} x\,x^{-1-\eps}\,R_n(x)=-\frac{R_n(0)}{\epsilon}+\int_0^1
  \mathrm{d} x\,x^{-\epsilon}\,\frac{R_n(x)-R_n(0)}{x}\quad \mbox{ with } \;R_n(0)=V_n\;.
\end{align}
The cancellation of the poles between $\frac{V_n}{\eps}$ and
$\frac{R_n(0)}{\eps}$ in the non-inclusive case will only work if 
\begin{equation}
\lim_{p_{n+1}\to 0}J(p_1 \ldots p_n, p_{n+1})=J(p_1 \ldots p_n, 0)\;.
\end{equation}
This is a non-trivial condition for the definition of an observable,
for example a jet algorithm, and is called {\it infrared safety}.
The formulation above is tailored to the soft limit where all components of $p_{n+1}$ go to zero; however, an analogous condition must hold if two momenta become collinear. Therefore, more generally, 
 if we define
differential cross sections $\mathrm{d}\sigma/\mathrm{d} X$,
we have $J(p_1 \ldots p_n)=\delta(X-\chi_n(p_i))$, where
$\chi_n(p_i)$ is the definition of the observable, based on $n$
partons.
Infrared safety requires $\chi_{n+1}(\{p\},p_i)\to \chi_{n}(\{p\})$ if $p_i$ becomes soft, or $\chi_{n+1}(\{p\},p_i,p_j)\to \chi_{n}(\{p\},p_i+p_j)$ if the momenta $p_i$ and $p_j$ become collinear to each other.

\subsubsection{Subtraction of IR singularities}
 
In less inclusive cases, and/or in the presence of initial-state singularities, a subtraction procedure has to be applied to obtain finite matrix elements that can be integrated with Monte Carlo methods. At NLO, subtractions schemes such as Catani-Seymour subtraction~\cite{Catani:1996vz,Catani:2002hc,Czakon:2009ss} and FKS subtraction~\cite{Frixione:1995ms} have been established and automated~\cite{Gleisberg:2007md,Frederix:2008hu,Frederix:2009yq}. Beyond NLO, automated subtraction methods are still subject of ongoing research, see \secref{sec:realNNLO} for more details.

At NLO, the general procedure is to include a local counterterm $\rd \sigma^{\mathrm{A}}$ such that
\begin{equation}
    \sigma_{\mathrm{NLO}} = \int_n \rd \sigma^{\mathrm{B}} +\int_n \rd \sigma^{\mathrm{V}} + \int_{n+1} \rd \sigma^{\mathrm{A}} + \int_{n+1} \left [\rd \sigma^{\mathrm{R}} - \rd \sigma^{\mathrm{A}} \right ],
\end{equation}
where $\rd \sigma^{\mathrm{A}}$ must have the same unintegrated singular behaviour as $\rd \sigma^{\mathrm{R}}$. By construction, the difference $\rd \sigma^{\mathrm{R}}\,-\,\rd \sigma^{\mathrm{A}}$ should be integrable in four dimensions such that it can be integrated numerically. Moreover, the subtraction term should be constructed such that the integration over the one-parton subspace (due to the extra emission) can be done analytically, and the IR divergences can be cancelled explicitly. In this case, the contributions to the NLO cross section can be organised as \cite{Catani:1996vz}
\begin{equation}
    \sigma_{\mathrm{NLO}} = \int_n \rd \sigma^{\mathrm{B}} +\int_n \left[ \rd\sigma^{\mathrm{V}} + \int_1 \rd \sigma^{\mathrm{A}} \right]_{\epsilon=0} + \int_{n+1} \left[ (\rd \sigma^{\mathrm{R}})_{\epsilon=0} - (\rd \sigma^{\mathrm{A}})_{\epsilon=0} \right].
\end{equation}
Under these conditions, the remaining phase-space integrals over the resolved particles are finite in four dimensions and can be sampled and integrated with Monte Carlo techniques. 

The discussion so far concerns IR divergences due to final-state radiation. As mentioned already, there can also be IR divergences originating from collinear emissions from the initial-state partons.  In processes with hadronic initial states, they are not cancelled against contributions from the virtual corrections, they are instead absorbed through redefinitions of the parton distribution functions. The general structure is to include a collinear subtraction counterterm $\rd \sigma^{\mathrm{C}}$ such that the NLO cross section is 
\begin{equation}
\label{eq:hadronic_NLO}
    \sigma_{\mathrm{NLO}} = \int_n \rd \sigma^{\mathrm{B}} +\int_n \rd\sigma^{\mathrm{C}} + \int_n \left[ \rd\sigma^{\mathrm{V}} + \int_1 \rd \sigma^{\mathrm{A}} \right]_{\epsilon=0} + \int_{n+1} \left[ (\rd \sigma^{\mathrm{R}})_{\epsilon=0} - (\rd \sigma^{\mathrm{A}})_{\epsilon=0} \right].
\end{equation}
In deep inelastic scattering, for example, the collinear counterterm contribution from a parton of type $a$, coming from a parent hadron with momentum $p^{\mu}$, is \cite{Catani:1996vz}
\begin{equation}
    \rd\sigma^{\mathrm{C}}_a(p) = -\frac{\alpha_s}{2\pi} \frac{1}{\Gamma(1-\epsilon)} \sum_b \int_0^1 \rd z \left[-\frac{1}{\epsilon} \left(\frac{4\pi\mu^2}{\muf^2}\right)^{\epsilon} P_{ab}(z) + K_{ab}(z) \right] \rd \sigma^{\mathrm{B}}_b(zp),
    \label{eq:coll_sub}
\end{equation}
where $P_{ab}(z)$ are the Altarelli-Parisi splitting functions~\cite{Altarelli:1977zs} and $K_{ab}(z)$ is a finite term depending on the factorisation scheme. Similarly as for UV renormalisation, there are various schemes corresponding to different definitions of the finite part. Taking $K_{ab}(z) = 0$ corresponds to the $\overline{\mathrm{MS}}$ scheme.


\subsubsection{Soft gluon emission}

Soft gluon emission is very important in QCD.
In contrast to the collinear case,
soft gluons are insensitive to the spin of the partons. The only
feature they are sensitive to is the colour charge.

\begin{figure}
    \centering
    \begin{minipage}{.28\textwidth}
\includegraphics[width=\textwidth]{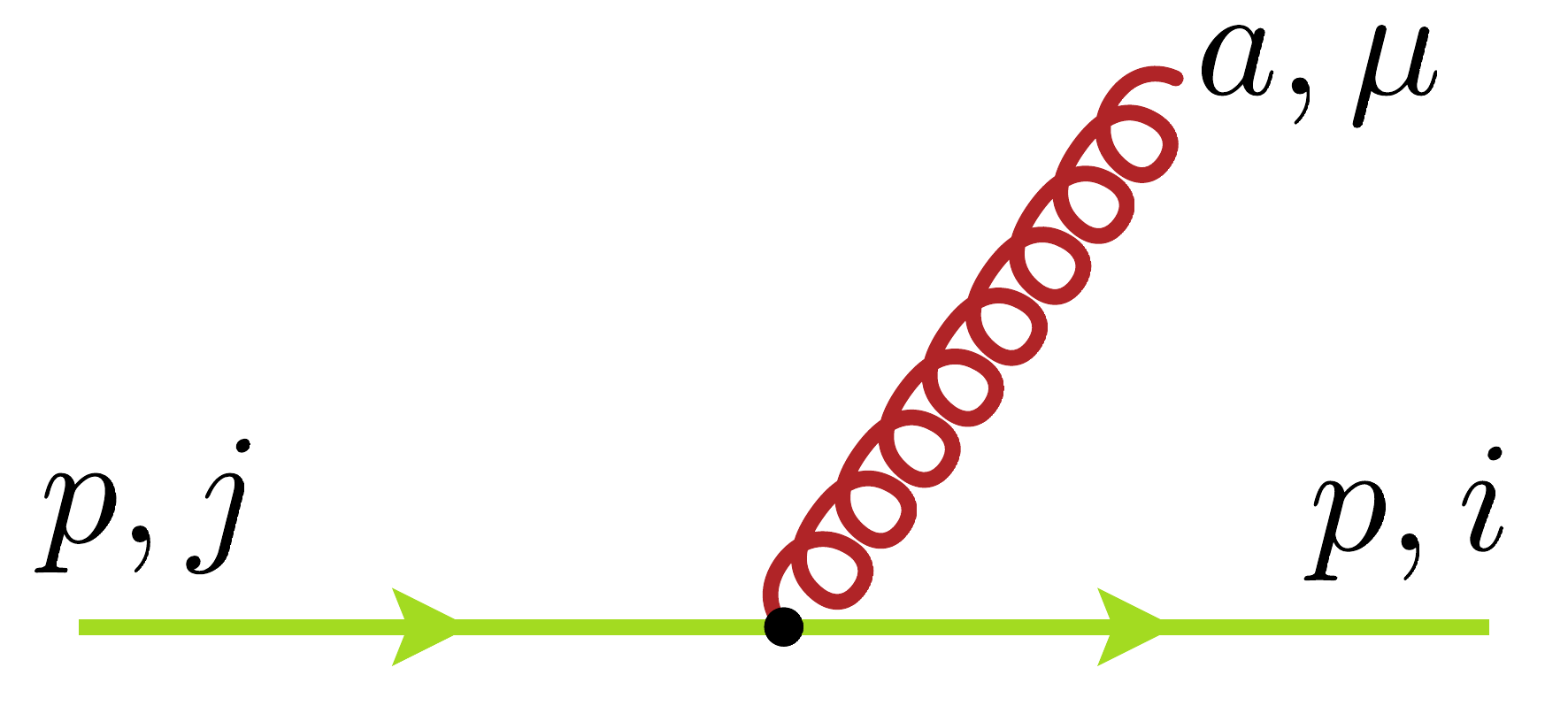}
\end{minipage}%
\begin{minipage}{.11\textwidth}
\centering
\scalebox{1.2}{$\, = \, \, \, g_s \, t^a_{ij} \, 2p^{\mu}$}
\end{minipage}%
\hspace{12mm}
\begin{minipage}{.28\textwidth}
\includegraphics[width=\textwidth]{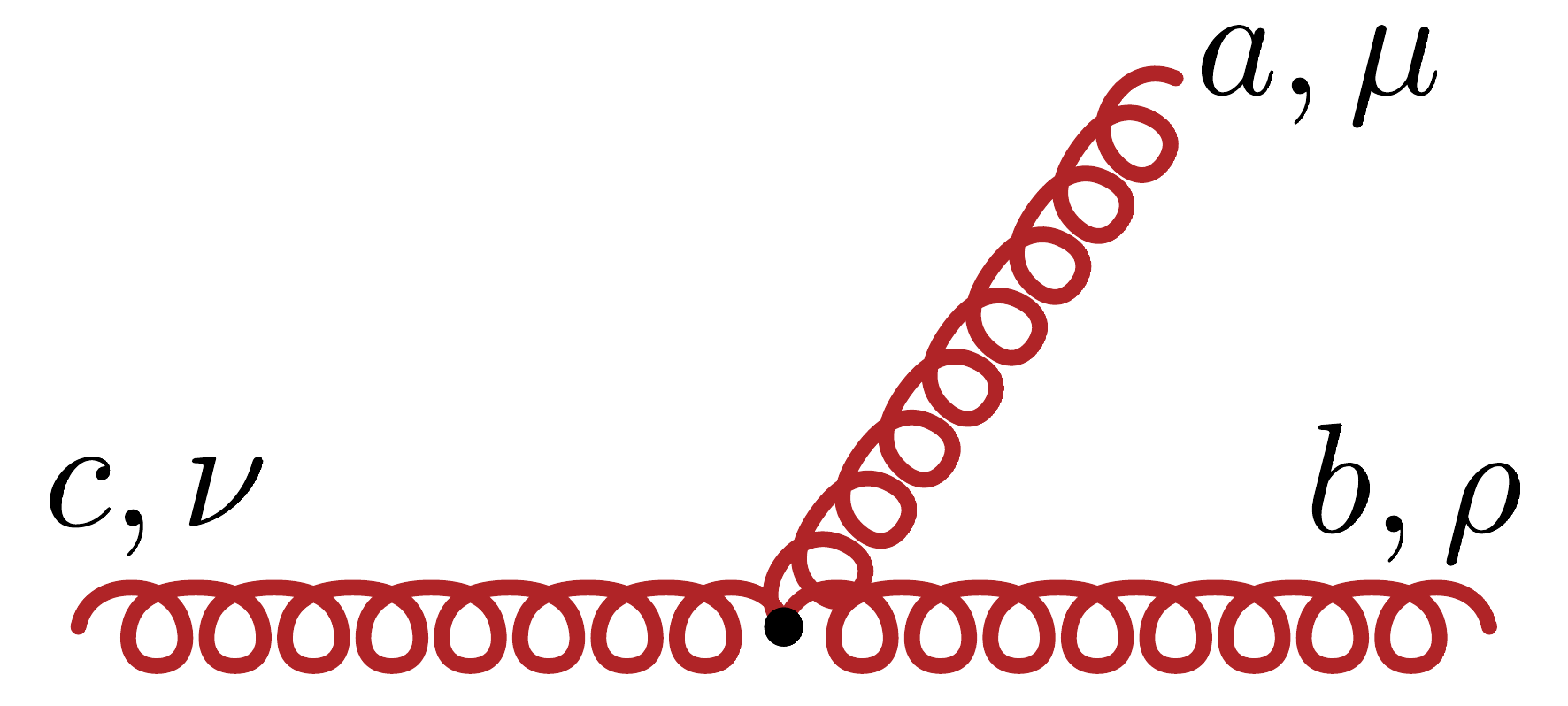}
\end{minipage}%
\begin{minipage}{.15\textwidth}
\centering
\scalebox{1.2}{$ =\, \, \, i g_s \, f^{abc} \, 2p^{\mu} \, g^{\nu\rho}$}
\end{minipage}%
    \caption{The Feynman rules for gluon emission in the soft limit.}
  \label{fig:softFeynman}
\end{figure}

To see this, consider the amplitude for the second row in
Fig.~\ref{fig:real_virtual}, with momentum $k$ and colour index $a$
for the gluon, and momenta and colour indices $p,i$ ($\bar{p},j$) for the
quark (antiquark). The amplitude for massless quarks is given by
\begin{equation}
{\cal M}_{ij}^{a,\mu}=t^a_{ij}\,g_s\,\mu^\varepsilon\bar{u}(p)\slashed{\epsilon}(k)\frac{\slashed p\,+\slashed k}{(p+k)^2}\Gamma^\mu v(\bar{p}) - t^a_{ij}\,g_s\,\mu^\varepsilon\bar{u}(p)\,\Gamma^\mu\frac{\slashed{\bar{p}}\,+\slashed k}{(\bar{p}+k)^2}\slashed{\epsilon}(k) v(\bar{p})\;,
\label{eq:pre-soft}
\end{equation}
where $\Gamma^\mu$ describes a general interaction vertex with the photon, in our case $\Gamma^\mu=\gamma^\mu$
(it can in principle represent an arbitrarily complicated vertex form factor).
Now we take the soft limit, which means that all components of $k$ are
much smaller than $p$ and $\bar{p}$, thus neglecting factors of $\slashed k$ in the
numerator and $k^2$ in the denominator. This leads to the following expression in the soft limit, using also the Dirac equation:
\begin{align}
{\cal M}^{a,\mu}_{ij,soft}&=g_s\,\mu^\varepsilon\,t^a_{ij}\,\bar{u}(p)\,\Gamma^\mu \,v(\bar{p})\left(\frac{2\epsilon(k)\cdot
p}{2p\cdot k} - \frac{2\epsilon(k)\cdot
\bar{p}}{2\bar{p}\cdot k}  \right)
=g_s\,\mu^\varepsilon\, J_{ij}^{a,\nu}(k)\epsilon_\nu(k)\,{\cal M}^\mu_{Born}\;\;,\;\;
{\cal M}^\mu_{Born}=\bar{u}(p)\Gamma^\mu v(\bar{p})\;.
\label{eq:soft}
\end{align}
We see that the amplitude factorises completely into the product of
the Born amplitude and the {\it soft gluon current}~\cite{Catani:2000pi}
\begin{equation}
J_{ij}^{a,\nu}(k)=\sum_{r=p,\bar{p}}\,\tilde{T}_{ij}^a\,\frac{r^\nu}{r\cdot k}\;,
\end{equation}
In our example $\tilde{T}_{ij}^a=t^a_{ij}$ for $r=p$ and $\tilde{{T}}_{ij}^a=-t^a_{ij}$ for $r=\bar{p}$.
This type of factorisation actually holds for an arbitrary number of soft
gluon emissions~\cite{Dixon:2008gr,Becher:2009cu,Gardi:2009qi,Becher:2009qa,Dixon:2009ur}, and can be obtained using the ``soft Feynman rules''
shown in Fig.~\ref{fig:softFeynman}.


\subsubsection{Collinear singularities}

\begin{figure}[htb]
  \centering
 \includegraphics[width=0.7\textwidth]{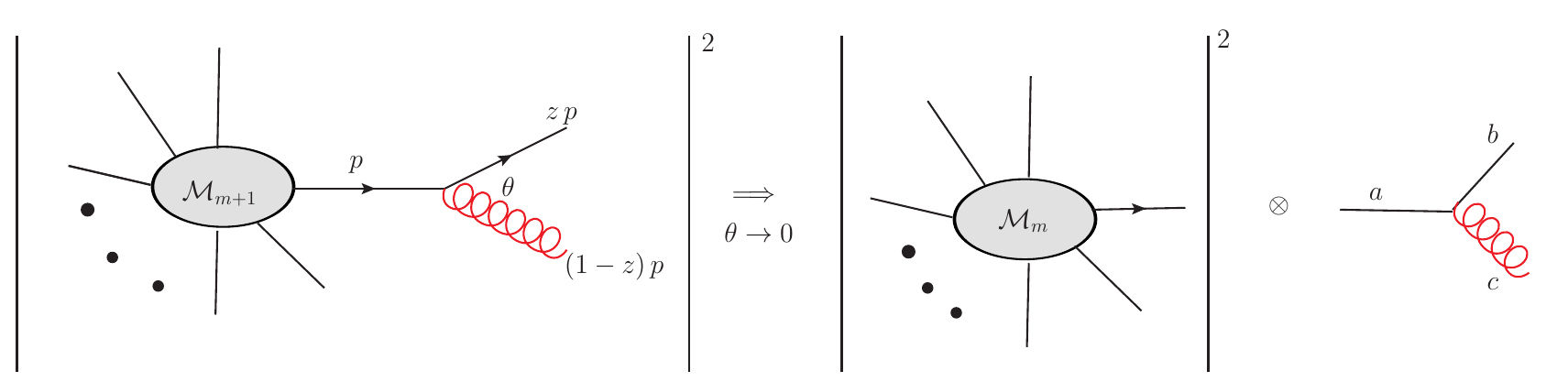}
  \caption{Factorisation in the collinear limit. }
  \label{fig:collinear_fact}
\end{figure}

Let us come back to the amplitude for the real radiation given in
\equref{eq:pre-soft}.
In a frame where $p=E_p(1,\vec{0}^{(d-2)},1)$ and
$k=k_0(1,\vec{0}^{(d-3)}\sin\theta,\cos\theta)$,
the denominator $(p+k)^2$ is given by 
\begin{align}
(p+k)^2=2k_0E_p\,(1-\cos\theta)\;\;\to 0 \mbox{ for }\left\{
\begin{array}{cc}
k_0\to 0&\mbox{(soft)}\\
\theta\to 0 &\mbox{(collinear)}
\end{array}
\right.
\end{align}
Note that if the quark line was massive, $p^2=m^2$, we would have
$$(p+k)^2-m^2=2k_0E_p\,(1-\beta\cos\theta)\,,\,\beta=\sqrt{1-m^2/E_p^2}$$
and thus the collinear singularity would be absent. This is why collinear singularities are
sometimes also called {\it mass singularities}: the propagator
can only develop a collinear divergence if the splitting partons are massless,
while the soft singularity is present irrespective of the mass of the quark radiating a gluon.
In the collinear limit, we
also have a form of factorisation, shown schematically in Fig.~\ref{fig:collinear_fact}.

\begin{figure}[htb]
  \centering
 \includegraphics[width=0.33\textwidth]{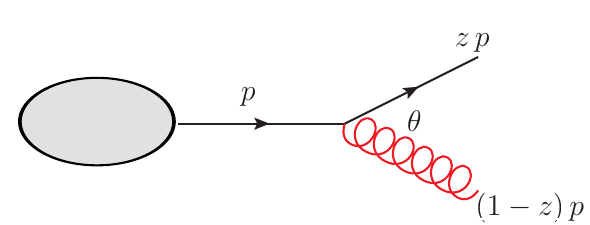}
  \caption{Gluon emission leading to the splitting function $P_{q\to qg}(z)$.}
  \label{fig:collinear_limit}
\end{figure}

\noindent The universal factorisation behaviour of an amplitude depending on $m+1$ external particles in the limit where two of them become collinear can be described as
\begin{align}
|\mathcal{M}_{m+1}|^2\,\mathrm{d} \Phi_{m+1}\to |\mathcal{M}_{m}|^2\mathrm{d}
\Phi_{m}\,\frac{\als}{2\pi}\,\frac{\mathrm{d} k_\perp^2}{k_\perp^2}\,\frac{\mathrm{d}\phi}{2\pi}\,\mathrm{d} z\,P_{a\to bc }(z)\;,
\label{eq:collinear_fact}
\end{align}
where we have used  the so-called {\em Sudakov parametrisation}:
\begin{align}
  &k^\mu=(1-z)\, p^\mu + \beta\,n^\mu + k_\perp^\mu\;,
\end{align}
with $n^\mu$ being a light-like vector satisfying $p\cdot n \not=0$ and $k_\perp\cdot n=0$,  and $\beta$ being determined by the requirement that $k$ must be light-like:
\begin{align}
k^2=0=2(1-z)\, \beta\, p\cdot n- k_\perp^2\Rightarrow \beta=\frac{k_\perp^2}{2\,p\cdot n\,(1-z)}\;.
\end{align}
Note that the phase space can also be written in a factorised form in the soft and collinear limits. \\
The function $P_{a\to bc }(z)$ is the {\em Altarelli-Parisi splitting function} already introduced in \equref{eq:coll_sub}, describing the splitting of parton $a$ into partons $b$ and $c$, 
and $z$ is the momentum fraction of the original parton $a$ carried by parton $b$ after emission of parton $c$.
For example, for collinear gluon emission off a quark, depicted in \figref{fig:collinear_limit},
the corresponding  Altarelli-Parisi splitting function for $z<1$ is given by 
\begin{equation}
P_{q\to qg}(z)\equiv P_{q/q}(z)=C_F\,\frac{1+z^2}{1-z}\;,
\end{equation}
another commonly used notation is $P_{qq}(z)$. 


%% file: PDFs.tex

 Parton distribution functions are discussed in detail in a dedicated chapter~\cite{Lorce:2025aqp},
 however we mention the most important features here for self-consistency.

 With the collinear initial state singularities absorbed into the PDFs at a factorisation scale $\muf$, the functions $f_{a/h}(x_a)$ defined in \equref{eq:cross_section} become scale dependent. 
 This gives us something like a renormalisation group equation, which
means that we can calculate how the PDFs evolve as the scale $\muf$
is changed. In other words,
while the PDFs themselves are non-perturbative objects,  their {\em scale dependence}   can be calculated in perturbation
 theory, which means that   we can
measure the PDFs in one process at a certain scale and then use them in
another process at a different scale. 
Defining $t=\ln\left(Q^2/\muf^2\right)$, we have
 \begin{equation}
\frac{\partial}{\partial
  t}\,f_{q_i}(x,t)=\int_x^1\frac{\mathrm{d}\xi}{\xi}\,P_{q_i/q_j}\Big(\frac{x}{\xi},\als(t)\Big)\,f_{q_j}(\xi,t)\;,\label{APallo}
\end{equation}
where $f_{q_i}\equiv f_{q_i/h}$ denotes the PDF for a quark of flavour $i$ and the hadron label $h$ has been omitted for ease of notation.
The splitting functions $P_{q_i/q_j}$, or ``splitting kernels" in \equref{APallo} can be generalised to higher orders and calculated as a power series in $\als$,
\begin{equation}
P_{q_i/q_j}(x,\als) =
\frac{\als}{2\pi}\,P^{(0)}_{ij}(x)+\Big(\frac{\als}{2\pi}\Big)^2\,P^{(1)}_{ij}(x)+\Big(\frac{\als}{2\pi}\Big)^3\,P^{(2)}_{ij}(x)+{\cal
O}(\als^4)\label{pqqex}\,.
\end{equation}
\equref{APallo} holds for parton distributions which are {\it non-singlets} under
the flavour group: either a single flavour or a combination $q_{\mathrm{ns}}=f_{q_i}-f_{q_j}$ with $q_i,q_j$ being a quark or antiquark
of any flavour. 
More generally, the DGLAP equation is a $(2n_f+1)-$dimensional
matrix equation in the space of quarks, antiquarks and gluons,
\begin{equation}
\frac{\partial}{\partial t}\left(\begin{array}{l}
f_{q_i}(x,t)\\
f_{g}(x,t)\end{array}\right)=\sum_{q_j,\bar{q}_j}\int_x^1\frac{d\xi}{\xi}\left(\begin{array}{ll}P_{q_i/q_j}(\frac{x}{\xi},\als(t))&P_{q_i/g}(\frac{x}{\xi},\als(t))\\
P_{g/q_j}(\frac{x}{\xi},\als(t))&P_{g/g}(\frac{x}{\xi},\als(t))\end{array}\right)
\left(\begin{array}{l}
f_{q_j}(\xi,t)\\
f_{g}(\xi,t)\end{array}\right)\;.
\label{eq:DGLAP}
\end{equation}
Eqs.~(\ref{APallo}) and (\ref{eq:DGLAP}) are called {\it DGLAP evolution equations}, named after Dokshitzer \cite{Dokshitzer:1977sg}, Gribov, Lipatov
\cite{Gribov:1972ri} and Altarelli, Parisi \cite{Altarelli:1977zs}.
They are among the most important equations in perturbative QCD.

Note that, because of charge conjugation invariance and $SU(n_f)$ flavour
symmetry, the splitting functions $P_{q/g}$ and $P_{g/q}$ are independent of the
quark flavour and the same for quarks and antiquarks.
Defining the singlet distribution
\begin{equation}
\Sigma(x,t)=\sum_{i=1}^{n_f}\,[\,f_{q_i}(x,t)+f_{\bar{q}_i}(x,t)\,]
\end{equation}
and taking into account the considerations above, Eq.~(\ref{eq:DGLAP}) simplifies to 
\begin{align}
\frac{\partial}{\partial t}\left(\begin{array}{l}
\Sigma(x,t)\\
g(x,t)\end{array}\right)=\int_x^1\frac{d\xi}{\xi}\left(\begin{array}{ll}P_{q/q}(\frac{x}{\xi},\als(t))&2n_f\,P_{q/g}(\frac{x}{\xi},\als(t))\\
P_{g/q}(\frac{x}{\xi},\als(t))&P_{g/g}(\frac{x}{\xi},\als(t))\end{array}\right)
\left(\begin{array}{l}
\Sigma(\xi,t)\\
g(\xi,t)\end{array}\right)\,.
\label{APsinglet}
\end{align}
%
The leading order splitting functions including the regulating contributions at $x=1$ are given by
\begin{align}
P_{q/q}^{(0)}(x)&=C_F\Big\{\frac{1+x^2}{(1-x)_+}+\frac{3}{2}\,\delta(1-x)\Big\}\\
P_{q/g}^{(0)}(x)&=T_R\Big\{x^2+(1-x)^2\Big\}\quad T_R = \frac{1}{2}\\
P_{g/q}^{(0)}(x)&=C_F\Big\{\frac{1+(1-x)^2}{x}\Big\}\\
P_{g/g}^{(0)}(x)&=2N_c\Big\{\frac{x}{(1-x)_+}+\frac{1-x}{x}+x(1-x)\Big\} +\delta(1-x)\left(\frac{11}{6}N_c-\frac{2}{3}n_f T_R\right)\;.
\end{align}


%% file: 4c_jets.tex
Jets and event shapes are discussed in detail in Ref.~\cite{Stagnitto:2025air}, therefore we will limit ourselves to the basic concepts here.

\subsubsection{Jet cross sections and jet algorithms}

\begin{figure}[htb]
\centering
\hspace{-1cm}
\begin{minipage}{0.49\textwidth}\centering
    \includegraphics[width=0.85\textwidth]{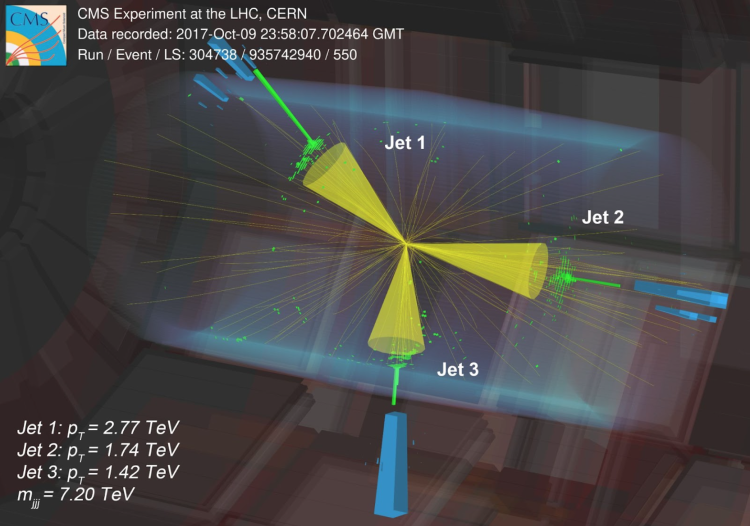}
\end{minipage}
\hspace{-1cm}
\begin{minipage}{0.49\textwidth}\centering
   \includegraphics[width=0.85\textwidth]{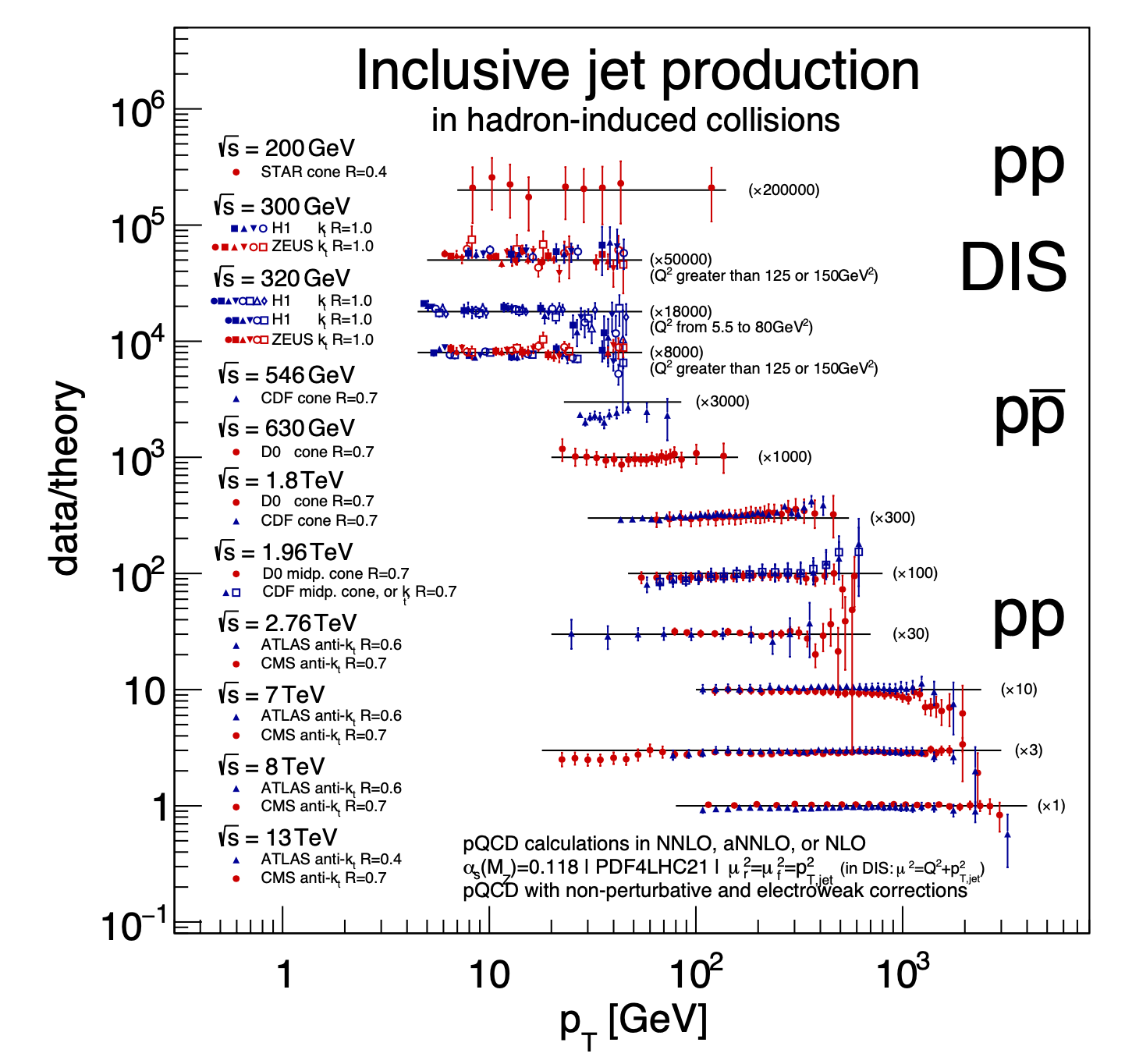}
\end{minipage}
\caption{Left:
   Three-jet event recorded by the CMS experiment, figure taken from the CERN image gallery. Right: Ratios of cross-section measurements to predictions in perturbative QCD for inclusive jet production at central (pseudo-)rapidity as a function of the jet transverse momentum at different colliders and energies, figure taken from Ref.~\cite{Gross:2022hyw}.}
  \label{fig:charm_3jets_CMS}
\end{figure} 
Jets can be pictured as clusters of particles  which
are close to each other in phase space, or, from an experimental point of view, in the detector. 
In Fig.~\ref{fig:charm_3jets_CMS} (left), an event consisting of three highly energetic jets recorded by the CMS experiment is shown.
As coloured particles do not exist unconfined, jets are primarily composed of charged and neutral mesons and baryons, 
small energy fractions
of electrons and muons are also present, originating from  heavy hadron decays. Nowadays, jets have been measured over a very large energy range at different colliders, see Fig.~\ref{fig:charm_3jets_CMS} (right).

Historically, one of the first suggestions to define jet cross sections was by
Sterman and Weinberg~\cite{Sterman:1977wj}. In their definition, a final state is classified as two-jet-like if
all but a fraction $\varepsilon$ of the total available energy $E$ is contained
in two cones of opening angle $\delta$.
The two-jet cross section is then obtained by integrating the matrix elements
for the various quark and gluon final states over the appropriate region of phase space
determined by $\varepsilon$  and $\delta$. 
The two-jet cross section thus depends on the values for
$\varepsilon$ and $\delta$. If they are very large, even extra
radiation at a relatively large angle $\theta<\delta$ will be
``clustered'' into the jet cone and almost all events will be
classified as 2-jet events. If $\varepsilon$ and $\delta$ are very small, the 2-jet cross section starts to diverge, because ``one parton'' is not an observable, it cannot be distinguished from ``one parton plus soft and/or collinear radiation''. 

The Sterman-Weinberg jet definition based on cones is not very practical
to analyse multijet final states. 
Modern jet algorithms are based on sequential recombination algorithms.
A better alternative is for example the following~\cite{Bethke:1991wk}:
\begin{enumerate}
\item Starting from $n$ particles, for all pairs $i$ and $j$ calculate
  $(p_i+p_j)^2$. 
\item If  ${\rm min}(p_i+p_j)^2 <\,\ycut\,Q^2$ then define a new
  ``pseudo-particle'' $p_J=p_i+p_j$, which decreases $n\to n-1$.
$Q$ is the center-of-mass energy in $e^+e^-$ collisions, or a typical hard scattering energy in hadronic collisions, and $\ycut$ is the jet resolution
parameter.
\item If $n=1$, stop, else repeat the step above.
\end{enumerate}
After this algorithm, all partons are clustered into jets.
This simple algorithm is sometimes called {\sc Jade}-algorithm because it has been used first at the {\sc Jade} experiment at {\sc petra} ({\sc desy}).
With this definition one finds at ${\cal O}(\als)$:
\begin{align}
\sigma^{2jet}&=\sigma_0\left(1-C_F\,\frac{\als}{\pi}\,\left[\ln^2\ycut+\frac{3}{2}\ln\ycut+\mbox{ finite}\right]\right)\;.
\end{align}
Algorithms which are particularly useful for hadronic initial states
are e.g. the so-called Durham-$k_T$
algorithm~\cite{Bethke:1991wk} or the anti-$k_T$ algorithm~\cite{Cacciari:2008gp}.
Both algorithms are based on a  distance measure
\begin{equation}
d_{ij} = \mathrm{min}\left(p_{T,i}^{2p},p_{T,j}^{2p}\right)\,\frac{\Delta R_{ij}^2}{R^2}\;,
\end{equation}
where $R$ is a radius parameter, $\Delta R_{ij}^2=\Delta y^2_{ij}+\Delta\phi_{ij}^2$ is the distance  in rapidity and azimuthal angle between particles $i$ and $j$, and the parameter $p$ is 1 for the $k_T$ algorithm, $0$ for the Cambridge-Aachen~\cite{Dokshitzer:1997in} algorithm and $-1$ for the anti-$k_T$ algorithm.
The distance $d_{ij}$ is calculated for all combinations of pairs of particles.
The pair with the lowest 
$d_{ij}$ is replaced by a pseudo-particle whose four-momentum is given by the 
sum of the four-momenta of particles $i$ and $j$.
Summing the 4-momenta to form the pseudo-particle is also called ``E-recombination 
scheme''. Note that the combined 4-momentum is not light-like anymore.
The clustering procedure is repeated as long as pairs with invariant 
mass fraction below a predefined resolution parameter
$\ycut$  are found. Once the clustering is terminated, the 
remaining (pseudo-)particles are the jets.

\begin{figure}[htb]
\centering
\hspace{-1cm}
\begin{minipage}{0.49\textwidth}\centering
    \includegraphics[width=0.75\textwidth]{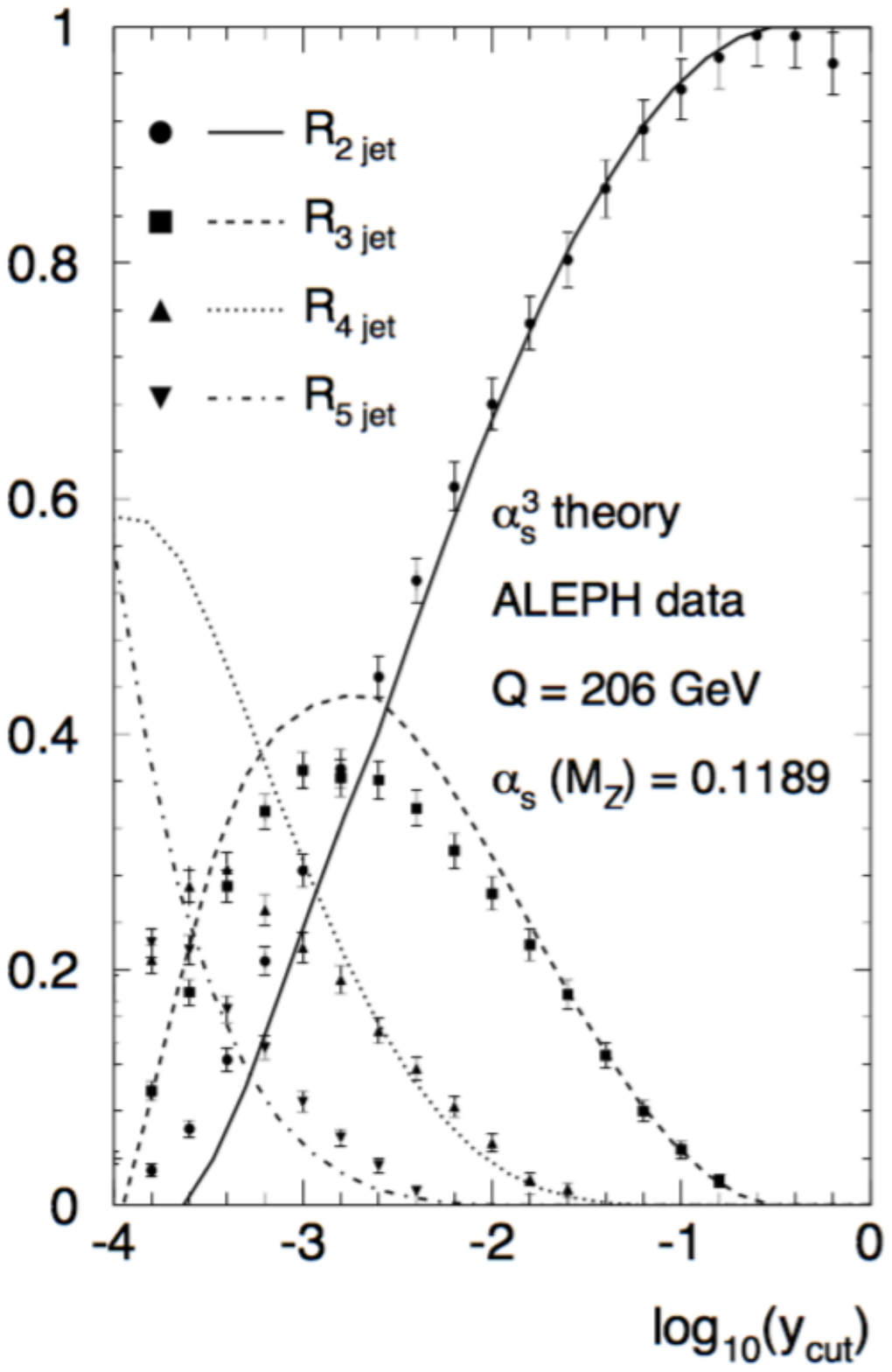}
\end{minipage}
\hspace{-1cm}
\begin{minipage}{0.49\textwidth}\centering
    \includegraphics[width=0.98\textwidth]{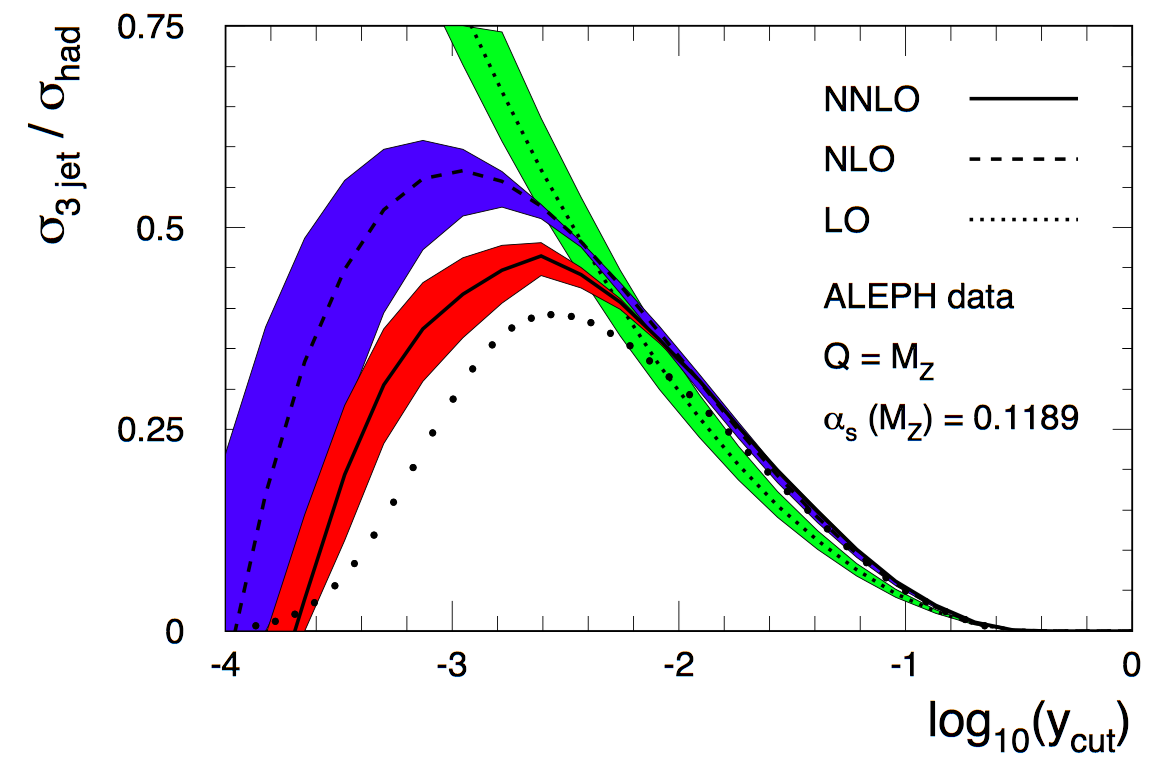}
\end{minipage}
\caption{Left:
   Jet rates as a function of the jet resolution parameter
    $y_{\rm{cut}}$~\cite{ALEPH:2003obs}. Right: higher order corrections to the
    3-jet rate~\cite{GehrmannDeRidder:2008ug}.}
  \label{fig:jets_ycut}
\end{figure}

It is evident that a large 
value of $\ycut$ will ultimately result in 
the clustering all particles 
into only two jets, while higher jet multiplicities will become more and 
more frequent as $\ycut$ is lowered. In experimental jet measurements, 
one therefore  studies the jet rates (n-jet cross sections normalised to 
the total hadronic cross section) as function of the jet resolution 
parameter $\ycut$. Fig.~\ref{fig:jets_ycut} (left) shows the jet rates
as a function of $\ycut$, compared to {\sc aleph} data. Fig.~\ref{fig:jets_ycut} (right) shows
predictions up to NNLO for the 3-jet rate as a function of
$\ycut$. Note that in this figure, for small values of $\ycut$, the  3-jet rate at LO diverges  (green band)
because  only three partons are present
at LO and therefore there is no room for extra radiation.
As an isolated parton is not an observable, the cross section diverges
in this limit. At higher orders, this
situation gradually improves by extra radiation being allowed. However, resummation or parton showering would be needed to achieve a better description of the very low $\ycut$ region.
At the LHC, the most commonly used jet algorithm is the  {\em  anti-$k_T$ algorithm}~\cite{Cacciari:2008gp}. More details about jet algorithms can be found in Refs.~\cite{Salam:2010nqg,Marzani:2019hun}.
Of course, it is very important that jet algorithms are infrared safe. 
%

\subsubsection{Event shapes}

Jets are not the only observables that can be defined based on
hadronic tracks in the detector. Other very useful
observables are so-called {\em event shapes}, which describe certain geometric features of an event. They are particularly useful at lepton colliders, since the full kinematic information can be reconstructed from the final-state momenta.

A particularly well-studied observable 
is {\em thrust}, which describes how ``pencil-like'' an
event looks.  
Thrust $T$ is defined by
\begin{equation} 
T=\max_{\vec{n}}\, 
\frac{\sum_{i=1}^{m}\,\left|\vec{p}_i\,\cdot\,\vec{n}\right|}{\sum_{i=1}^{m}\left|\vec{p}_{i}\right|}\;,
\label{eq:thrustdef}
\end{equation}
where $\vec{n}$ is a three-vector (the direction of the thrust axis)
such that $T$ is maximal. The particle three-momenta $\vec{p}_{i}$ are
defined in the  centre-of-mass frame. Therefore, the above definition only holds for lepton colliders where the partonic 
centre-of-mass energy is fixed. At hadron colliders, the definition of event shapes such as thrust is still possible, but in this case it is based on transverse momenta.
$T$ is an example of a
measurement function $J(p_1,\ldots,p_m)$. It is infrared safe because neither $p_j \to 0$, nor
replacing $p_i$ with $z p_i+(1-z) p_i$ change $T$. 

%
\begin{figure}[htb]
  \centering
  \includegraphics[width=0.5\textwidth]{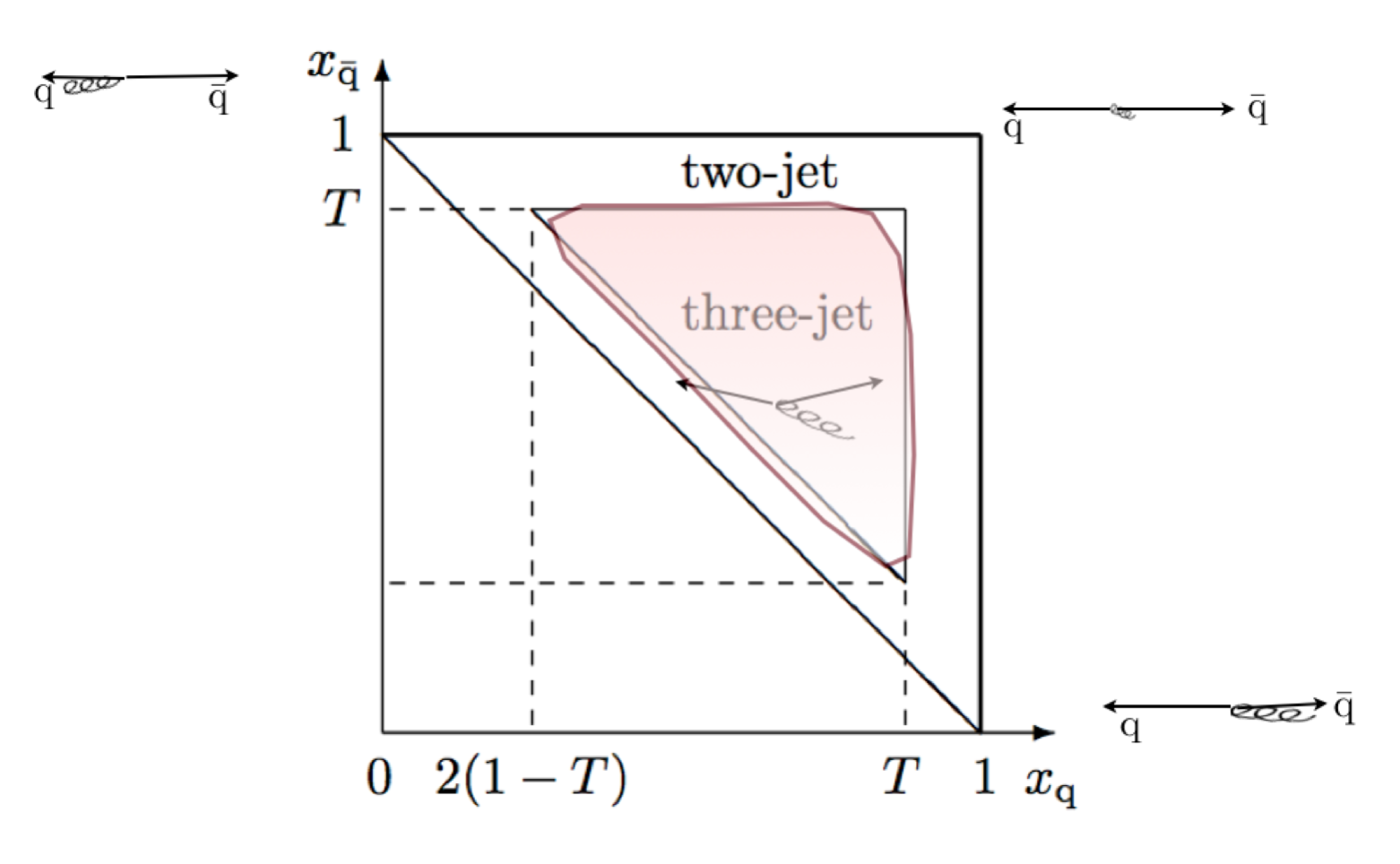}
  \caption{Dalitz-plot showing the allowed 2-jet and 3-jet regions and thrust values. Figure from Refs.~\cite{Dissertori:2003pj,DissertoriLecture}.}
  \label{fig:Dalitz_thrust}
\end{figure}
\figref{fig:Dalitz_thrust} shows the collinear and soft regions in a Dalitz-plot, where $x_i$ denote the energy fractions, defined by
\begin{align}
  x_q=2\frac{E_q}{\sqrt{s}}\;, \; x_{\bar{q}}=2\frac{E_{\bar{q}}}{\sqrt{s}}\;,\; x_g=2\frac{E_g}{\sqrt{s}}\;\;,\;\;
  x_q+x_{\bar{q}}+x_g=2\;.
 \end{align}
At leading order it is
possible to perform the phase space integrations analytically, to obtain
\begin{equation} 
\frac{1}{\sigma} \frac{{\rm d} \sigma}{{\rm d} T} = C_{F}
\frac{\alpha_{{\rm s}}}{2\pi}
\left[\frac{2 \big(3 T^2-3T+2\big)}{T(1-T)}
\ln\left(\frac{2T-1}{1-T}\right)
- 3(3T-2)\frac{2-T}{1-T}\right]
\,.
\label{eq:thrust}
\end{equation}
We see that the perturbative prediction for the thrust distribution becomes
singular as $T\to 1$. In addition to the factor of $1-T$ in the
denominator,  there is also a logarithmic divergence $\sim\ln(1-T)$.
The latter is characteristic for event shape
distributions.
For an event shape $Y$ with $Y\to 0$ in the two-jet limit (so for
example $Y=1-T$), the behaviour  at $n^{\rm{th}}$ order in perturbation theory is~\cite{Catani:1992ua}
$$\frac{1}{\sigma}\frac{d\sigma^{(n)}}{dY}\simeq \als^n \frac{1}{Y}\ln^{2n-1}(\frac{1}{Y})\;.$$ These logarithms spoil the
convergence of the perturbative series and should be resummed if we
want to make reliable prediction near  the phase space region where
$Y\to 0$, see also \secref{sec:resummation}.

\begin{figure}[htb]
  \centering
  \begin{minipage}{0.49\textwidth}\centering
    \includegraphics[width=0.85\textwidth]{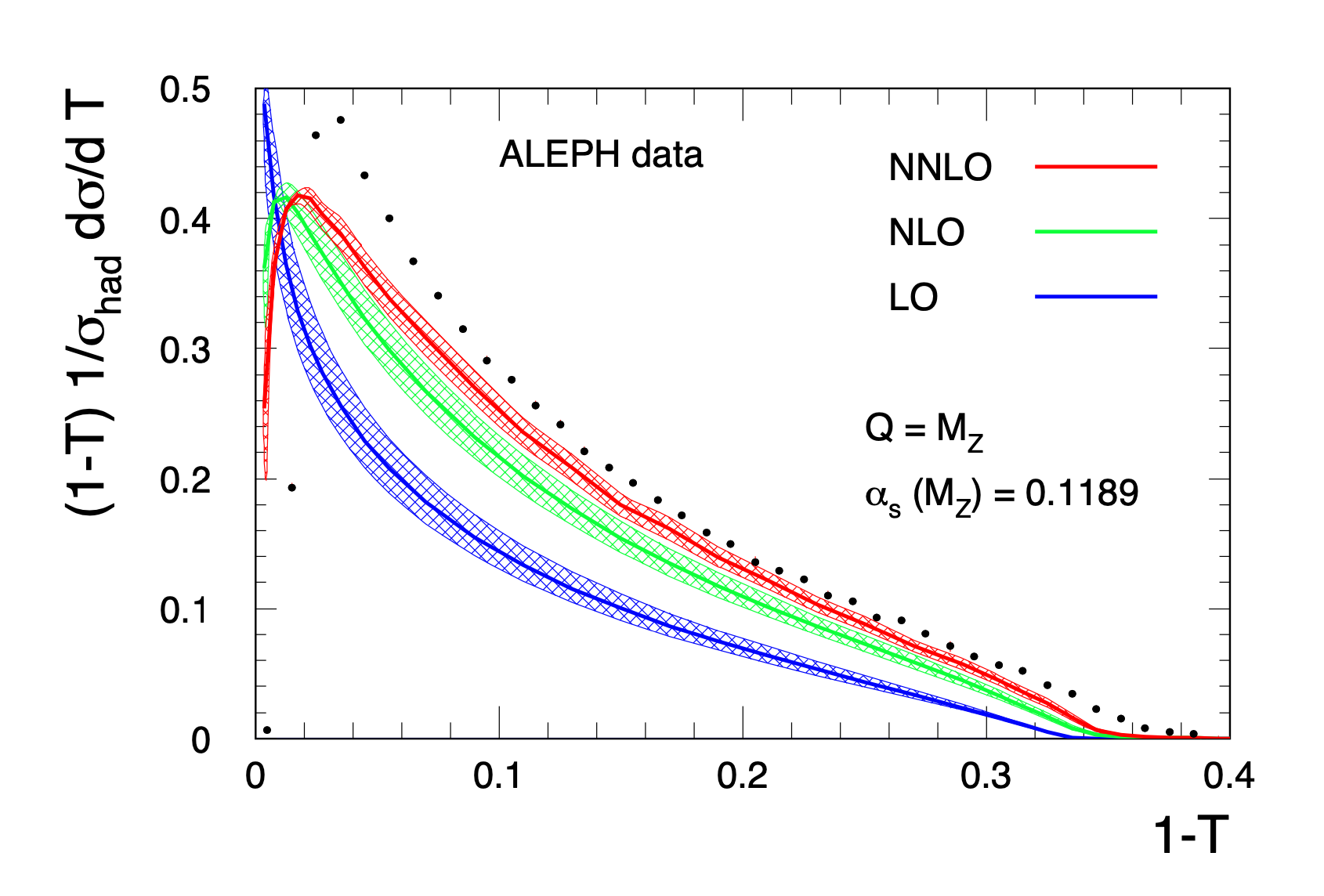}  
  \end{minipage}
  \begin{minipage}{0.49\textwidth}\centering
     \includegraphics[width=0.75\textwidth]{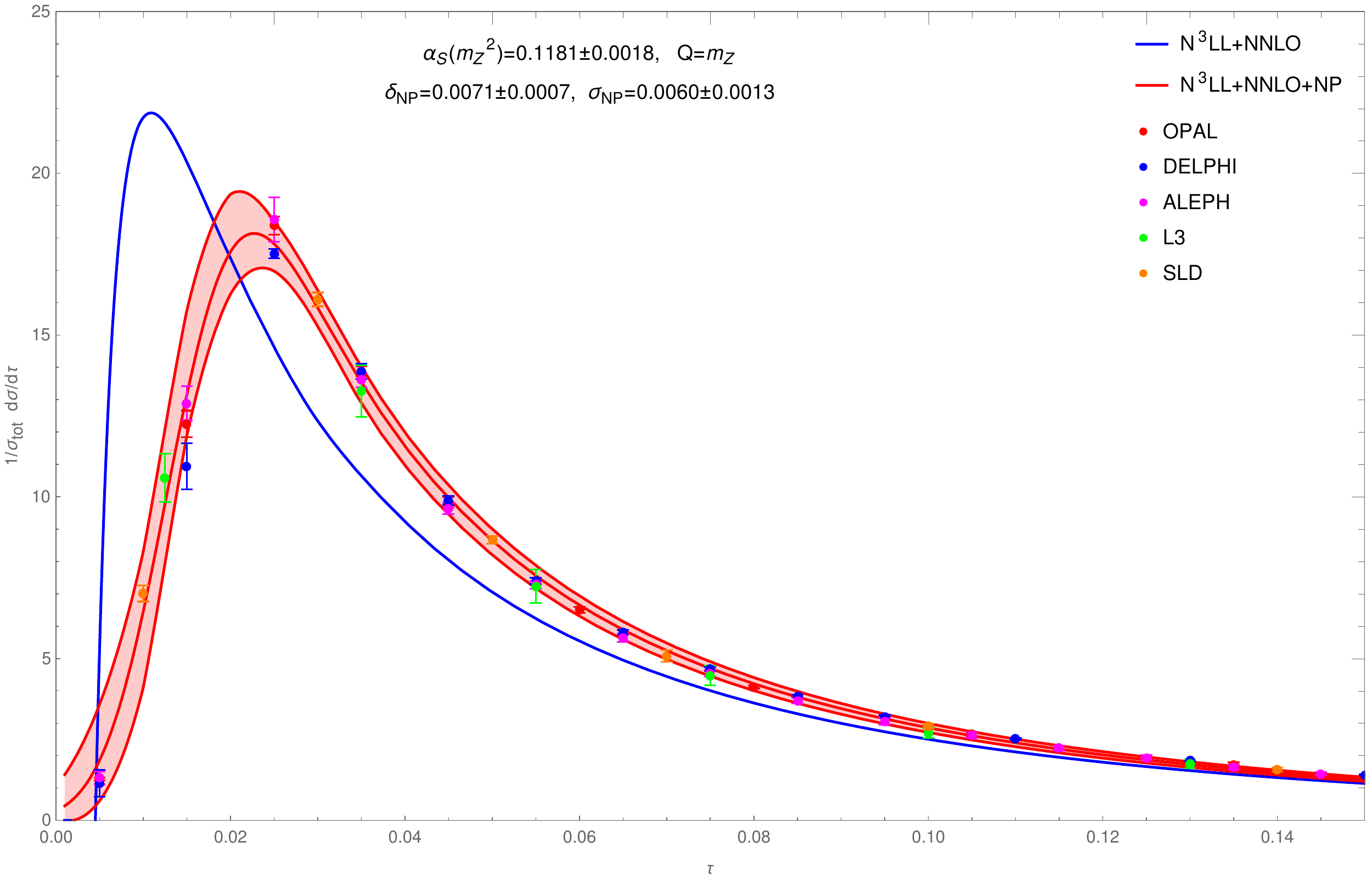}
  \end{minipage}
  \caption{Left: The thrust distribution up to NNLO in QCD, compared to
    {\sc aleph} data. Figure from Ref.~\cite{Gehrmann-DeRidder:2007vsv}.
  Right: The thrust distribution including resummation and non-perturbative corrections, compared to LEP data. Figure from Ref.~\cite{Aglietti:2025jdj}. 
}
  \label{fig:thrust_eerad3}
\end{figure} 
\figref{fig:thrust_eerad3} (left) shows the thrust distribution up to
NNLO precision in QCD. This is an observable where both resummation
and power corrections $\sim (\Lambda/Q)^p$ need to be included to describe the data well
over the whole kinematic range, as can be seen from Fig.~\ref{fig:thrust_eerad3} (right). 

As the availability of perturbative higher-order corrections increased rapidly in recent years, estimating the non-perturbative corrections gets more and more important, also beyond the context of event shape observables, see e.g. Refs.~\cite{Farren-Colloty:2025amh,Hoang:2025uaa,Agarwal:2025dvo,Makarov:2024ijn,Chen:2024nyc,Nason:2023asn,Manohar:1994kq,Dokshitzer:1995qm} for work in this direction.

%% file: 4d_scaledep.tex

Let us consider an observable $R$,  calculated in perturbation theory to order $\alpha_s^{N+k}$, depending on $\mu$ only through $\als(\mu)$.
\begin{equation}
R^{(N)}(\als(\mu))=\sum_{n=0}^N C_n \, \alpha_s^{n+k}(\mu)\;,\label{eq:RN}
\end{equation}
where $k$ is the power of $\als$ of the leading order cross section.
From the perturbative solution of the RGE we can derive how the physical
quantity $R^{(N)}(\als(\mu))$, 
truncated at  order $N$ in perturbation theory, changes
with the renormalisation scale $\mu$:
\begin{align}
\frac{\mathrm{d}}{\mathrm{d} \log(\mus)}\,R^{(N)}(\als(\mu)) =\beta(\als)\frac{\partial R^{(N)}}{\partial \als} \sim \als^{N+1}(\mu)\;,
\end{align}
because $\beta(\als)=-b_0\als^2+{\cal O}(\als^3)$.
This means that, the more higher order coefficients $C_n$
we can calculate, the weaker the dependence of the result on the unphysical
scale $\mu$ will be.
Therefore, the dependence on the scale is used to estimate the
uncertainty of a result calculated to a certain order in perturbation
theory.
 
If the scale dependence
of an observable is given through $\als(\mu)$, we can use the renormalisation group equation
to move from a result at a scale $\mu_0$ to a
result at a different scale.
For the observable $R$,  known to order $\als^{N+k}$,
we can use the requirement $\mathrm{d} R/\mathrm{d} \ln\left(\frac{\mu_r^2}{\mu_0^2}\right)=0$ and \equref{eq:RGE} to derive how $R$ changes with a change of scale, leading to
\begin{align}
R = \alpha_s^{k}(\mu_r) \left\{C_0 
+ \left(C_1 + b_0 C_0\ln\left(\frac{\mu_r^2}{\mu_0^2}\right)\right)
\alpha_s(\mu_r) + {\cal O}(\als^{2})\right\}\;.
\end{align}
Variations of  $\mu_r$ will change the $C_0$-part of the ${\cal O}(\als^{k+1})$ term,
however the magnitude of $C_1$ can only be
determined by direct calculation. The analogous pattern persists at higher orders.
As the logarithms involving the renormalisation scale are known, this can be used to reduce the scale dependence of perturbative predictions, as has been suggested already long time ago~\cite{Stevenson:1982wn,Brodsky:1982gc,Brodsky:2011ig,DiGiustino:2020fbk}. 

In hadronic collisions there is another scale, the factorisation scale
$\muf$, which comes from the factorisation of initial-state infrared singularities.
It also needs to be taken into account when assessing the
uncertainty of a theoretical prediction.
Varying  both $\mur$ and $\muf$ simultaneously in the same
direction can lead to accidental cancellations and hence
an underestimation of the perturbative uncertainties.
Therefore, in the presence of both $\mur$ and $\muf$, usually so-called
{\em 7-point scale variations} are performed, which means
$\mu_{R,F}=c_{R,F}\,\mu_0$, where $c_R,c_F\in \{2,1,0.5\}$ and where the
extreme variations $(c_R,c_F)=(2,0.5)$ and $(c_R,c_fF=(0.5,2)$ are omitted.

Furthermore, the behaviour of the scale uncertainty bands can depend
sensitively on the definition of the central scale $\mu_0$, see Refs.~\cite{Czakon:2016dgf,Currie:2018xkj} for examples.
A convenient choice is a scale where the higher-order corrections are
small, i.e. a scale showing good ``perturbative stability''.

\begin{figure}[htb]
\centering
\includegraphics[width=0.49\textwidth]{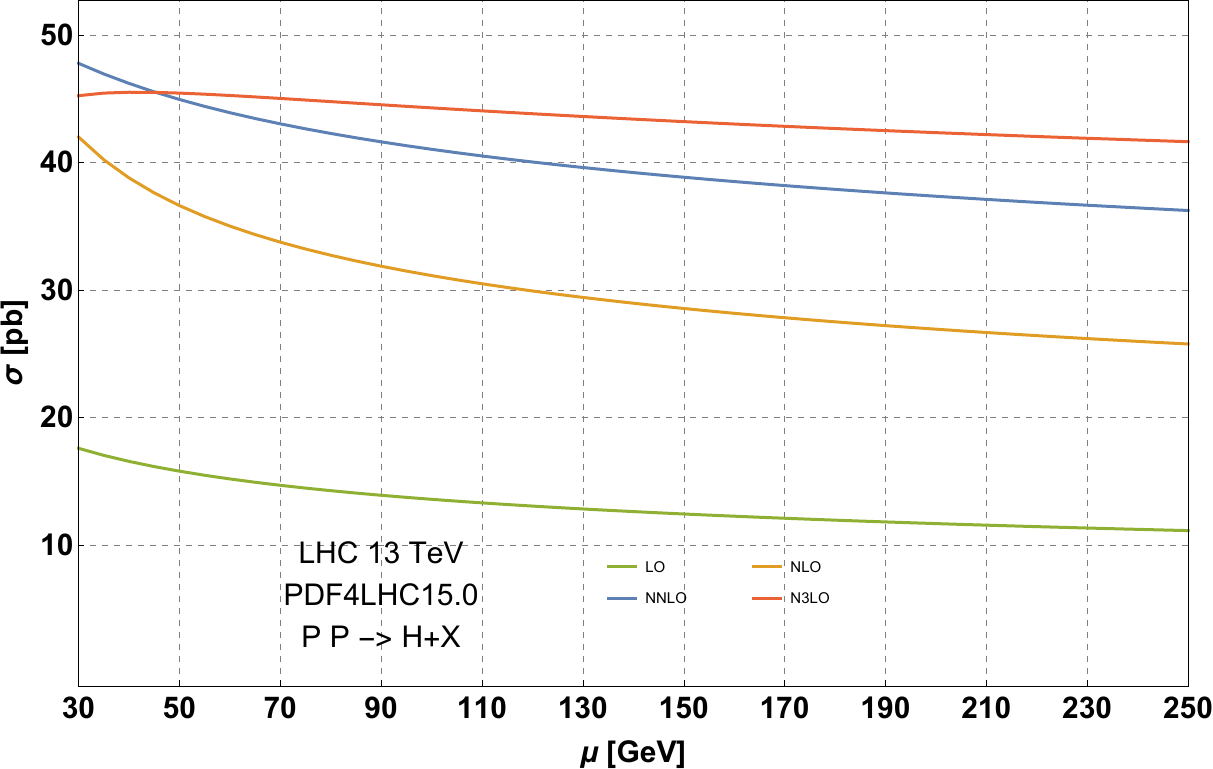}
\vspace*{3mm}
\includegraphics[width=0.44\textwidth]{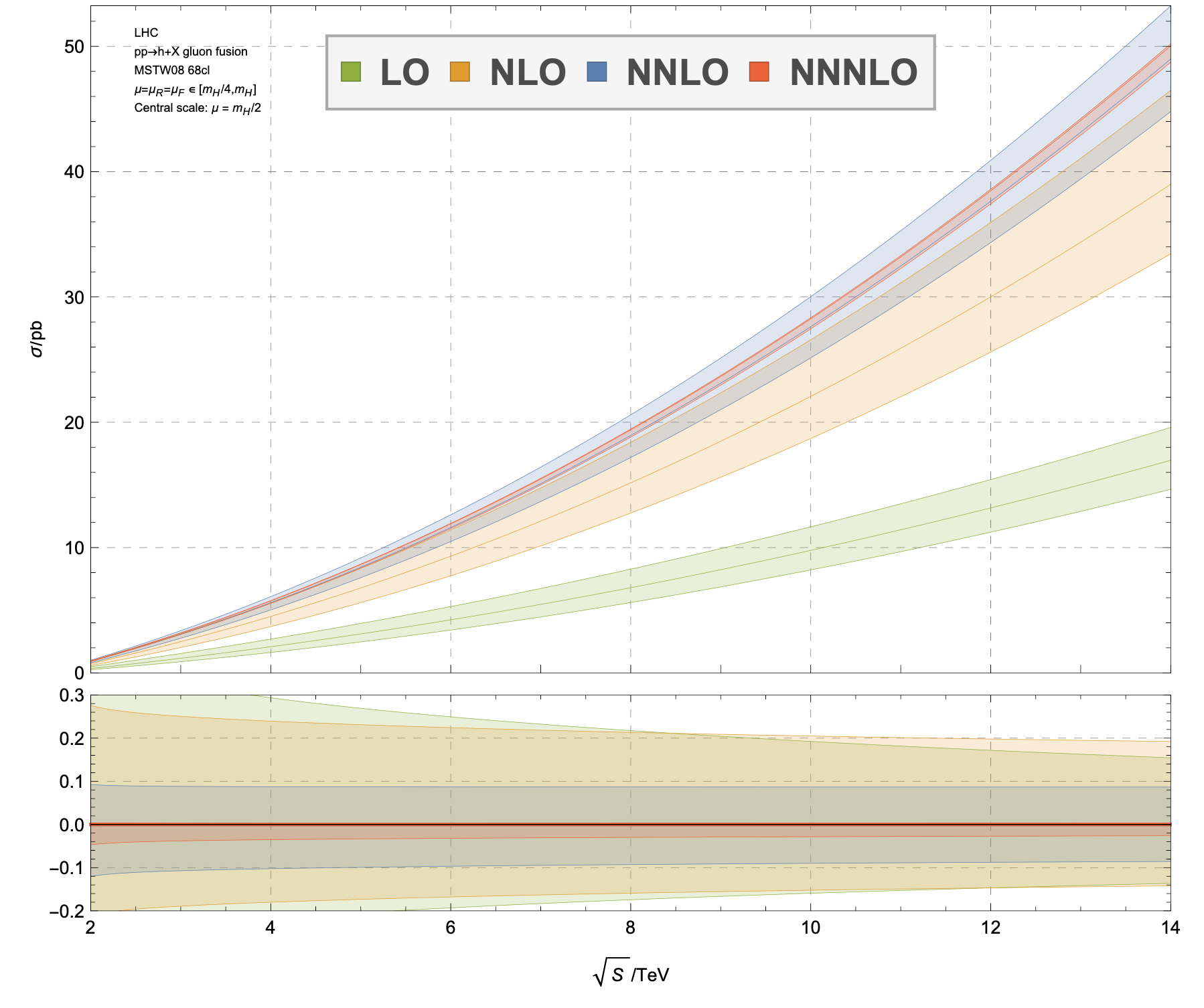}
\caption{Left: Higgs production in gluon fusion, stabilisation of the scale
  dependence at higher perturbative orders, figure from Ref.~\cite{Mistlberger:2018etf}.
  Right: Scale uncertainty bands for Higgs production in gluon
  fusion, figure from
  Ref.~\cite{Anastasiou:2015vya}.}\label{fig:ggH_scalebands}
\end{figure}

Let us now consider an example where such scale variations do not
capture the true uncertainties, and the scale
uncertainty bands obtained from 7-point scale variations do not (fully)
overlap between the different orders. One such example is Higgs
boson production in gluon fusion.
Fig.~\ref{fig:ggH_scalebands} (left) shows that only at very high perturbative order, at N$^3$LO, 
a satisfactory stabilisation of the scale dependence is reached,
and that the higher order corrections are very large.
The  scale uncertainty bands are shown in
Fig.~\ref{fig:ggH_scalebands} (right), where it is obvious that the LO scale variation band would be a very poor measure of the uncertainty due to missing higher orders.
Among the reasons for the large K-factors (i.e. the relative size of the higher order corrections), in particular the NLO
K-factor, are large colour factors and new partonic channels opening up. For the case of inclusive Higgs boson production, the large corrections are also related to the analytic continuation of the gluon form factor to time-like momentum transfer, see Ref.~\cite{Ahrens:2008qu}.
 
Recently, methods utilising Bayesian inference on the known perturbative orders have been suggested to model the size of missing higher orders~\cite{Cacciari:2011ze,Bagnaschi:2014wea,Bonvini:2020xeo,Duhr:2021mfd}. Another method is to obtain an uncertainty estimate for a considered process based on
the scale variations of a set of QCD reference processes~\cite{Ghosh:2022lrf}.
A very recent approach is based on theory nuisance parameters as a way to parametrise unknown higher-order terms, see e.g. Refs.~\cite{Tackmann:2024kci,Lim:2024nsk}, or to use concepts of information theory~\cite{Assi:2025ibi}. 

%% file: 5_state_of_art.tex
In the following, we will give a brief review of the state of the art
in the calculation of perturbative QCD corrections.
We will focus mainly on the calculation of multi-loop
scattering amplitudes and the handling of IR divergences.
For details about resummation and parton showers we refer to  Refs.~\cite{Stagnitto:2025air,Reuter:2025imz}.

\subsection{Multi-loop amplitudes}
\label{sec:multi-loop}

\begin{figure}[t]
\centering
\begin{subfigure}[t]{0.27\linewidth}
    \centering
    \includegraphics[width=\linewidth]{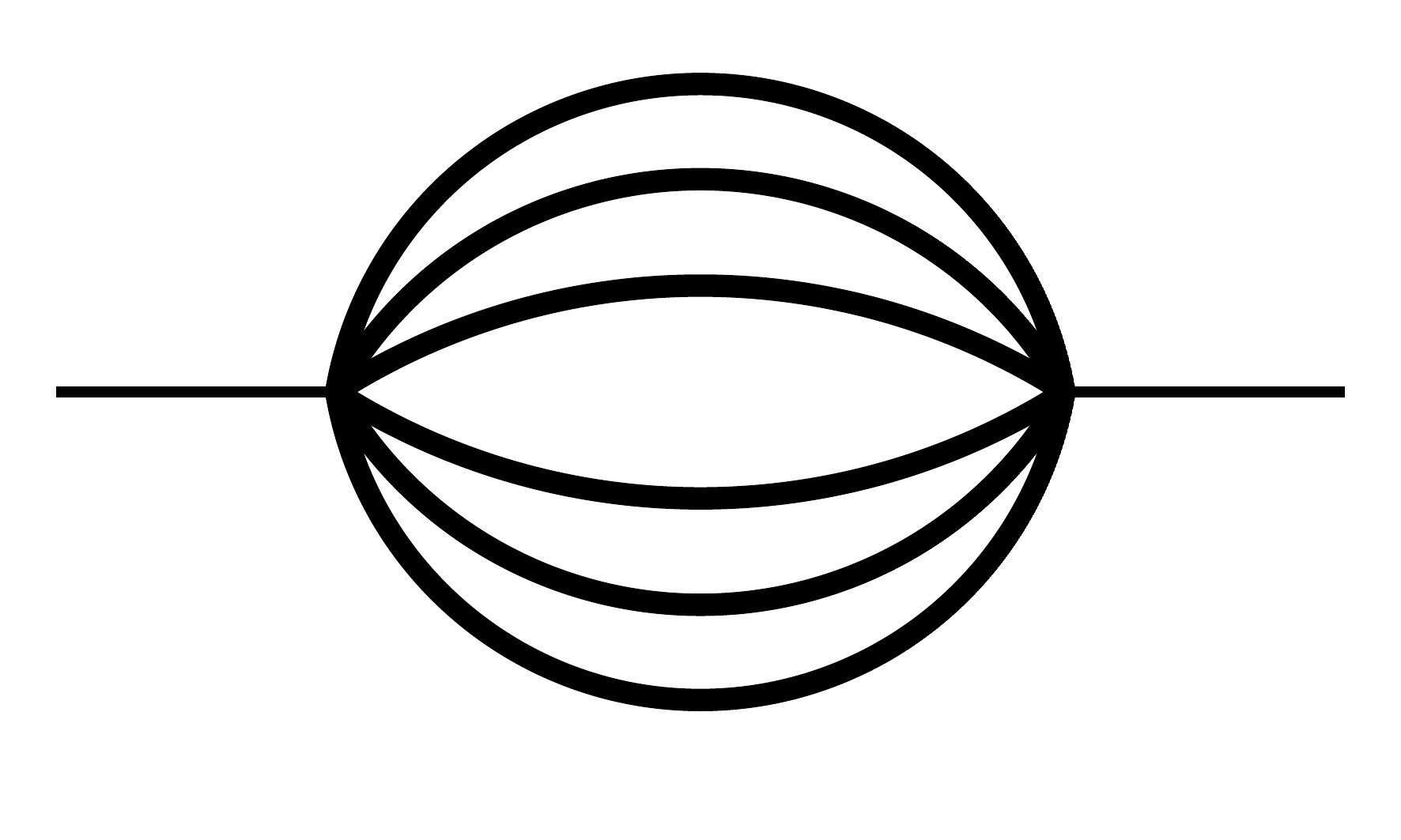}
    \caption{2-point, 5-loop}
\end{subfigure}
\begin{subfigure}[t]{0.28\linewidth}
    \centering
    \includegraphics[width=\linewidth]{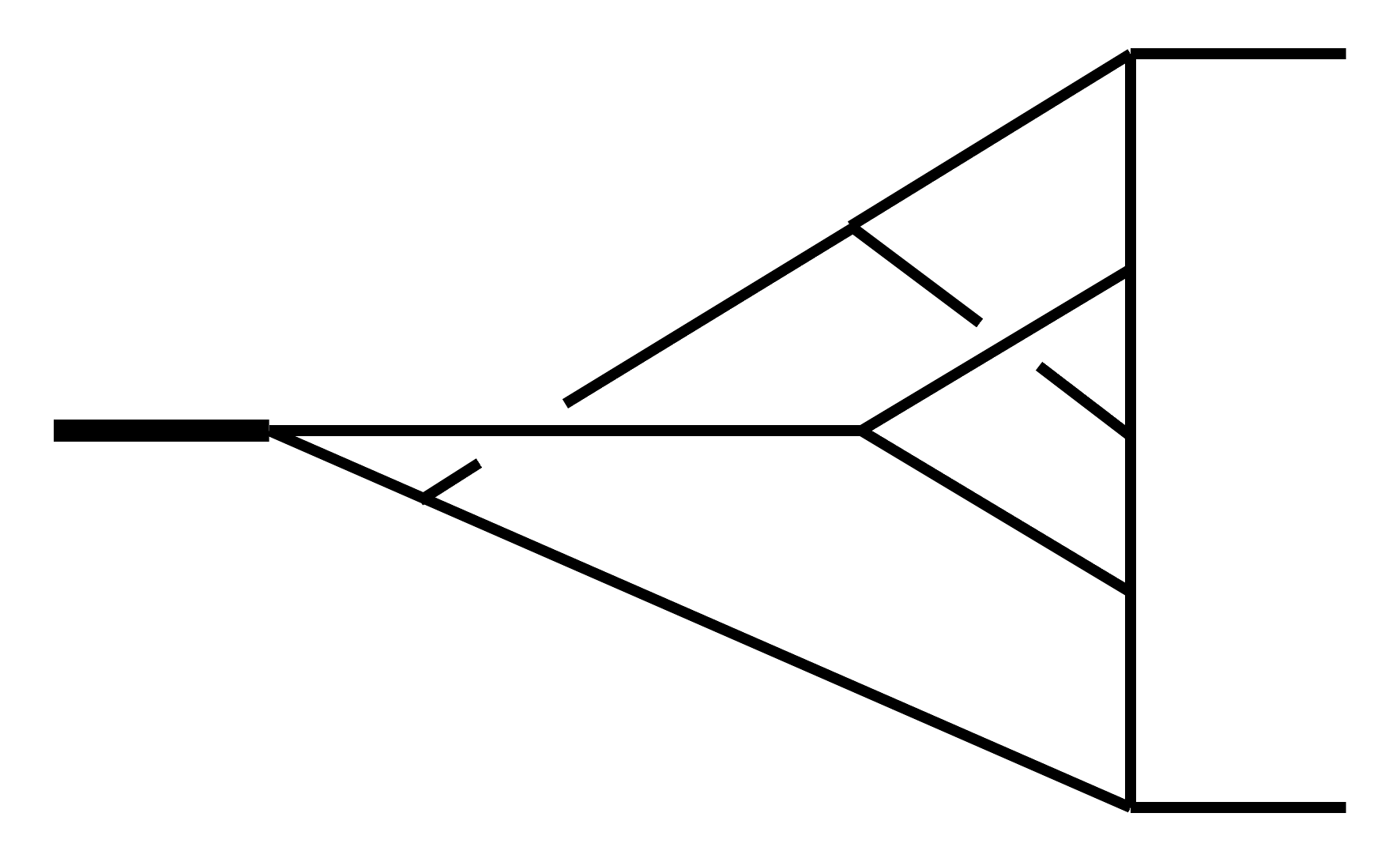}
    \caption{3-point, 4-loop, non-planar}
\end{subfigure}
\begin{subfigure}[t]{0.28\linewidth}
    \centering
    \includegraphics[width=\linewidth]{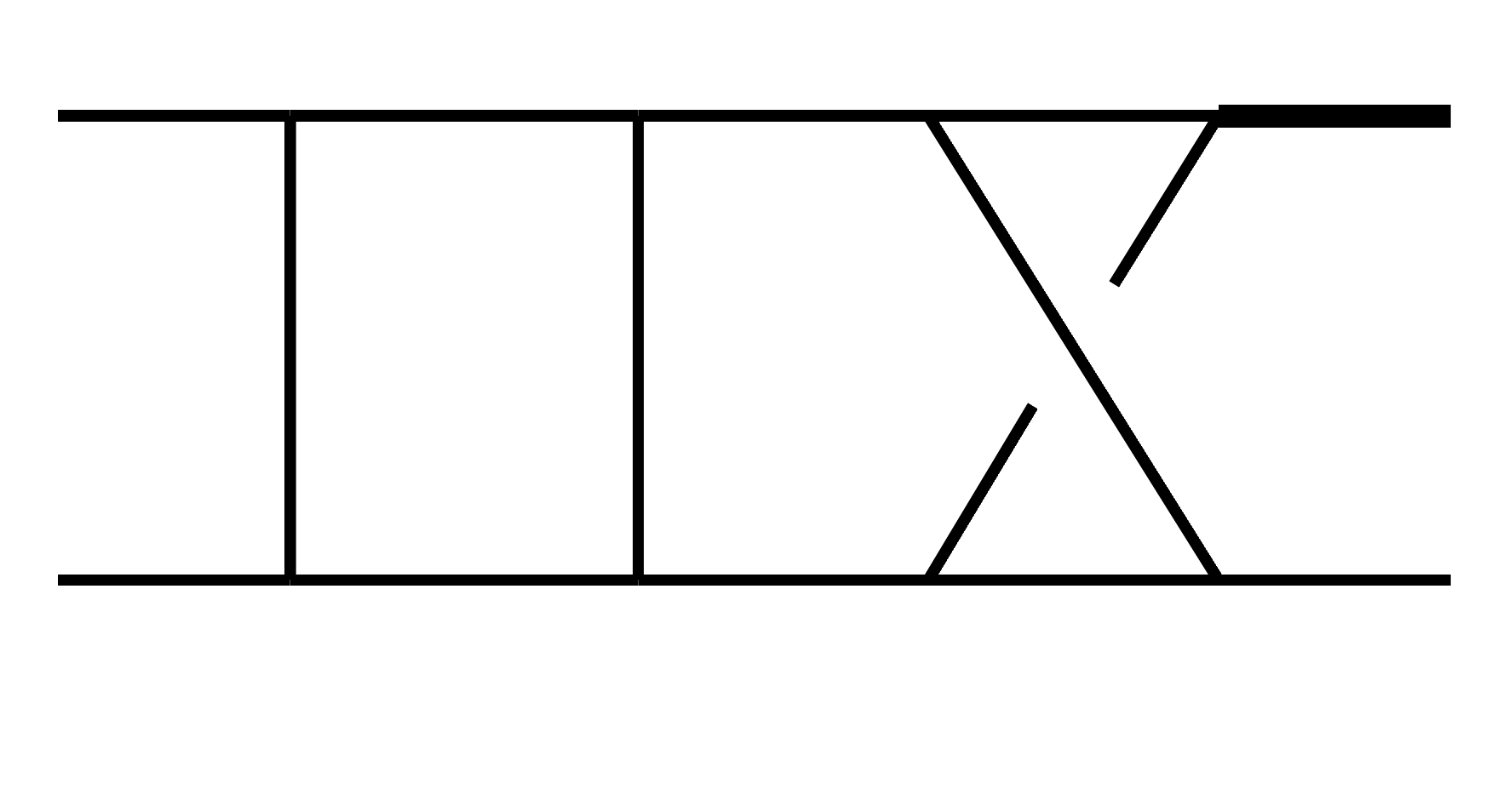}
    \caption{4-point, 3-loop, non-planar}
\end{subfigure}
\begin{subfigure}[t]{0.25\linewidth}
    \centering
    \includegraphics[width=\linewidth]{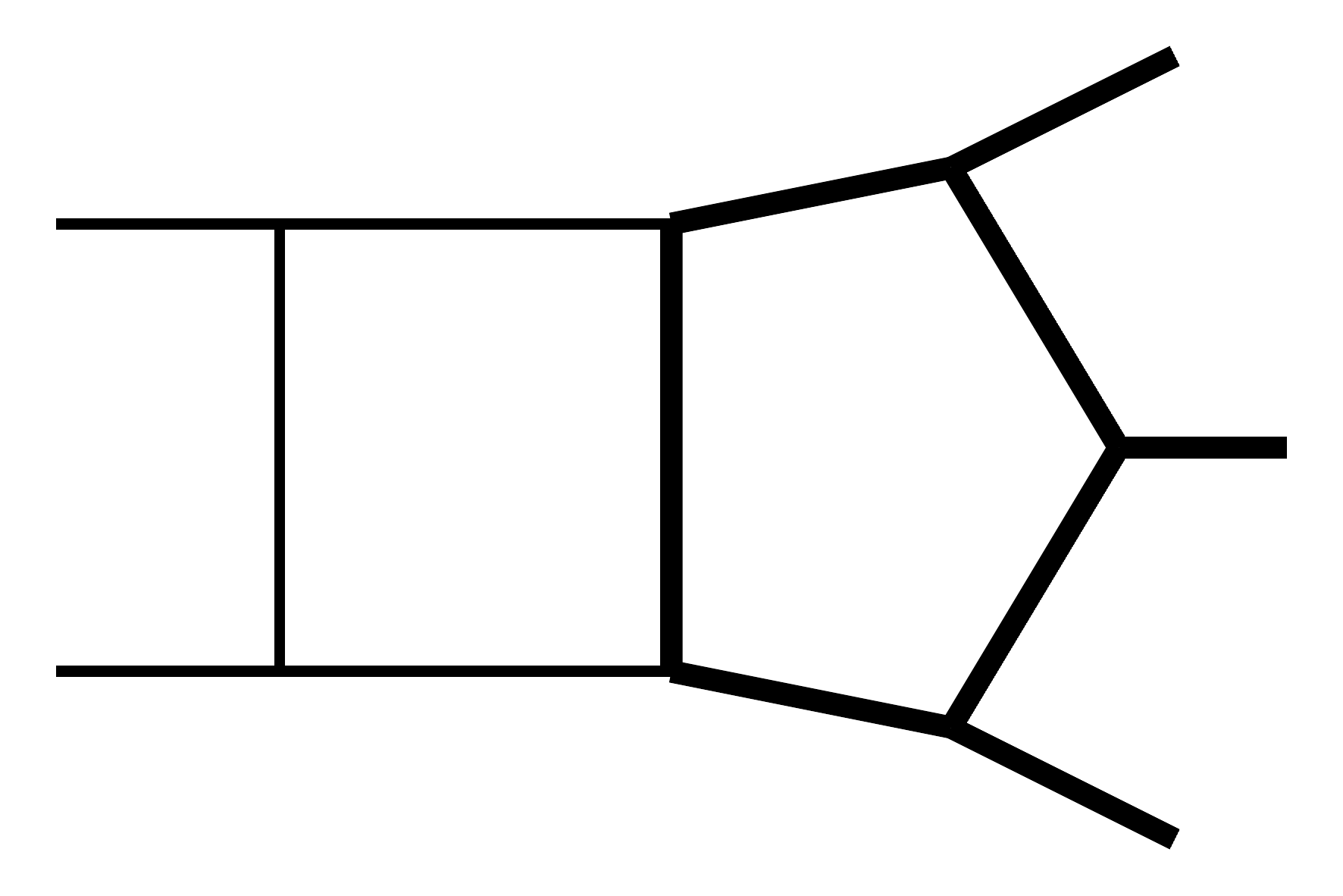}
    \caption{5-point, 2-loop}
\end{subfigure}
\hspace{2mm}
\begin{subfigure}[t]{0.25\linewidth}
    \centering
    \includegraphics[width=\linewidth]{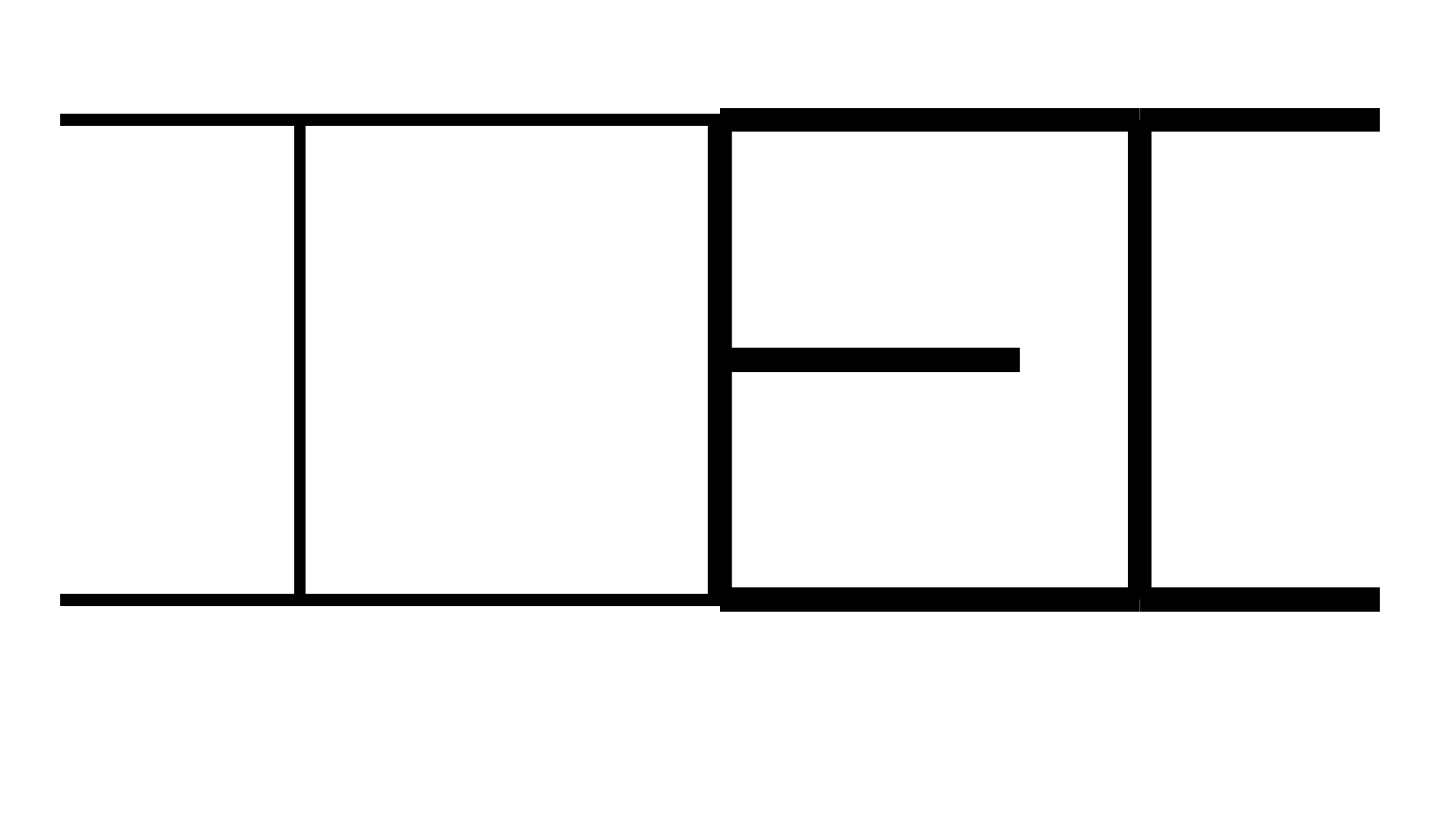}
    \caption{5-point, 2-loop, non-planar}
\end{subfigure}
\hspace{4mm}
\begin{subfigure}[t]{0.28\linewidth}
    \centering
    \includegraphics[width=\linewidth]{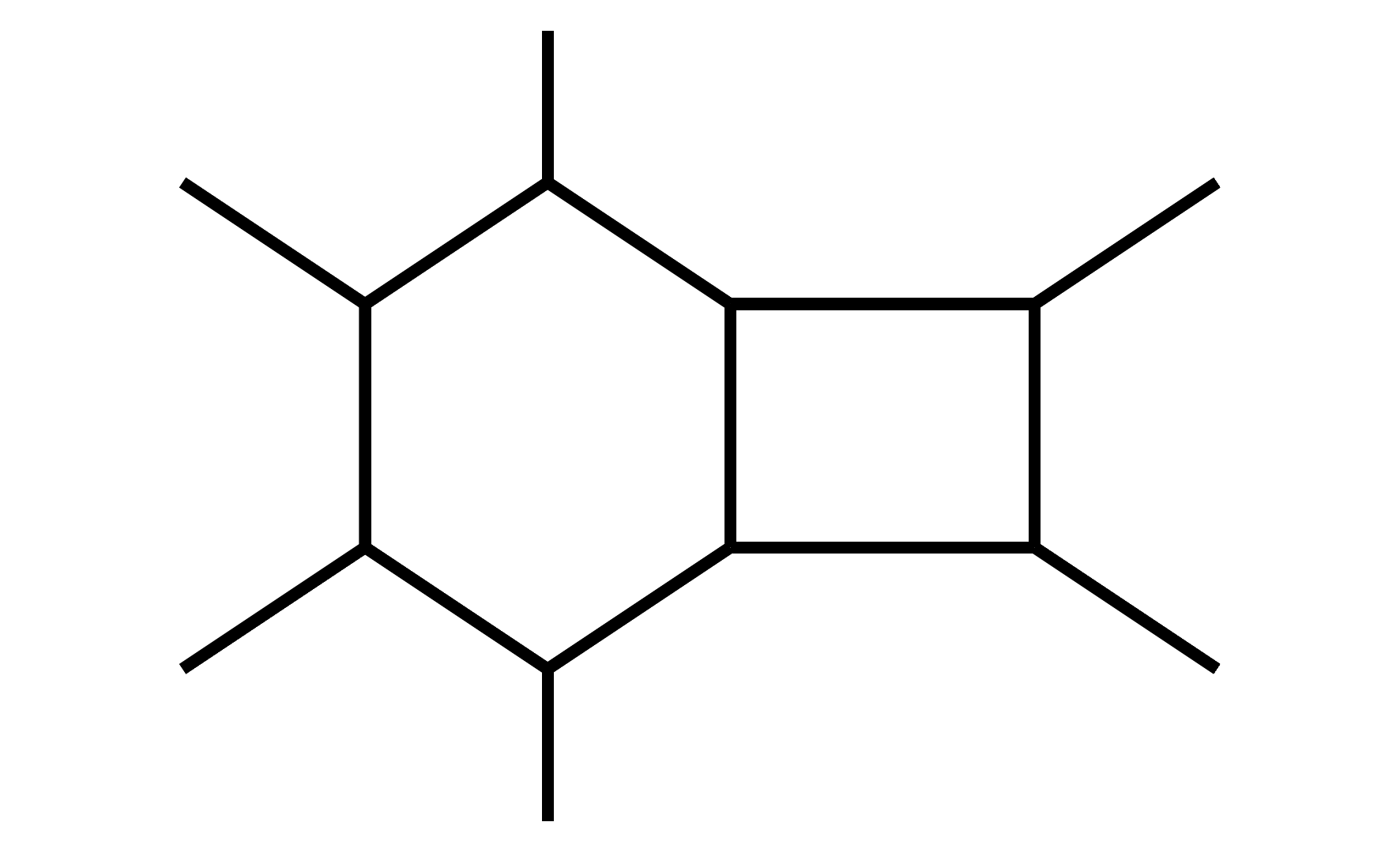}
    \caption{6-point, 2-loop}
\end{subfigure}
\caption{Representative examples of multi-loop topologies. 
A rule of thumb for the state of the art is roughly that $N+L=7$, where $N$ is the number of external legs and $L$ is the number of loops, but the exact status depends strongly on the number of internal masses and massive or off-shell legs (drawn as bold lines). }
\label{fig:2-loop_topologies}
\end{figure}

For loop amplitudes, the complexity is rising not only with the number
of loops, but also with the number of kinematic scales (related to the number of external legs and their virtuality) and mass
scales. Therefore, the current multi-loop frontier with regard to matrix elements for collider phenomenology are amplitudes for $2\to 2$ scattering at 3-loop level with 
one off-shell leg, such as Higgs boson plus jet production in gluon fusion in the heavy top limit~\cite{Gehrmann:2024tds,Chen:2025utl}, see Fig.~\ref{fig:2-loop_topologies} (c) for a representative topology. Some results for 3-loop diagrams with two off-shell legs are also available~\cite{Long:2024bmi,Canko:2024ara,Davies:2025ghl}.
For massless $2\to 2$ scattering, the 3-loop amplitudes for $q\bar{q}\to \gamma\gamma$~\cite{Caola:2020dfu}, $gg\to \gamma\gamma$~\cite{Bargiela:2021wuy},
$q\bar{q}\to q^\prime\bar{q}^\prime$~\cite{Caola:2021rqz}, $q\bar{q}\to gg$~\cite{Caola:2022dfa}, $gg\to gg$~\cite{Jin:2019nya,Caola:2021izf} and photon+jet production~\cite{Bargiela:2022lxz} have been calculated.

Other landmarks at 3-loop level are e.g. the calculation of 3-loop splitting functions~\cite{Vogt:2004mw,Ablinger:2014nga,Ablinger:2017tan,Blumlein:2024tjw}, gluon fusion Higgs boson production in the heavy top limit at N$^3$LO~\cite{Anastasiou:2015vya,Dulat:2018bfe,Mistlberger:2018etf,Cieri:2018oms}, also at the level of fiducial cross sections~\cite{Chen:2021isd,Billis:2021ecs},  Higgs boson production at NNLO with full top quark mass dependence~\cite{Czakon:2021yub}, see also~\cite{Czakon:2023kqm,Czakon:2024ywb,Niggetiedt:2023uyk}, or vector boson production (Drell-Yan process) at N$^3$LO~\cite{Duhr:2020sdp,Duhr:2021vwj,Chen:2021vtu,Chen:2022lwc}.
Higgs boson pair production in gluon fusion has been calculated at N$^3$LO in the heavy top limit~\cite{Chen:2019lzz,Chen:2019fhs,AH:2022elh} and in vector boson fusion in the structure function approach~\cite{Dreyer:2018qbw}. These calculations reduce the scale uncertainties typically to the level of a few percent, such that other uncertainties, such as the PDF+$\als$ uncertainties, or uncertainties related to the treatment of the heavy quark masses, or missing higher-order electroweak corrections become dominant.
For more details see e.g. Refs.~\cite{Caola:2022ayt,Huss:2025nlt}.

At the two-loop frontier, pentagon amplitudes with both massive propagators and massive final-state particles are the main challenge, where (partial) results exist  for $pp\to t\bar{t}j$~\cite{Badger:2024dxo}, $pp\to t\bar{t}W$~\cite{Becchetti:2025qlu,Buonocore:2023ljm}, $pp\to b\bar{b}W$~\cite{Buonocore:2022pqq}, $pp\to Zb\bar{b}$~\cite{Mazzitelli:2024ura}, $pp\to b\bar{b}H$~\cite{Badger:2024mir} and $pp\to t\bar{t}H$~\cite{FebresCordero:2023pww,Wang:2024pmv,Agarwal:2024jyq,Devoto:2024nhl}. Example diagrams are shown in Fig.~\ref{fig:2-loop_topologies} (d) and (e). The availability of 2-loop pentagon functions for the massless case~\cite{Chicherin:2020oor} and for the case with one off-shell leg~\cite{Chicherin:2021dyp,Abreu:2023rco} have driven developments such as the flagship results for 3-jet production~\cite{Czakon:2021mjy}, $Wb\bar{b}$ production~\cite{Badger:2021nhg,Hartanto:2022qhh} and $Zb\bar{b}$ production~\cite{Mazzitelli:2024ura} at NNLO in hadronic collisions,  or the 
analytic results for all massless 2-loop five-point helicity amplitudes, including all colour structures~\cite{Agarwal:2023suw,DeLaurentis:2023nss,DeLaurentis:2023izi} and for $Vjj$ production including leptonic decays of the vector boson~\cite{DeLaurentis:2025dxw}.
Results for massless 2-loop 6-point amplitudes (Fig.~\ref{fig:2-loop_topologies} (f)) are also starting to emerge~\cite{Abreu:2024fei,Henn:2025xrc}.

Beyond three loops, the available results are mostly based on four-loop three-point or five-loop two-point integrals, see topologies (b) and (a) in Fig.~\ref{fig:2-loop_topologies}, respectively. 
Four-loop results are e.g. ingredients for PDF evolution, such as contributions to four-loop splitting functions, structure functions or anomalous dimensions~\cite{Agarwal:2020nyc,Agarwal:2021him,Agarwal:2023yos,Guan:2024hlf,Falcioni:2024qpd,Kniehl:2025ttz,Goyal:2025bzf,Kniehl:2025jfs},  quark and gluon form factors~\cite{Chetyrkin:1997iv} entering e.g. Higgs production in gluon fusion in the soft-virtual approximation~\cite{Das:2020adl,Lee:2022nhh,Chakraborty:2022yan}, or heavy quark matching coefficients and contributions to $B$-meson decays~\cite{Fael:2022frj,Egner:2024azu}.
Examples for five-loop results are calculations for the beta function~\cite{Baikov:2016tgj,Luthe:2016ima,Herzog:2017ohr,Luthe:2017ttg,Chetyrkin:2017bjc},
ingredients for anomalous dimensions~\cite{Herzog:2018kwj,Mishra:2023acr}, or contributions to the electron anomalous magnetic moment~\cite{Volkov:2024yzc,Aoyama:2024aly}.
For a recent overview about five-loop results in perturbative QCD we refer to Ref.~\cite{Maier:2024rng}.

\subsection{Infrared subtraction schemes beyond NLO}
\label{sec:realNNLO}

\begin{figure}[h]
\centering
\includegraphics[width=0.6\textwidth]{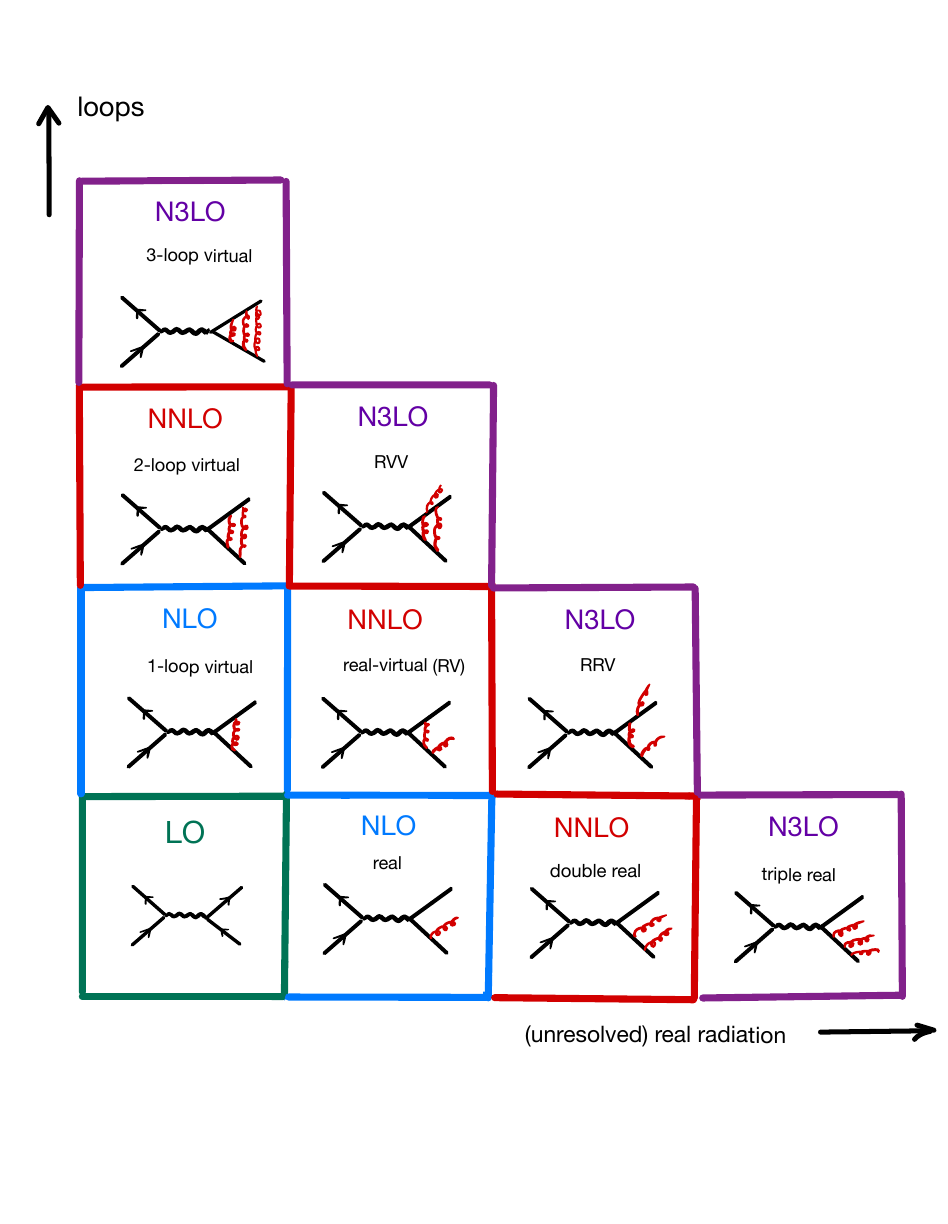}
\caption{ Schematic building blocks of an N$^x$LO calculation. The higher order diagrams are only representatives of their class, the number of diagrams grows rapidly with the perturbative order.}
\label{fig:anatomy_higher_orders}
\end{figure}

According to the KLN theorem, IR singularities due to soft radiation and final state collinear radiation must cancel in inclusive cross sections. However, in order to produce fully differential results, and in the presence of kinematic cuts, the integrands describing the radiation of extra partons (i.e. extra relative to the Born kinematics) need to be rendered finite before carrying out the phase-space integration. How to do this at NLO, where only one extra parton can be unresolved, has been described in Section~\ref{sec:IR}. Beyond NLO, the structure is more involved because at N$^x$LO, up to $x$ partons can become unresolved. This is illustrated in Fig.~\ref{fig:anatomy_higher_orders}.

The schemes that have been devised to treat unresolved real radiation at NNLO can be broadly divided into two categories, which are often called ``subtraction'' and ``slicing''. In the former category, expressions describing the amplitude  in singular limits are subtracted (mostly locally in phase space), and added back in a form where the integration over the unresolved phase space has been carried out in dimensional regularisation, such that the IR poles become manifest and can be cancelled against other explicit poles.
The main subtraction methods are antenna subtraction~\cite{GehrmannDeRidder:2005cm,Currie:2013vh}, as used in the programs {\sc NnloJet}~\cite{NNLOJET:2025rno} and {\sc Eerad3}~\cite{Gehrmann-DeRidder:2014hxk,Aveleira:2025svg}, ColorFul subtraction~\cite{DelDuca:2015zqa,DelDuca:2024ovc}
sector-improved residue subtraction~\cite{Czakon:2010td,Czakon:2011ve,Boughezal:2011jf,Czakon:2014oma},
Nested soft--collinear subtraction~\cite{Caola:2017dug,Caola:2018pxp,Delto:2019asp,Devoto:2023rpv} and local analytic sector subtraction~\cite{Magnea:2018hab, Magnea:2018ebr,Magnea:2020trj,Chargeishvili:2024xuc,Bertolotti:2025clg}.

Slicing methods partition the phase space into regions based on a slicing parameter (such as transverse momentum $q_T$~\cite{Catani:2007vq,Catani:2019hip} or $N$-jettiness~\cite{Stewart:2010tn, Boughezal:2015dva,Gaunt:2015pea}). The public NNLO code library {\sc Matrix}~\cite{Grazzini:2017mhc} is based on $q_T$-subtraction, the NNLO codes contained in the library {\sc Mcfm}~\cite{Campbell:2019dru,Campbell:2022gdq} are mainly based on $N$-jettiness. Similarly for the code {\sc Geneva}~\cite{Alioli:2025hpa}, which in addition contains parton shower matching.

The slicing parameter divides the space into resolved (hard) and unresolved (soft/collinear) regions. Therefore, slicing methods are based on non-local subtraction: instead of subtracting IR singularities point-by-point in phase space, slicing removes entire regions of phase space that contain singularities, making it fundamentally non-local.
Integrals below this cutoff can be computed using resummation techniques or Soft-Collinear Effective Theory (SCET), exploiting the universal behaviour of IR singularities in QCD, while those above can be treated with methods known from lower orders in perturbation theory (usually NLO). This possibility to ``recycle'' known elements is a great advantage of this method. The N$^3$LO calculations mentioned above are all based on slicing methods.
However, extensions of local subtraction methods to N$^3$LO are also under construction~\cite{Chen:2025ojp,Chen:2025kez,Magnea:2024jqg}.

The slicing parameter $\tau_{\rm{cut}}$ acts as an infrared cutoff, which needs to be relatively small. This introduces large cancellations between logarithms of $\tau_{\rm{cut}}$. The results are only accurate up to corrections suppressed by powers of the slicing parameter, so-called ``(perturbative) power corrections''.
Therefore, controlling the power corrections is important to improve the reliability and numerical convergence of this method~\cite{Moult:2016fqy,Boughezal:2016zws,Moult:2017jsg,Boughezal:2018mvf,Ebert:2018lzn,Cieri:2019tfv,Agarwal:2025dvo}.

The ``projection-to-Born"~\cite{Cacciari:2015jma,Dreyer:2016oyx,Chen:2021isd} method is particularly suited for processes where the remapping of the unresolved momenta does not affect the produced boson(s).
It can also be used to improve the stability of slicing methods~\cite{Campbell:2024hjq}.

Reviews about recent developments in IR subtraction schemes can be
found e.g. in Refs.~\cite{Huss:2025nlt,Caola:2022ayt}.
  

\subsection{Beyond fixed order in perturbation theory}
\label{sec:resummation}

There are kinematic regions that are poorly described by fixed-order QCD. This is typically the case when large logarithms arise, due to phase-space constraints or disparate kinematic scales. If $\alpha_s$ is accompanied by large logarithms, the perturbative series in $\alpha_s$ no longer converges.

For example, near partonic thresholds, i.e. when the final state is produced near the minimal available energy, the phase space for soft-gluon emissions is severely restricted, which leads to large logarithms $\sim \ln(1-\frac{Q^2}{s})$, where $Q^2$ is the invariant mass squared or virtuality of the produced particle and $\sqrt{s}$ is the available energy. Similarly, $Z$-boson production in hadronic collisions leads to large logarithms of the form $\ln\left(\frac{p_T^2}{M_Z^2}\right)$ since the $Z$-boson gets its transverse momentum $p_T$ from recoil against soft gluons. At fixed order, the limit $p_T\to 0$ is divergent.
The cross section differential in $p_T$ at order $\alpha_s$ can schematically be written as
\begin{align}
\frac{d\sigma^{\rm{NLO}}}{d p_T}&=c_0^0\,\delta(p_T)+\alpha_s\left(c_0^1\,\delta(p_T)+c_1\frac{1}{p_T}+c_2\frac{\ln(p_T)}{p_T}\right)\;,
\end{align}
which is divergent for $p_T\to 0$. However, as the pattern of soft gluon radiation in QCD is known and factorises to all orders, the radiation of $n$ soft gluons can be summed to all orders~\cite{Catani:1989ne,Dixon:2008gr}. After integration over the soft phase space it 
 leads to the series representation of an exponential function, such that the resummed expression has the schematic form
\begin{align}
\frac{d\sigma^{\rm{resum}}}{d p_T}&=c_0\exp{\left[-\alpha_sc_2\ln^2(p_T)+\ldots\right]}\;.
\end{align}
 The exponential factor is called {\em Sudakov} factor~\cite{Sudakov:1954sw,Catani:1989ne}.
 It also forms the basis of parton showers. 

For an observable $R$ normalised to its Born level, the perturbative series, which usually has the form
\begin{align}
R&=1+\als (L^2+L+1)+\als^2 (L^4+L^3+L^2+L+1)+\ldots\;,
\end{align}
where $L$ is a large logarithm, can be re-organised as
\begin{align}
R&=1+C(\als)\,\exp{\left[ \sum_na_n\als^n L^{n+1}+\sum_nb_n\als^n L^{n}+\sum_nc_n\als^n L^{n-1}+ +\ldots\right]}\;.
\end{align}
Keeping only the first term $\sim \als^n L^{n+1}$ is called ``leading log (LL)'' resummation, keeping the first and the second is called ``next-to-leading log (NLL)'' resummation, and so on.

Ideally, resummed calculations are matched to fixed-order calculations, such that all kinematic regions are described well.
To achieve this, the fixed-order and the resummed results are added and then the resummed result, expanded to the order in $\als$ of the fixed-order calculation, is subtracted to avoid double counting, for example, at NLO:
\begin{align}
R^{\rm matched}=R^{\rm{NLO}}+R^{\rm{resum}}-R^{\rm{resum}}\Big|_{\rm{expanded \,to \,}{\cal O}(\als)}\;.
\end{align}
For more details on resummation we refer to the chapter by {\it G.~Stagnitto}~\cite{Stagnitto:2025air}.
 